\RequirePackage{fix-cm}

\RequirePackage{amsmath}
\documentclass[smallextended]{svjour3}       
\smartqed  
\usepackage[title]{appendix}
\usepackage{graphicx}
\usepackage{chngcntr}
\usepackage{amssymb}
\usepackage{amsmath}

\usepackage{times}
\usepackage{graphics}
\usepackage{subfig}
\usepackage{romannum}
\usepackage{verbatim}
\usepackage{epstopdf}
\usepackage{amsfonts}
\usepackage{pifont}
\usepackage{pgf}
\usepackage{pgffor}
\usepackage{subfig}
\usepackage{hyperref}
\usepackage{enumerate}
\usepackage{breqn}
\usepackage{tikz}
\usepackage{multirow}
\graphicspath{ {./Figures/} }

\usepackage{nomencl}
\usepackage{amssymb}

\usepackage{wrapfig}
\usepackage{lscape}
\usepackage{rotating}
\usepackage{adjustbox}
\usepackage{afterpage}

\newcommand{\Sp}{\mathbb{S}}
\usepackage{silence}
\usepackage{caption}
\usepackage{blindtext}
\journalname{Theor. Comput. Fluid Dyn.}
\begin{document}

\title{Model-based multi-sensor fusion for reconstructing  wall-bounded turbulence
}



\titlerunning{Model-based multi-sensor fusion for turbulence reconstruction}        

\author{Mengying Wang         \and
         C. Vamsi Krishna       \and \\
         Mitul Luhar        \and
         Maziar S. Hemati
}


\institute{M. Wang \at
              Aerospace Engineering and Mechanics, University of Minnesota, Minneapolis, MN 55455  \\
              \email{wang5772@umn.edu}           
           \and
           C.V. Krishna \at
           Aerospace and Mechanical Engineering, University of Southern California, Los Angeles, CA 90089 \\
           \email{vchinta@usc.edu}
            \and
            M. Luhar \at
            Aerospace and Mechanical Engineering, University of Southern California, Los Angeles, CA 90089 \\
           \email{luhar@usc.edu}
            \and
            M.~S. Hemati \at
              Aerospace Engineering and Mechanics, University of Minnesota, Minneapolis, MN 55455 \\
              Tel.: +612-625-6857\\
              Fax: +612-626-1558\\
              \email{mhemati@umn.edu}
}

\date{Received: date / Accepted: date}

\maketitle

\begin{abstract}
Wall-bounded turbulent flows can be challenging to measure within experiments due to the breadth of spatial and temporal scales inherent in such flows. Instrumentation capable of obtaining time-resolved data (e.g., Hot-Wire Anemometers) tends to be restricted to spatially-localized point measurements; likewise, instrumentation capable of achieving spatially-resolved field measurements (e.g., Particle Image Velocimetry) tends to lack the sampling rates needed to attain time-resolution in many such flows. In this study, we propose to fuse measurements from multi-rate and multi-fidelity sensors with predictions from a  physics-based model to reconstruct the spatiotemporal evolution of a wall-bounded turbulent flow. A ``fast'' filter is formulated to assimilate high-rate point measurements with estimates from a linear model derived from the Navier-Stokes equations. Additionally, a ``slow'' filter is used to update the reconstruction every time a new field measurement becomes available. By marching through the data both forward and backward in time, we are able to reconstruct the turbulent flow with greater spatiotemporal resolution than either sensing modality alone. We demonstrate the approach using direct numerical simulations of a turbulent channel flow from the Johns Hopkins Turbulence Database. A statistical analysis of the model-based multi-sensor fusion approach is also conducted.
\end{abstract}

\section{Introduction}

An ability to predict and control turbulence would benefit numerous engineering and scientific applications.
Yet, such efforts are often compromised by the sensing and diagnostic requirements of resolving 
the breadth of spatiotemporal scales underlying turbulence.
Instruments capable of resolving the fastest velocity fluctuations in time---e.g.,~Hot-Wire Anemometers~(HWA), 
Laser-Doppler Anemometers~(LDA)---are often restricted to localized measurements at a single point in space.
On the other hand, Particle Image Velocimetry~(PIV) systems capable of acquiring spatially-resolved field 
measurements tend to lack the temporal resolution required for higher Reynolds number flows.
In this study, we formulate a framework for reconstructing the spatiotemporal evolution of turbulent flows by
fusing multi-rate and multi-fidelity measurements from these different sensing modalities 
together with predictions from simple physics-based models.

The notion of flow estimation and reconstruction from available measurements is not a new one.
%
%
Numerous efforts have considered linear stochastic estimation~(LSE) to extract coherent structures from various measurement sources~\cite{adrian1979conditional,adrian1994stochastic,guezennec1989stochastic}.
For example, LSE has been used in the investigation of wall-bounded turbulence and cavity flow using wall-based pressure and shear stress measurements~\cite{encinar2019logarithmic,murray2003estimation,naguib2001stochastic}.
A Proper Orthogonal Decomposition~(POD)-based LSE has been utilized to address challenges of multi-scale, nonlinear, and higher order estimation~\cite{bonnet1994stochastic,mokhasi2009predictive,nguyen2010proper}.
Time-delay variants of the LSE-POD approaches have been investigated for real-time flow estimation ~\cite{durgesh2010multi} and flow control~\cite{taylor2004towards,ukeiley2008dynamic}.
Recently, LSE and spectral POD algorithms have been combined into a unified framework that enables modal analysis of fluid flows from non-time-resolved PIV data in conjunction with unsteady pressure measurements~\cite{zhang2020}.
Similarly, modal analysis based on the resolvent framework has been shown to be suitable for estimating space-time statistics of flows~\cite{towne2020resolvent} and also useful for estimation of turbulent flows from wall-measurements~\cite{amaral2020}.

Some recent investigations have considered machine learning methods and super-resolution data reconstruction techniques for reproducing turbulent flows~\cite{fukami2019,fukami2021machine,jiang2020}.
Once trained on high-quality data, these methods have been shown capable of reproducing the underlying flow field using remarkably coarse measurements.
Previous studies have also considered purely statistical data fusion methods for flow reconstruction.
In~\cite{van2015bayesian}, a ``model-free'' maximum a posteriori~(MAP) algorithm was proposed for fusing low-temporal-high-spatial resolution data 
with high-temporal-low-spatial resolution data for turbulent flow reconstruction.
However, it is important to note that this work did not leverage a model to perform dynamic estimation.
Indeed, model-based dynamic estimation can improve reconstruction performance because it offers an additional information source in the form of model predictions.

Dynamic estimation has been investigated in prior works on flow reconstruction, especially in the context of band-limited oscillator-type flows.
Tu et al.~\cite{tu2013integration} established a dynamic estimator to reconstruct the wake of a thick flat plate at low Reynolds numbers.
This was done using a data-driven model of the flow in conjunction with a Kalman smoother that was used to fuse model predictions with non-time-resolved PIV and time-resolved point-sensor measurements.
Recent efforts have focused on using physics-based models grounded in the resolvent formalism to predict and reconstruct flows~\cite{mckeon2010critical,illingworth2018estimating,towne2020resolvent}.
The use of low-complexity physics-based models for flow reconstruction has the benefit of ensuring that the estimated flow satisfies physical constraints (e.g., conservation of mass and momentum).
Recent studies have also shown that it is possible to perform a data-driven refinement of the linearized Navier-Stokes equations~(NSE) to account for second-order statistics~\cite{jovanovic2005componentwise,zare2017colour}.

In previous work~\cite{krishna2019fusion,krishna2020reconstructing}, we have shown that even simple linear time-invariant models can be used to reconstruct the spatiotemporal evolution of wall-bounded turbulence from non-time-resolved 2-dimensional 2-component~(2D2C) field measurements, as would be obtained from a PIV system.
For example, linear models grounded in Rapid Distortion Theory~(RDT)~\cite{hunt1990rapid,savill1987recent} 
are capable of predicting the flow evolution over a time period much shorter than a typical eddy turnover time.
This is a consequence of the fact that persistent features of the local eddy structure change slowly over the time of distortion, making higher-order nonlinear effects negligible over a fast time-scale~\cite{hunt1990rapid}.
As a result, we found that it is possible to use linear RDT models to reconstruct the time evolution of wall-bounded turbulent flows between snapshots of a non-time-resolved PIV system~\cite{krishna2019fusion,krishna2020reconstructing}. 
The model-based flow reconstruction was further improved by propagating the snapshots both forward and backward in time, then using an appropriately designed spatiotemporal weighting scheme to fuse the model predictions.
%

In this paper, we build upon our previous study on model-based turbulent flow reconstruction with an aim to improve performance even further.
In particular, we investigate the potential for improving the quality of turbulent flow reconstruction by introducing additional time-resolved measurements at distinct spatial locations in the flow field.
To do so, we propose a model-based multi-sensor fusion approach for wall-bounded turbulence reconstruction.
The framework fuses model-predictions with time-resolved point measurements using a ``fast filter'' that can be applied both forward and backward in time.
Predictions from the fast filter are fused with non-time-resolved field measurements on a slower time scale, corresponding to the sampling rate of the slow measurements.
We show that both filtering tasks can be conducted with conventional Kalman filtering algorithms formulated in conjunction with the RDT-based dynamics. This also serves to improve the flow reconstruction based on noisy measurements. 
The model-based multi-sensor fusion approach is investigated using DNS data of a turbulent channel flow from the Johns Hopkins Turbulence Database~(JHTDB)~\cite{graham2016web}.
In order to fully evaluate the approach, we also investigate the role of (1)~placement of the ``fast'' point sensors, and (2)~design of the weighting function used for the forward-backward fusion on the resulting reconstruction.

The paper proceeds as follows: In Section~\ref{Sec:Method}, we present the ingredients needed for the reconstruction algorithm: the RDT-based dynamics model, the sensor measurement models, and the forward-backward filtering algorithms and corresponding weighting functions for data fusion. Section~\ref{Sec:DataEva} describes the benchmark dataset and the metrics that will be used to evaluate performance of the reconstruction. Section~\ref{Sec:discussion} 
reports the results of Monte Carlo simulations for different fusion approaches and sensor placement strategies.  
Finally, we provide concluding remarks of our study in Section~\ref{Sec:conclusion}.


\section{Methods} \label{Sec:Method}
\subsection{Model Formulation}\label{Sec:RDT model}

\begin{figure}
    \centering
    \includegraphics[width = 0.8\textwidth]{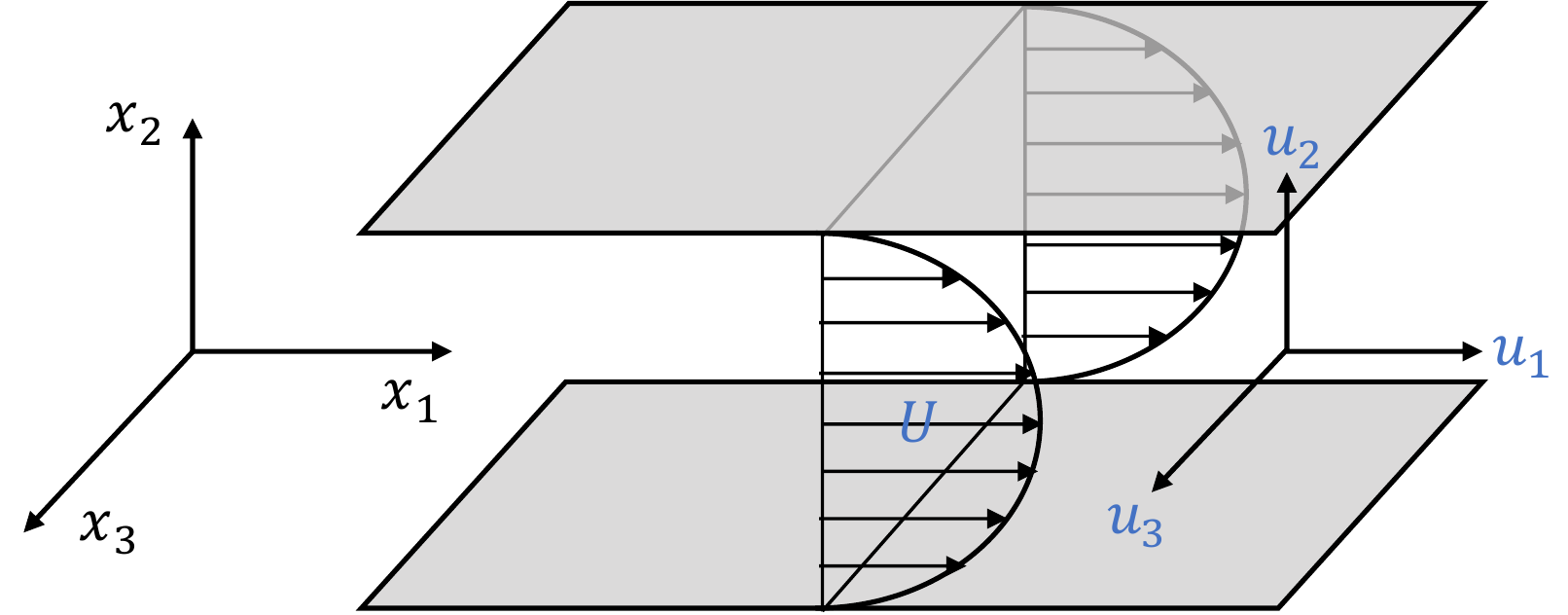}
    \caption{Geometry of the three dimensional wall-bounded turbulent channel flow.}
    \label{fig: 3Dflow}
\end{figure}

An essential element in reconstructing the turbulent flow between two consecutive non-time-resolved snapshots is a model for approximating the evolution of the flow.
%
%
Here, we use a linear model grounded in Rapid Distortion Theory~(RDT), which is based on the assumption that nonlinear flow interactions can be neglected on time-scales that are much shorter than a typical eddy turnover time~\cite{hunt1990rapid}.
%
This allows the equations of motion to be linearized for predictions of the flow evolution over short time-horizons.
%
Beginning with the incompressible Navier-Stokes equations~(NSE) and considering the dynamics of turbulent velocity fluctuations $\boldsymbol{u} = (u_1,u_2,u_3)$ about a mean flow $\boldsymbol{U} = (U(x_2),0,0)$---see Fig.~\ref{fig: 3Dflow}---we have
\begin{align}
  \frac{\partial \boldsymbol{u}}{\partial t}+\boldsymbol{U}\cdot \nabla \boldsymbol{u}+\boldsymbol{u}\cdot \nabla \boldsymbol{U} &=- \nabla p+\frac{1}{\textrm{Re}_\tau} {\nabla}^2\boldsymbol{u}+ (\textrm{NL})\\
    \nabla \cdot \boldsymbol{u} &= 0
\end{align}
where $p$ represents pressure fluctuations and $\textrm{Re}_\tau$ denotes the friction Reynolds number. 
By the assumptions of RDT, the nonlinear term NL can be neglected when the sampling period of the slow measurement system is less than a typical eddy turnover time~\cite{hunt1990rapid}. 
For additional simplicity, we also relax the incompressibility constraint.
Since most commonly used PIV systems can only capture planar two-dimensional two-component snapshots, we further simplify this linear model by neglecting the out-of-plane flow and  pressure gradient terms. Then, the simplified two-dimensional flow model will be:
\begin{equation}
    \begin{aligned}
        \underbrace{\frac{\partial u_1}{\partial t} +  U\frac{\partial u_1}{\partial x_1}}_{\text{advection}} &= \underbrace{\frac{1}{\textrm{Re}_\tau}\left(\frac{\partial^2 u_1}{\partial x_1^2} + \frac{\partial^2 u_1}{\partial x_2^2} \right)}_{\text{diffusion}} -  \underbrace{u_2\frac{\partial U}{\partial x_2}}_{\text{coupling}},\\
        \underbrace{\frac{\partial u_2}{\partial t} +  U\frac{\partial u_2}{\partial x_1}}_{\text{advection}} &= \underbrace{\frac{1}{\textrm{Re}_\tau}\left(\frac{\partial^2 u_2}{\partial x_1^2} + \frac{\partial^2 u_2}{\partial x_2^2} \right)}_{\text{diffusion}}.
    \end{aligned}
    \label{eqn:RDT}
\end{equation}
For the numerical simulation, Eq.~\ref{eqn:RDT} are numerically integrated in time using an explicit Euler method. A first-order upwinding scheme is used for the convection terms, and a second-order central differencing scheme is used for the diffusion and coupling terms.

This simplified model can be used to predict the evolution of the flow both before and after a given snapshot, simply by propagating the dynamics either backward or forward in time, respectively. 
%
For instance, given a snapshot at time $t=0$, numerical integration of
Eq.~\ref{eqn:RDT} can be used to predict the forward evolution of the flow up to the time $t=T$ when a new snapshot is available.
Similarly, the flow can be propagated backward in time from $t=T$ to $t=0$ by utilizing the transformation $\tau = T-t$ in Eq.~\ref{eqn:RDT}, such that the new time variable has value $\tau = 0$ at the end of the interval.
This transformation yields a negative convection velocity, mean shear, and effective viscosity in the backward dynamics. 
The negative viscous diffusion steepens spatial gradients in velocity, which eventually leads to blowup in the model prediction, especially when additional error is introduced by finite precision numerics. However, the blowup in the backward reconstruction due to this negative diffusion is not prominent for a short prediction horizon, as analyzed in a benchmark study~\cite{krishna2020reconstructing}.
Indeed, by propagating forward and backward, it is possible to formulate physically motivated weighting schemes to fuse forward and backward estimates of the flow field to improve accuracy in the reconstruction between PIV snapshots~\cite{krishna2020reconstructing}.

To facilitate model-based multi-sensor fusion, we discretize Eq.~\ref{eqn:RDT} in both space and time to obtain a finite-dimensional discrete-time state-space representation of the flow:
%
\begin{equation}
    q(t+\Delta t) = A_+ q(t)+w(t)
    \label{eqn:sssys}
\end{equation}
where $q = (u_1, u_2)$ denotes the $N$-dimensional flow state,
$A_+$ denotes the linear propagator for the forward dynamics, $\Delta t$ denotes the time increment, and $w(t)$ denotes model uncertainty or process noise. We assume the process noise $w(t)$ is zero-mean, Gaussian distributed white noise with covariance $Q$, i.e., $w\sim \mathcal{N}(0, Q)$. Although this is a crude modeling assumption, it will simplify the formulation of the filter, which can be carefully tuned to achieve adequate reconstruction performance.

\begin{figure}
    \centering
    \includegraphics[width=\linewidth]{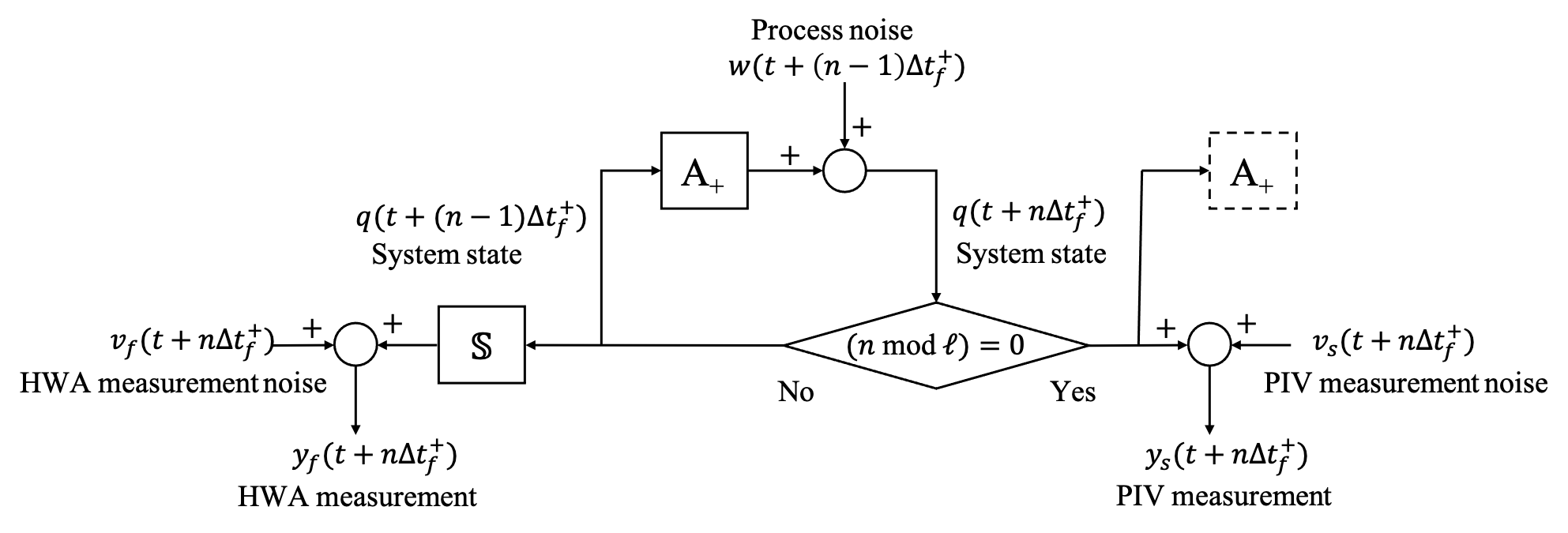}
    \caption{Block diagram of the multi-output state-space system in Eq.~\ref{eqn:system}} 
    \label{fig:BlockD}
\end{figure}

In addition, we define output equations to map the flow state at a given instant to the sensor measurements. Here, we are interested in modeling sensors with disparate sampling rates~\textemdash~a slow-in-time field measurement~(e.g.,~PIV) with sampling period $\Delta t_s^{+}$, and a fast-in-time point measurement~(e.g.,~HWA) with sampling period $\Delta t_f^{+}$, where the superscript ``$+$'' denotes the normalization with respect to friction velocity and kinematic viscosity of the turbulent flow.
For simplicity, we assume the slow and fast time scales are integer multiples of each other, i.e., $\Delta t_s^{+}=\ell \Delta t_f^{+}$ where $\ell$ is an integer. This yields the fluid dynamics and sensor-output equations,
\begin{subequations}
\begin{align}
        q(t+n\Delta t_f^{+}) &= A_+ q(t+(n-1)\Delta t_f^{+})+w(t+(n-1)\Delta t_f^{+}), \label{eqn:sysDynamics}\\
        y_f(t+n\Delta t_f^{+}) &= \Sp q(t+n\Delta t_f^{+}) + v_f(t+n\Delta t_f^{+}), \label{eqn:HWAio}\\
        y_s(t+n\Delta t_f^{+})&= q(t+n\Delta t_f^{+})+v_s(t+n\Delta t_f^{+}),\,  \text{ when } (n \, \text{mod}\, \ell)= 0 \label{eqn:PIVio}
\end{align}
\label{eqn:system}
\end{subequations}
where $n$ is the time-index for the fast system. Additionally, $y_f \in \mathbb{R}^m$ denotes a vector of ``fast'' velocity fluctuation measurements~\textemdash~herein referred to as ``fast'' measurements, for simplicity~\textemdash~taken at $m \ll N$ discrete points in space, determined by a spatial sub-sampling matrix $\Sp \in \mathbb{R}^{m\times N}$ composed of zeros and ones. 
These measurements are assumed to be contaminated by additive zero-mean Gaussian white noise $v_f\sim \mathcal{N}(0, R_f)$, where $R_f$ denotes the measurement noise covariance. Slow measurements are modeled in Eq.~\ref{eqn:PIVio}, where $y_s$ denotes the non-time-resolved field measurements~\textemdash~herein referred to as ``slow'' measurements, for simplicity~\textemdash~which are also assumed to be contaminated with zero-mean Gaussian white noise with covariance $R_s$: $v_s\sim \mathcal{N}(0, R_s)$. 
Note that ``mod'' denotes the remainder operator, so that the output equation Eq.~\ref{eqn:PIVio} reflects that new field measurements are only obtained at slow sampling times. 
The block diagram of the discrete-time state space system is shown in Fig.~\ref{fig:BlockD}.


\subsection{Model-based multi-sensor fusion framework}\label{Sec:KF}
\begin{figure}[h!]
        \begin{minipage}{\textwidth}
        \centering
        (a)~Available measurements
        \vfill
        \includegraphics[width=0.5\linewidth]{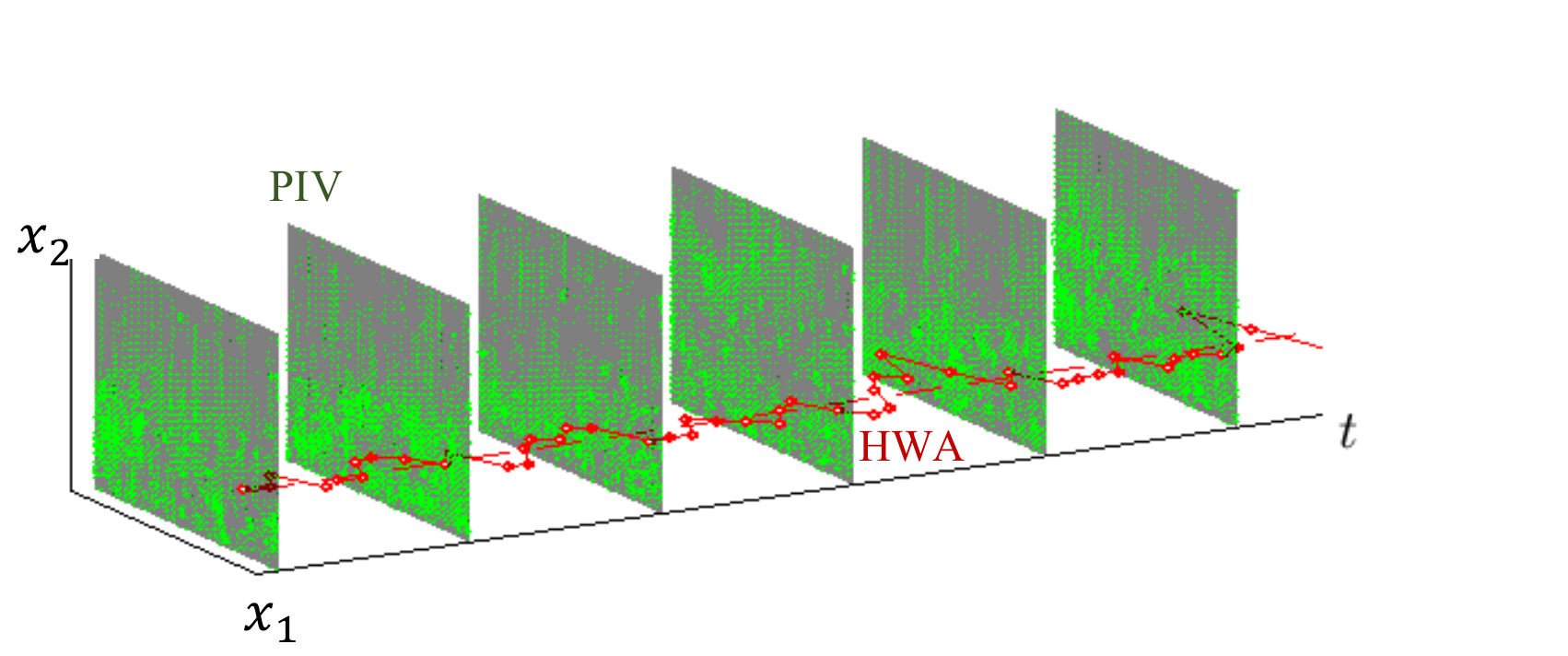}
        \end{minipage}\vfill
        \begin{minipage}{0.5\textwidth}
        \centering
        (b)~Forward fast filter
        \includegraphics[width=\linewidth]{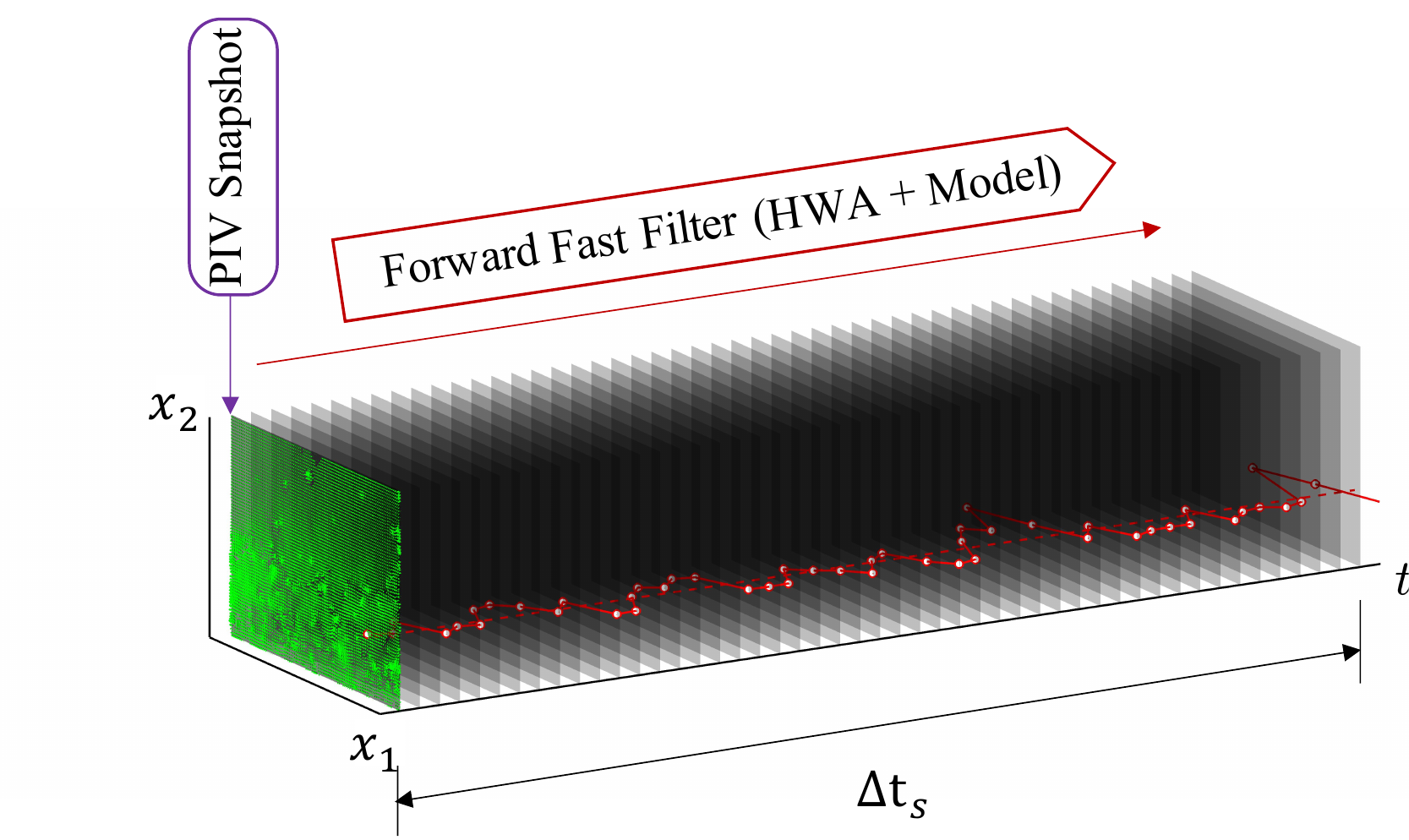}
        \end{minipage}\hfill
        \begin{minipage}{0.5\textwidth}
        \centering
        (c)~Backward fast filter
        \includegraphics[width=\linewidth]{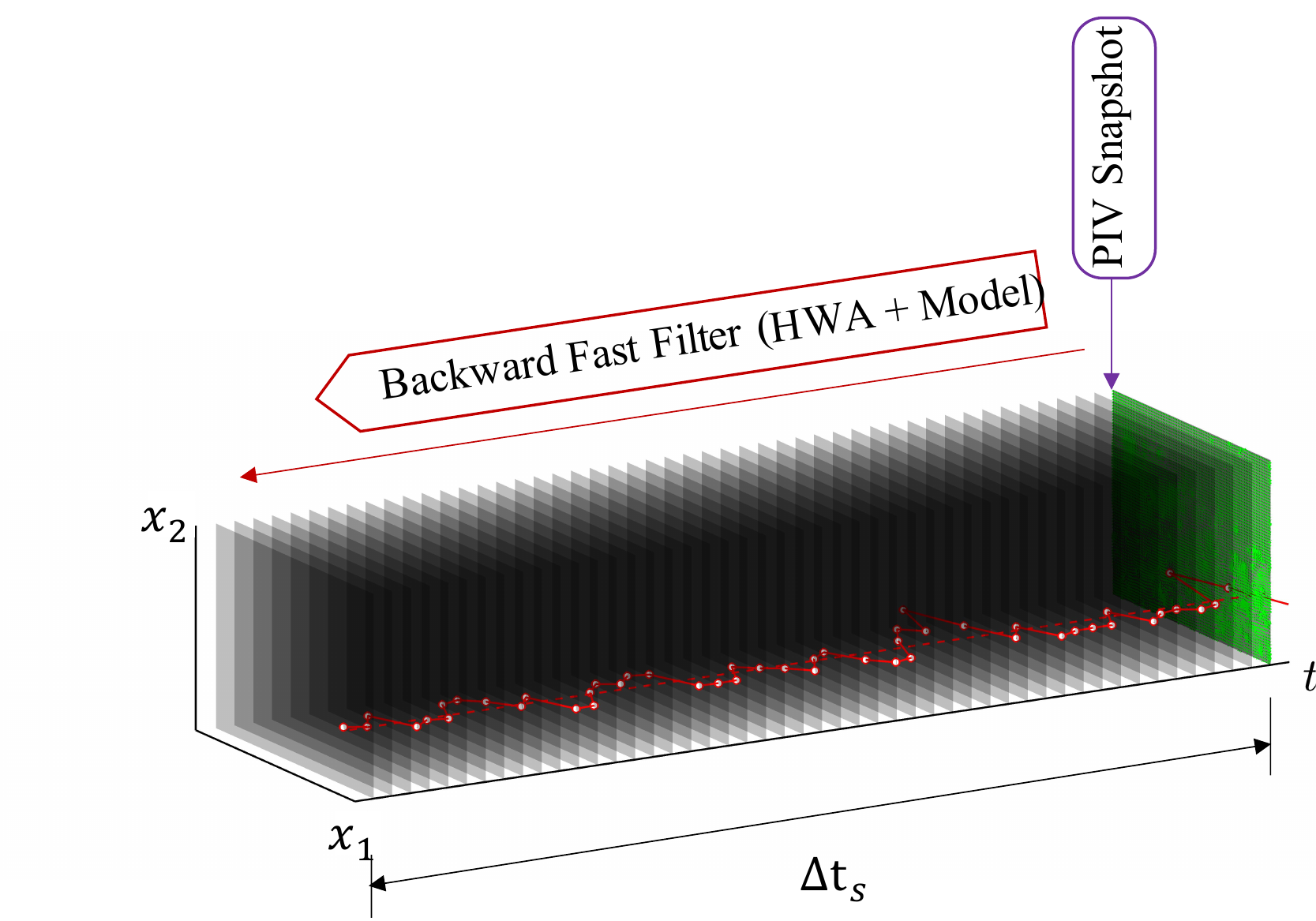}
        \end{minipage}
        \vfill
        \begin{minipage}{\textwidth}
        \centering
        (d)~Forward-backward fusion algorithm
        \includegraphics[width=\linewidth]{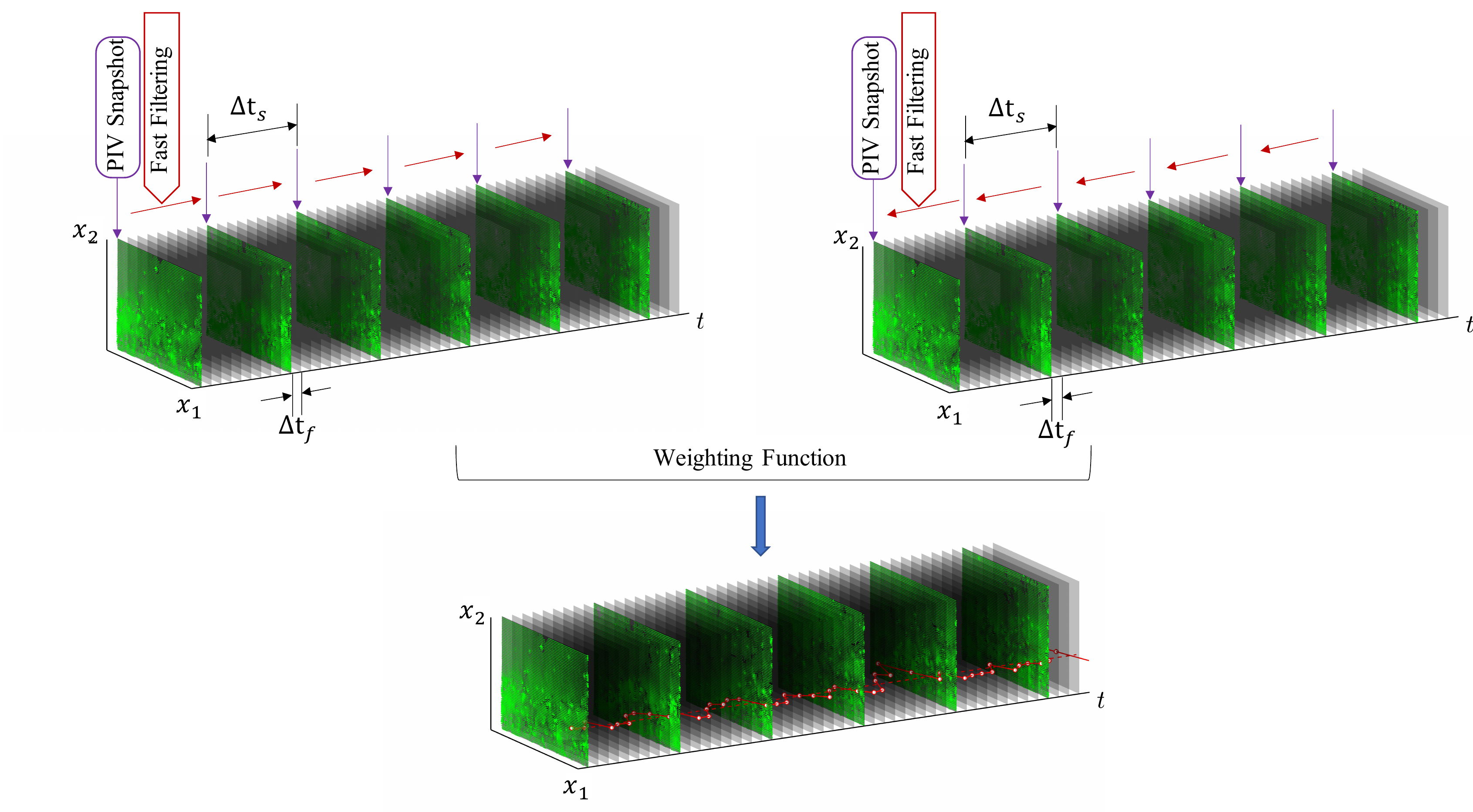}
        \end{minipage}
    \caption{Overview of the proposed model-based multi-sensor fusion approach. (a) shows the multi-rate measurement setup. The green meshed squares represent the slow measurements~(e.g., from PIV) and the red track represents a sample signal from a fast measurement at a fixed location in the flow~(e.g., from HWA).~(b) and ~(c) show the ``fast'' filter within one interval forward and backward in time, respectively. ~(d) shows the fusion of estimates from the forward and backward filters together using a weighting function to reconstruct the flow. }
    \label{fig:framework}
\end{figure}

With the dynamics and sensor models, we are now positioned to formulate a multi-sensor fusion approach. The general idea is to fuse the fast measurements with the model in Eq.~\ref{eqn:system} using a ``fast filter''. Then, a ``slow filter'' is formulated to fuse the fast filter estimates with the slow measurements. Since the model equations in Eq.~\ref{eqn:system} are linear models with zero-mean Gaussian distributed noise, a Kalman filter can be used in this fusion approach, as will be described momentarily. Further, the fast and slow filters can be applied both forward and backward in time in post-process. A schematic of the overarching fusion algorithm is shown in Fig.~\ref{fig:framework}.

Consider the reconstruction forward in time as an example. We use the first snapshot as the initial condition and apply the RDT model to propagate the system state. We fuse the fast measurement and the model prediction to yield an estimate from the fast filter, and when the next snapshot is available, a slow fusion algorithm is applied to integrate the fast estimate with the slow field measurement. The backward fusion method can be applied similarly with the final snapshot as the initial condition. 
Then, a weighted sum of the forward and backward filter estimates is used to reconstruct the flow with higher fidelity. 

A filtering algorithm computes the marginally estimated distribution of the state from past measurements. The Kalman filter deals with the particular case where the dynamic model is linear with the zero-mean Gaussian distributed process and measurement noise~\cite{kalman19601960}.
The RDT-based model derived above meets these conditions, and so we will design a Kalman filter to handle multi-rate multi-fidelity measurements to improve the estimates of the flow state predicted by the RDT model. During the time period between two consecutive snapshots, we apply a fast Kalman filter to fuse the fast measurement with the model generated state. At each time a new snapshot is collected, we use a slow Kalman filter to fuse the slow measurement with the prediction of the state governed by the model. 
The method we use in this study can be viewed as a special case of time-varying Kalman filter, based on time-invariant dynamics and a time-varying measurement equation. We summarize the multi-rate forward Kalman filtering algorithm as follows:
\begin{equation*}\label{eqn:SlowKF}
    \begin{split}
    \text{Slow filter, when} \, (n \, \text{mod}\, \ell)= 0\\
        \hat{q}_+(n|n-1)&=A_+\hat{q}_+(n-1|n-1)\\
        \hat{q}_+(n|n)& =\hat{q}_+(n|n-1)+K_s(n)(y_s(n)-\hat{q}_+(n|n-1))\\
        P(n|n-1) &= A_+P(n-1|n-1)A_+^T+Q\\
        K_s(n) &= P(n|n-1)(P(n|n-1)+R_s)^{-1}\\
        P(n|n)&=P(n|n-1)-K_s(n)P(n|n-1),
    \end{split}
    \end{equation*}
    \begin{equation*}\label{eqn:FastKF}
      \begin{split}
    \text{Fast filter, when} \,(n \, \text{mod}\, \ell)\neq 0:\\
        \hat{q}_+(k|k-1)&=A_+\hat{q}_+(n-1|n-1)\\
        \hat{q}_+(n|n)& =\hat{q}_+(n|n-1)+K_f(n)(y_f(n)-\Sp \hat{q}_+(n|n-1))\\
        P(n|n-1) &= A_+P(n-1|n-1)A_+^T+Q\\
        K_f(n) &= P(n|n-1)\Sp ^T(\Sp P(n|n-1)\Sp ^T+R_f)^{-1}\\
        P(n|n)&=P(n|n-1)-K_f(n)\Sp P(n|n-1),
    \end{split}  
\end{equation*}
where $n$ denotes the new time index corresponding to the discretized version of the RDT model, $\hat{q}_+(n|n-1)$ denotes the \emph{a priori} state estimate, and $\hat{q}_+(n|n)$ denotes the \emph{a posteriori} state estimate. The $+$ subscripts on $A_+$ and $\hat{q}_+$ indicate that these variables are related to the forward estimates.
Note that both the slow Kalman filter and the fast Kalman filter estimate the same state $q$ and its covariance matrix $P$, and that the dynamics of the state do not change with time, only the output equations do.
Thus, the process uncertainty covariance matrix $Q$ in both filters is consistent. However, the measurement noise covariance depends on the measurement system: Recall that $R_s$ and $R_f$ denote the measurement noise covariance matrix for slow-in-time and fast-in-time measurements, respectively, and their distributions follow $v_f \sim \mathcal{N}(0,R_f)$ and $ v_s \sim \mathcal{N}(0,R_s)$ in Eq.~\ref{eqn:system}. For the backward Kalman filter, the \emph{a priori} state estimate becomes  $\hat{q}_-(n|n+1) = A_-\hat{q}_-(n+1|n+1)$, i.e., the dynamic evolves backward in time, where the operator $A_-$ is the backward RDT model.


Establishing the multi-rate Kalman filter is expected to improve the flow reconstruction compared to using a filter based on a single measurement source alone. 
However, performance will depend on the specific weighting scheme used for the fusion of forward and backward estimates. 
A smoothing algorithm is usually used to reduce the large errors associated with forward and backward filtering at later times.
Traditionally, a fixed-interval smoothing algorithm can be used directly. However, the process noise in our model accounts for uncertainties due to the neglected nonlinear terms, including the nonlinear interactions in NSE. As such, we lack an exact characterization of this process noise, and the conventional smoother requires careful tuning.
In our previous study, a temporal weighting function and a spatiotemporal weighting function were applied to fuse the forward and backward model predictions with noteworthy performance~\cite{krishna2020reconstructing}. As such, we opt to use the same approach for fusion here, as described next.

Assume the evolution of the turbulent flow can be propagated forward in time starting from the first slow-in-time snapshot as well as backward in time starting from the second snapshot. A smoothing algorithm of the forward and backward propagation can be formulated to achieve an improved estimate of the flow profile by implementing a fixed-interval smoothing algorithm. The resultant reconstruction of the flow field can be written in the form of a weighted combination of these forward and backward estimates, as follows:
\begin{equation}
\hat{q} = G_{+} \hat{q}_{+}+ G_{-} \hat{q}_{-}
\label{eqn:WeightedSum}
\end{equation}
where $\hat{q}_{+}$ and $\hat{q}_{-}$ denotes the estimated outputs of the forward and backward Kalman filter, respectively, while $G_{+}$ and $ G_{-}$ are the weighting factors to be determined. In this study, we consider two types of weights to be applied in this smoothing problem: (1) temporal weights and (2) spatiotemporal weights~\cite{krishna2020reconstructing}. 

A temporal weighting scheme is formulated based on linear weighting in time between two consecutive slow-in-time snapshots as:
\begin{equation}
\begin{split}
    G_{+} &= G_{+}(t) = 1-\frac{t}{\Delta t_s^{+}}\\
    G_{-} &= G_{-}(t) = \frac{t}{\Delta t_s^{+}}.
\end{split}
\label{eqn:Tweights}
\end{equation}
Recalling that $\Delta t_s^{+}$ is the sampling time of the snapshots, which is also the interval period, we denote $t \in [0,\Delta t_s^{+}]$ as the time within the interval. The temporal weights $G_{+}$ and $ G_{-}$ ensure that the forward estimate is emphasized closer to the beginning snapshot and gradually yields its weight to the backward estimate towards the ending snapshot. 

\begin{figure}
    \centering
    \includegraphics[width=0.8\textwidth]{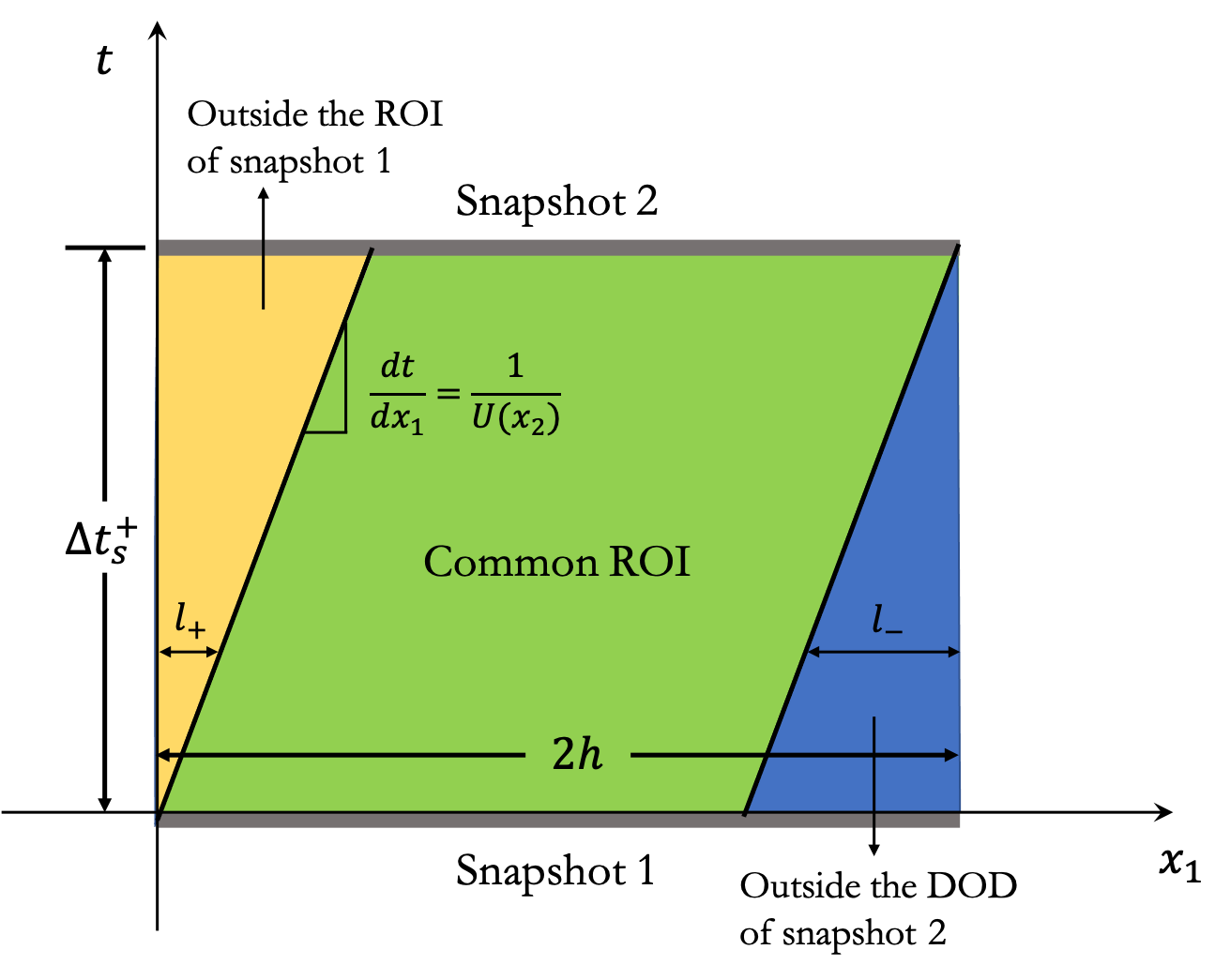}
    \caption{Schematic showing the Region of Influence~(ROI) and Domain of Dependence~(DOD) of two consecutive snapshots and their common ROI in the $x_1-t$ plane at a given wall-normal location. The forward and backward estimates can be combined in the green region, i.e., the common ROI.}
    \label{fig:rdtROI}
\end{figure}

In addition, a spatiotemporal weighting scheme can be formulated to account for the hyperbolic nature of the governing equations. Recall that Eq.~\ref{eqn:RDT} contains a hyperbolic advection term, which is dominant in wall-bounded turbulent flows. The flow estimate follows a propagation at a speed corresponding to the local mean velocity $U(x_2)$. This property yields two special regions in the $x_1-t$ plane: One is  excluded from the region of influence~(ROI) for the initial snapshot and the other is excluded from the domain of dependence~(DOD) for the ending snapshot, as shown in Fig.~\ref{fig:rdtROI}. The characteristic lines with slope $dt/dx_1 = 1/U(x_2)$ separate the ROI of snapshot 1 and the DOD of snapshot 2, and the common region in green illustrates the combination of forward and backward predictions. Note that the slope of the characteristic lines are determined by the local mean velocity $U(x_2)$, which also varies with the wall-normal direction $x_2$.
The spatiotemporal weighting scheme accounts for these effects as follows:
\begin{equation}
\begin{split}
G_{+}&= G_{+}(x_1,x_2,t) = \begin{cases}
0 & 0 \leq x_1<l_{+} \\
\left(1-\frac{t}{\Delta t_s^{+}}\right) & l_{+} \leq x_1 \leq L_{x_1} - l_{-} \\
1 & L_{x_1}-l_{-}<x_1 \leq L_{x_1}
\end{cases}\\
 G_{-}& = G_{-}(x_1,x_2,t) = 
\begin{cases}
1 & 0 \leq x_1 < l_{+} \\
\frac{t}{\Delta t_s^{+}} & l_{+} \leq x_1 \leq L_{x_1} - l_{-} \\
0 & L_{x_1}-l_{-}<x_1 \leq L_{x_1}
\end{cases}
\end{split}
\label{eqn:STweights}
\end{equation}
where $l_{+}=U(x_2)t$ and $l_{-}=U(x_2)(\Delta t_s^{+}- t)$ are calculated as the convective length scale corresponding to the local mean flow velocity $U(x_2)$. $L_{x_1}$ is the streamwise length of the snapshot window. More details about these two approaches can be found in~\cite{krishna2020reconstructing}. The reconstructions using these two approaches are compared in Section~\ref{Sec:discussion}.


\section{Data generation and evaluation metrics} \label{Sec:DataEva}
In this section, we present some prerequisites to conduct the reconstruction and evaluate its performance: Section~\ref{Sec:DNS} introduces the DNS data source serving as ground truth for the algorithm validation in this study. The measurement signals in this paper are also simulated by subsampling the data and artificially adding sensor noise. Section~\ref{Sec:EQ} presents the metrics for quantifying the reconstruction performance.

\subsection{DNS database}\label{Sec:DNS}

In this study, we consider the DNS data for a turbulent channel flow at $Re_\tau =u_\tau h / \nu = 1000$ from the JHTDB~\cite{graham2016web}, where $u_\tau$ is the friction velocity, $\nu$ is the kinematic viscosity, and $h$ is the half-height of the channel. We utilize the streamwise and wall-normal components of the velocity and subsample the data on a restricted domain of $n_{x_1}\times n_{x_2} = 65 \times 65$. The domain consists of a $h \times h$ planar window with a streamwise length $h$ and vertically spanning from the bottom wall to the centerline of the channel. The spatial resolution for this dataset is $\Delta x_1^+ =\Delta x_2^+ = h^+/64 \approx 16 $. This resolution is chosen to be consistent with a physical PIV system with analysis based on $16 \times16$ pixel segments with $50\%$ overlap as a benchmark case involving the use of a camera with 1 MP resolution. 
%
More details about the simulation parameters of the turbulence database can be found in \cite{graham2016web}. The velocity data is extracted at the time step $t^+ = 0.065$, based on the DNS time step. The superscript ``$+$'' denotes the normalization with respect to $u_\tau$ and $\nu$ such that the dimensionless $h^+ = Re_\tau$. Note that the plus mark ``$+$'' denotes normalization only when it appears as a superscript; when in the subscript, it indicates that the variables are related to the forward estimates, e.g., as in $G_+$ and $\hat{q}_+$. 

To sample the slow measurements, we add zero-mean Gaussian distributed noise to every 2-dimensional 2-component snapshot on the slow-sampling interval $\Delta t_s^{+}=6.24$.
The interval time $\Delta t_s^{+}$ is estimated for a PIV system capable of 100 Hz sampling rate~(i.e, sampling time $T=0.01\:s)$, corresponding to water flow with friction velocity $u_\tau = \sqrt{\Delta t_s^{+} \nu/ T} \approx 0.025\: ms^{-1}$.
As discussed in our previous work~\cite{krishna2020reconstructing}, when the total time period between the first and last snapshot is $\Delta t_s^{+} \geq 24$, the model-based reconstructions become less accurate. We choose the time interval for reconstruction to be $\Delta t_s^{+}=96 t^+ =6.24$ to satisfy this condition.
The noise level of the measurements is quantified by the Signal to Noise Ratio~(SNR), which is defined as the ratio between local data magnitude to the variance of the additive noise. For slow measurements, we use $SNR_s =20$, which is a conservative estimate based on experimental experience~\cite{raffel2018particle}.
 
The fast measurements are spatially subsampled from the DNS using the sparse sampling matrix $\Sp$.
The fast measurements are assumed to be of higher precision than the slow measurements, and contaminated with additive noise such that $SNR_f = 100$.
We apply three different approaches to generate the matrix $\Sp$: (1) Uniformly distributed~(UD) sensor placement, which requires no prior knowledge of the flow and places all the sensors uniformly across the observation window; (2) Model-uncertainty-based~(MU) sensor placement, which places sensors in regions where the largest model uncertainty occurs without filtering; and (3) pivoted QR sensor placement~\cite{manohar2018data}, which optimally chooses the sensor locations based on the dominant POD modes obtained from the slow measurements. More details about these approaches and the sensor locations are described in Appendix~\ref{Sec:Sensor}. We place 16 sensors in the simulations and compare the reconstructions as a representative case to evaluate these three placement strategies. In addition, we compare the reconstruction errors of QR placement with 9 sensors vs. 16 sensors to study the influence the number of the sensors has on the reconstruction.
The fast measurements are assumed to be temporally resolved with the same sampling time as the DNS snapshots, i.e., $\Delta t_f^{+} =t^+ = 0.065$. 

We collect the data from JHTDB with a total time duration $T^+ = 37.96$, yielding a total of 584 snapshots. We applied singular value decomposition to the first 200 snapshots to tailor the POD basis for QR pivoting approach in sensor placement as described in Appendix~\ref{Sec:Sensor}, and the remaining 384 snapshots are subsampled to simulate the measurements for reconstruction, as well as serving as ground truth for validation of the fusion algorithm. This results in four slow-time intervals for investigating the reconstruction.

\subsection{Reconstruction error evaluation}\label{Sec:EQ}
The Root Mean Square Error~(RMSE) of the reconstruction can be calculated as
\begin{equation}
    \epsilon(t) = \frac{(\int_{x_1=0}^{h} \int_{x_2=0}^{h} \left(\left(u_1- \hat{u}_1\right)^2 + \left(u_2 - \hat{u}_2 \right)^2\right) dx_1dx_2)^{1/2}}{(\int_{x_1=0}^{h} \int_{x_2=0}^{h} \left((u_1)^2 + (u_2)^2 \right) dx_1dx_2)^{1/2}}.
    \label{eqn:RMSE}
\end{equation}
where $\hat{u}_1$ and $\hat{u}_2$ are the reconstructed velocity fluctuations while $u_1$ and $u_2$ are the velocity fluctuation from the DNS ground truth. Recall that $x_2=0$ indicates the location at the lower wall of the channel, and $x_2 = h$ represents the centerline of the channel. The reconstruction accuracy is evaluated for forward estimates, backward estimates, and the forward-backward estimates using different weighting schemes as described in Section ~\ref{Sec:KF}. 

To analyze the impact of uncertainty from the measurement and process noise, we conduct Monte Carlo simulations and evaluate the mean and variance of the RMSE over multiple realizations. Following the same metrics, we can also evaluate the RMSE with respect to the streamwise and wall-normal velocity components separately, as follows:
\begin{equation}
    \begin{split}
        \epsilon_{x_1}(t) &= \frac{(\int_{x_1=0}^{h} \int_{x_2=0}^{h} \left(u_1- \hat{u}_1\right)^2  dx_1dx_2)^{1/2}}{(\int_{x_1=0}^{h} \int_{x_2=0}^{h} u_1^2  dx_1dx_2)^{1/2}}\\
        \epsilon_{x_2}(t) &= \frac{(\int_{x_1=0}^{h} \int_{x_2=0}^{h} \left(u_2- \hat{u}_2\right)^2  dx_1dx_2)^{1/2}}{(\int_{x_1=0}^{h} \int_{x_2=0}^{h} u_2^2  dx_1dx_2)^{1/2}}.
    \end{split}
\end{equation}
The equations above aim to evaluate the reconstruction errors for each component of the 2D velocity profiles. 
Specific details will be presented with the results in Section~\ref{Sec:discussion}.

\section{Results and Discussion}\label{Sec:discussion}
In this section, we evaluate the reconstruction performance of the proposed fusion algorithms and investigate the influence of implementation decisions (e.g.,~sensor placement, weighting functions).
Section~\ref{Sec:Rmse} compares the time evolution of the RMSE of the forward, backward, and fused estimates.
Section~\ref{Sec:VeloRec} compares these reconstructed flow fields to model prediction without sensor fusion and to the JHTDB ground truth. 
Section~\ref{Sec:pEffect} evaluates the influence of the number and placement of sensors on reconstruction accuracy.
In Section~\ref{Sec:StatPSD}, the reconstructed turbulence intensity and Reynolds stresses are compared to DNS to further evaluate the fusion performance. Unless otherwise state, the total number of sensors will be 16 for all results reported. Corresponding sensor locations are reported in the Appendix.
Table~\ref{tab:1} summarizes the different turbulent flow reconstruction techniques as well as the sensor placement approaches used in this study. Figures with the corresponding results are also listed in Table~\ref{tab:1}.
\begin{table}
\caption{Description of the different reconstruction and sensor placement techniques used in this study. 
}
\label{tab:1}       
\begin{tabular}{llp{.42\textwidth}l}
\hline\noalign{\smallskip}
& Technique & Description & Figure no.  \\
\noalign{\smallskip}\hline\noalign{\smallskip}
\multirow{10}{.1\textwidth}{Fusion approach}& RDT$_\pm^t$ & \raggedright Forward-backward model prediction fused using temporal weights Eq.~\ref{eqn:Tweights}& Fig.~\ref{fig:ModelCompRMSE},~\ref{fig:Stat}  \\
& RDT$_\pm^{st}$ & Forward-backward model prediction fused using spatiotemporal weights Eq.~\ref{eqn:STweights}& Fig.~\ref{fig:ModelCompRMSE},~\ref{fig:ModelComp}~-~\ref{fig:VelosnapRec},~\ref{fig:Stat} \\
& KF$_+$ & Forward time Kalman filtering & Fig.~\ref{fig:RMSEmc18}~-~\ref{fig:RMSEcomp},~\ref{fig:QRrec},~\ref{fig:sensorCompRMSE}  \\
& KF$_-$ & Backward time Kalman filtering & Fig.~\ref{fig:RMSEmc18}~-~\ref{fig:RMSEcomp} \\
& KS$_\pm^t$ &\raggedright Forward-backward Kalman smoothing fused using temporal weights Eq.~\ref{eqn:Tweights}& Fig.~\ref{fig:RMSEmc18}~-~\ref{fig:QRrec},~\ref{fig:sensorCompRMSE}~-~\ref{fig:Stat} \\
& KS$_\pm^{st}$ &\raggedright Forward-backward Kalman smoothing fused using spatiotemporal weights Eq.~\ref{eqn:STweights}& Fig.~\ref{fig:RMSEmc18}~-~\ref{fig:Stat} \\
\noalign{\smallskip}\hline\noalign{\smallskip}
\multirow{5}{.1\textwidth}{Sensor placement}& UD & Sensors are uniformly distributed. & Fig.~\ref{fig:sensorCompRMSE}~-~\ref{fig:Stat} \\
& MU & \raggedright Sensors are placed in the maximum model uncertainty region. & Fig.~\ref{fig:sensorCompRMSE}~-~\ref{fig:Stat}\\
& QR & Sensors are placed based on a pivoted QR approach~\cite{manohar2018data}. & Fig.~\ref{fig:RMSEmc18}~-~\ref{fig:Stat}  \\
\noalign{\smallskip}\hline
\end{tabular}
\end{table}

\subsection{Flow reconstruction RMSE comparison}\label{Sec:Rmse}

\begin{figure}
    \begin{center}
        \includegraphics[width=0.95\textwidth]{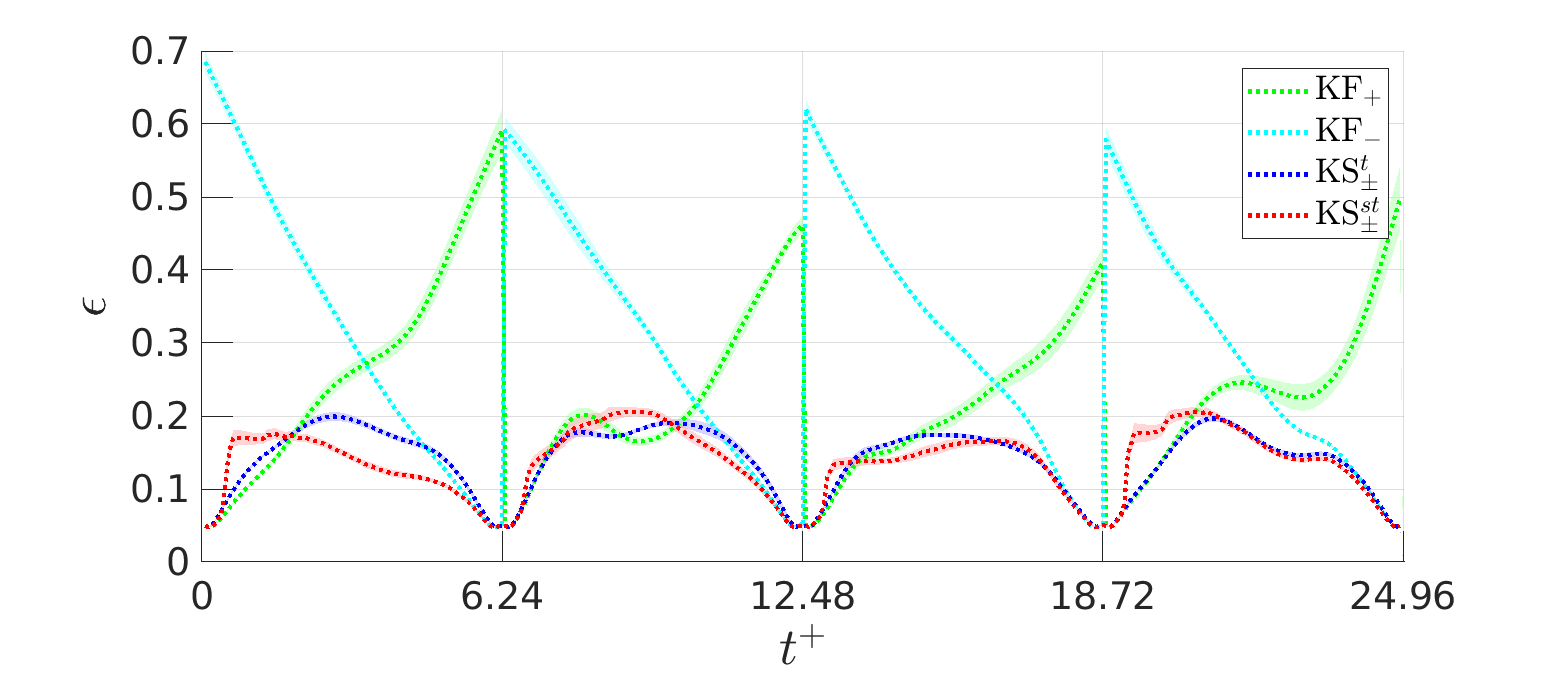}        
    \end{center}
    \caption{Reconstruction error for 4 consecutive intervals under 18 runs of Monte Carlo simulations with $\Delta t_s^{+} = 6.24$. The green dots denote the mean of forward estimate, and the cyan dots denote the mean of backward estimate. The temporal weighted fusion and spatiotemporal weighted fusion results are are plotted in blue and red, respectively. The color shades denote the 1-$\sigma$ with respect to each reconstruction.}
    \label{fig:RMSEmc18}
\end{figure}

We begin by assessing the reconstruction performance of the various fusion algorithms using Monte Carlo simulations.
The error statistics of the temporally weighted smoother~(KS$^{t}_\pm$) and spatiotemporally weighted smoother~(KS$^{st}_\pm$) are compared with those of the forward filter~(KF$_+$) and the backward filter~(KF$_-$).
%
The mean and 1-$\sigma$~(1 standard deviation) bounds from 18 independent realizations over four slow sampling intervals ($\Delta t_s^{+} =6.24$) are reported in Fig.~\ref{fig:RMSEmc18}.
These results correspond to the case of 16 fast sensors arranged according to the QR placement approach.
%
%
An initial RMSE of $\epsilon\approx0.05$ is observed at the beginning and end of each interval, and can be attributed to the noise floor determined by the sensor noise that contaminates the slow snapshot measurements.  
%
The maximum RMSE of the fused forward-backward reconstruction is found to be $\epsilon\approx0.2$ over each interval, regardless of the weighting scheme employed.
%
%

Comparing the forward estimate with the backward estimate, we observe that the backward prediction tends to exhibit a larger RMSE than the forward prediction over each interval. 
The same observation was made regarding backward predictions from the RDT model without fast sensors or fast filtering, as reported in our previous study~\cite{krishna2020reconstructing}.
As noted earlier, this can be explained by the fact that the backward RDT model contains a negative viscous diffusion term which leads to a steep increase in prediction error over time. 
As expected, the error in the forward filter increases toward the end of an interval, while the error in the backward filter increases toward the beginning of an interval.
The smoothing algorithms combine these forward and backward estimates in a manner that weighs the more reliable estimate---at a particular time instant or point in space---more heavily, thus yielding an improvement in the reconstruction performance over the entirety of each interval.

\begin{figure}
    \begin{center}
        \includegraphics[width=0.75\textwidth]{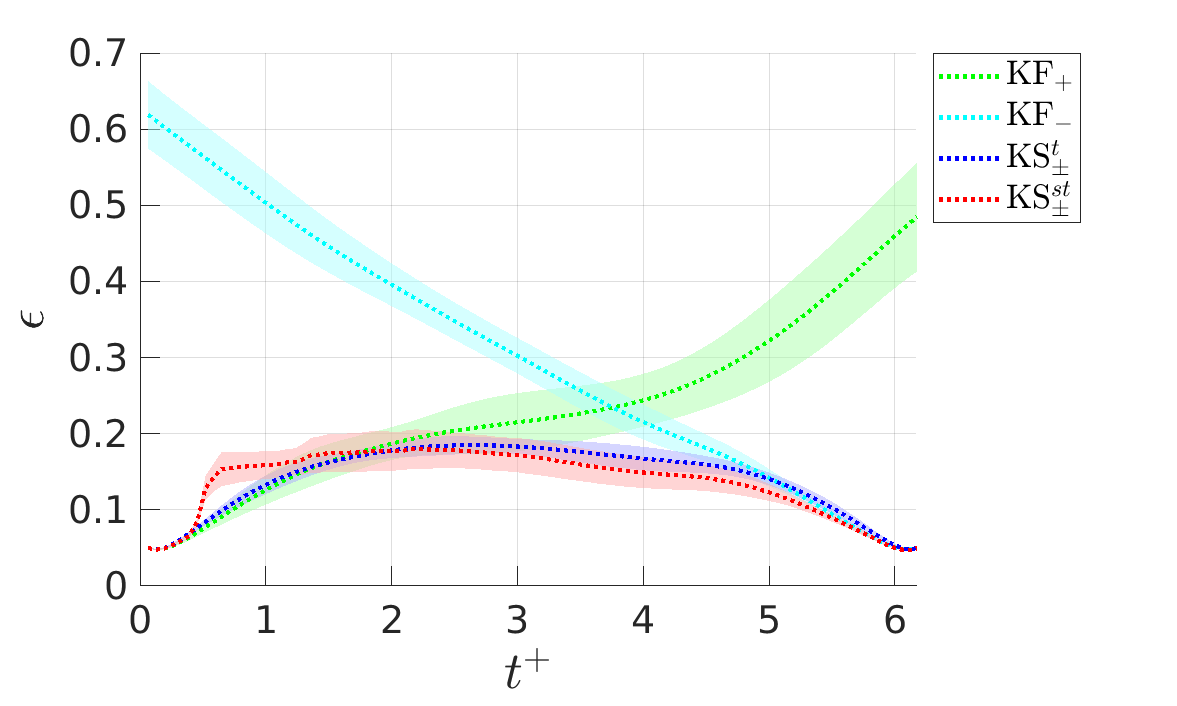}        
    \end{center}
    \caption{Mean and 1-$\sigma$ bounds of the RMSE averaged over 72 different simulations. The 1-$\sigma$ band of the interval-averaged simulation here is larger than that of consecutive intervals in Fig.~\ref{fig:RMSEmc18} because the flow profile also changes for different starting time.}
    \label{fig:RMSEmcQRall}
\end{figure}

From Fig.~\ref{fig:RMSEmc18}, it can also be observed that the reconstruction accuracy varies between the different intervals.  These differences are especially prominent in the estimates from the forward filter~(KF$^+$). This indicates that filter initialization plays a role in reconstruction performance, as is to be expected.
%
We plot the mean and 1-$\sigma$ bounds of interval-averaged RMSE in Fig.~\ref{fig:RMSEmcQRall}, based on the 18 realizations over the four intervals (i.e., 72 total realizations).
Accounting for the reconstruction performance in this way reveals a wider band in the estimation error variance.
However, the magnitude of the RMSE never exceeds $\epsilon\approx0.2$ for reconstructions based on fusion of the  forward and backward estimates.
This suggests that reconstruction performance is not restricted to the specific dataset or flow profile used in this study.

Another interesting observation is that smoothing with the temporal weighting scheme~(KS$^{t}_\pm$) performs slightly better than smoothing with the spatiotemporal weighting scheme~(KS$^{st}_\pm$), at least for the time-horizon and data considered here.
This can be seen in Fig.~\ref{fig:RMSEmcQRall}, where the temporal weighting scheme generates predictions with a narrower variance than the spatiotemporal weighting scheme.
In addition, the spatiotemporal weighting is also observed to exhibit a large error at $t^+\approx 0.5$.
These performance differences can be attributed to the fact that the spatiotemporal weighting scheme is based purely on the flow physics, not taking into consideration any specific properties concerning the specific configuration of point sensors.
%
Accounting for the placement of fast point sensors could improve performance here; however, obtaining and implementing such a weighting scheme would be substantially more complicated compared to a simple weighting scheme based on the ROI/DOD figures shown in Fig.~\ref{fig:rdtROI}.
This is precisely the reason that the spatiotemporal weighting scheme works best when applied to the uniformly distributed (UD) sensor placement, as we will see later in Section~\ref{Sec:pEffect}.
We can consider the temporal and spatiotemporal weighting schemes to be more ``universal'' in the context of multi-sensor fusion, since they are agnostic to the spatial positioning of point sensors in the domain.

\begin{figure}[h!]
    \begin{center}
        \subfloat[Streamwise RMSE]{
        \includegraphics[width=.4\textwidth]{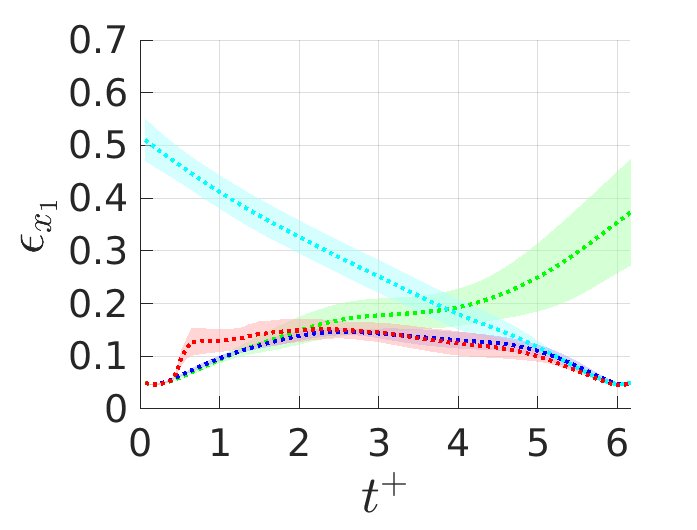}}
        \subfloat[Wall-normal RMSE]{
        \includegraphics[width=.52\textwidth]{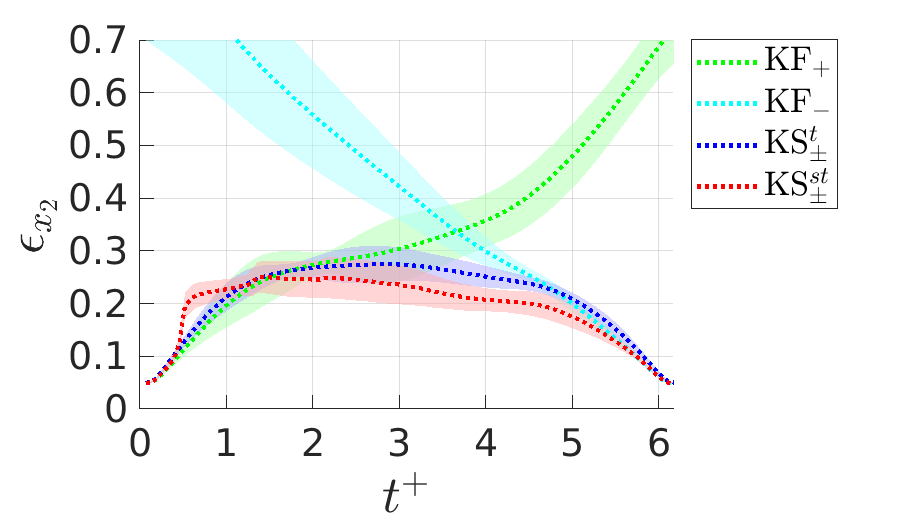}}        
    \end{center}
    \caption{
    Streamwise and wall-normal normalized reconstruction error $\epsilon_{x_1}$ and $\epsilon_{x_2}$. The mean and 1-$\sigma$ bounds of the error plots are also averaged over 72 different realizations.}
    \label{fig:RMSEcomp}
\end{figure}

The plots of $\epsilon_{x_1}(t)$ and $\epsilon_{x_2}(t)$ in the time domain are reported in Fig.~\ref{fig:RMSEcomp}.
The fact that these reconstruction errors are normalized with respect to the spatial integration of independent velocity components (i.e.,~$u_1$ and $u_2$) leads to different RMSE magnitudes between the two.
The maximum streamwise RMSE for both smoothers is approximately $\epsilon_{x_1}\approx0.15$, while the maximum wall-normal RMSE for both smoothers is approximately $\epsilon_{x_2}\approx0.3$.
%
We note that the reconstruction of the wall-normal velocity fluctuations $u_2$ is more sensitive to measurement noise than the reconstruction of the streamwise velocity fluctuations $u_1$.
This sensitivity to measurement uncertainty arises on account of the artificial viscosity that is introduced from discretizing the governing equations via finite differences. 
This artificial viscosity serves to dampen higher-frequency and smaller-scale components more than lower-frequencies and larger-scale components. 
For wall-normal velocity $u_2$, higher energy appears at the higher frequencies than for the streamwise velocity $u_1$;
thus, the attenuation is larger for wall-normal velocity $u_2$. 
This explains the fact that the wall-normal RMSE $\epsilon_{x_2}(t)$ is found to be larger than the streamwise RMSE $\epsilon_{x_1}(t)$.

\begin{figure}
\begin{minipage}[4int]{\textwidth}
        \centering
        (a)~RMSE for 4 consecutive intervals
        \vfill
        \includegraphics[width=\textwidth]{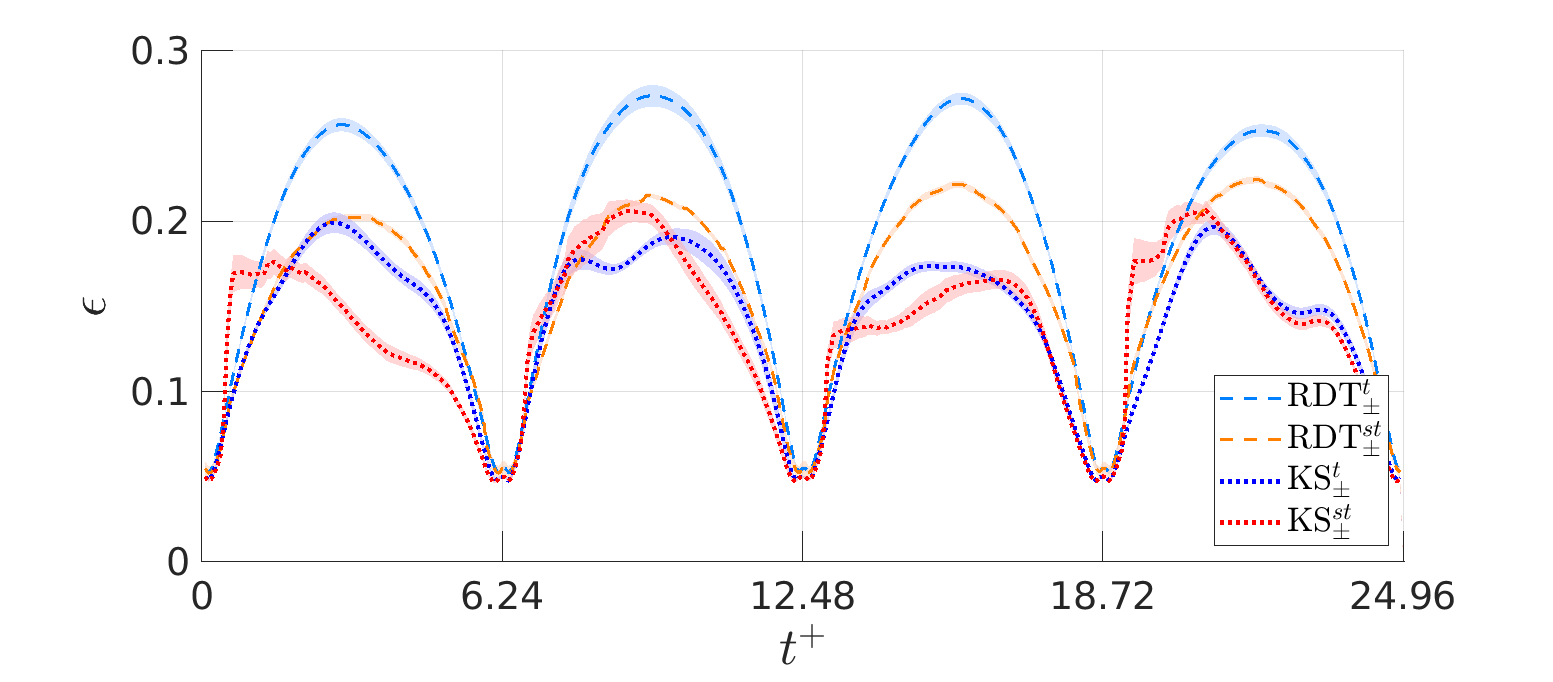}
        \end{minipage}\vfill
        \begin{minipage}[inter-averaged]{\textwidth}
        \centering
        (b)~Interval-averaged RMSE
        \vfill
        \includegraphics[width=0.7\textwidth]{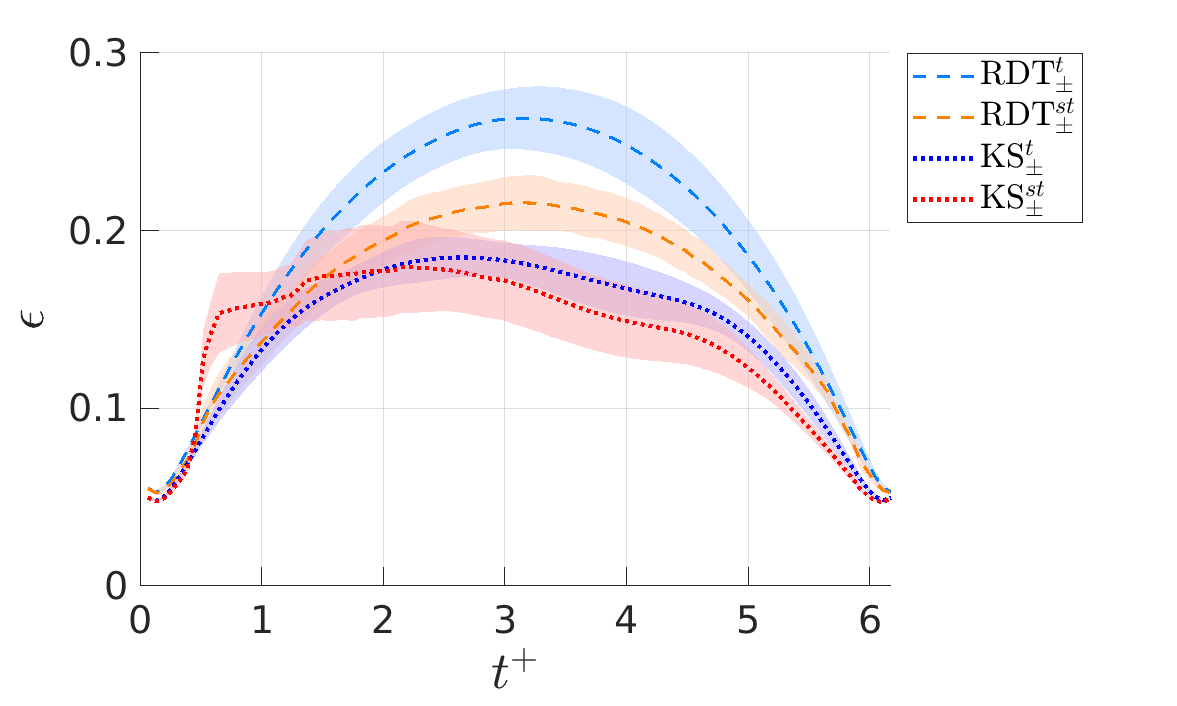}
\end{minipage}
    \caption{Mean and 1-$\sigma$ bounds RMSE from a Monte Carlo study of forward-backward multi-sensor fusion and model predictions. The RMSE in~(a) are averaged over 18 realizations and the interval-averaged RMSE are averaged over 72 simulations. }
    \label{fig:ModelCompRMSE}
\end{figure}

Up to this point, we have posited that multi-sensor fusion can improve the accuracy of the reconstructed flow compared with reconstructions from model predictions based on non-time-resolved field measurements alone.
To see that this is true, we compare the resulting RMSE for the multi-sensor fusion approach with those from model predictions, shown in in Fig.~\ref{fig:ModelCompRMSE}. 
The multi-sensor fusion approach---regardless of weighting scheme---yields a lower mean RMSE compared to the model-prediction-based reconstructions.
As an example, consider the third interval (i.e., $12.48\le t^+\le18.72$), the RMSE from temporal weighting of the forward-backward model predictions (RDT$_\pm^t$) reaches approximately $\epsilon\approx 0.27$, whereas multi-sensor fusion with the same temporal weighting (KS$_\pm^t$) yields a maximum RMSE of $\epsilon\approx0.17$---an improvement of about $40\%$.
Even for the other intervals, the percentage improvement is approximately $20\%$.
However, comparing results between the spatiotemporal weighting schemes is less encouraging.
Here, the performance gains achieved by multi-sensor fusion with spatiotemporal weighting (KS$_\pm^{st}$) are less prominent, but do outperform the spatiotemporal model prediction (RDT$_\pm^{st}$) in terms of the mean RMSE in each interval.
For model predictions, spatiotemporal weighting yields a lower maximum error than temporal weighting.
This is in contrast to the multi-sensor fusion results, for which the temporal and spatiotemporal weighting scheme does not have noticeable effect on reconstruction accuracy.
As discussed above, the spatiotemporal weighting scheme was not designed with the spatial locations of the fast measurements in mind.
It may be possible to achieve superior reconstruction accuracy with an alternative weighting scheme that takes the sensor arrangement into account.
%

Considering the statistics between all realizations and all intervals in a single Monte Carlo plot better highlights the aggregate performance between these approaches (see Fig.~\ref{fig:ModelCompRMSE}(b)).
An interesting observation here is that the RMSE of the model-prediction-based reconstruction tends to possess a smaller variance compared to the multi-sensor fusion result.
The model-prediction-based reconstructions only make use of noisy snapshot data at the beginning and end of each interval.
In contrast, the multi-sensor fusion approach makes use of these same noisy snapshots, but also noisy measurements from the fast sensors.
Although the fast measurements provide time resolved information for the fast filter, the sensor noise from these measurements introduces uncertainty into the estimate at each fast time-step.
This uncertainty results in a larger estimation variance, which compounds over time.



\subsection{Flow field reconstruction comparison}\label{Sec:VeloRec}
\begin{figure}[h!]
        \begin{minipage}[Streamwise]{0.44\textwidth}
        \centering
        (a) $u_1$, $\hat{u}_1$
        \vfill
        \includegraphics[width=\linewidth]{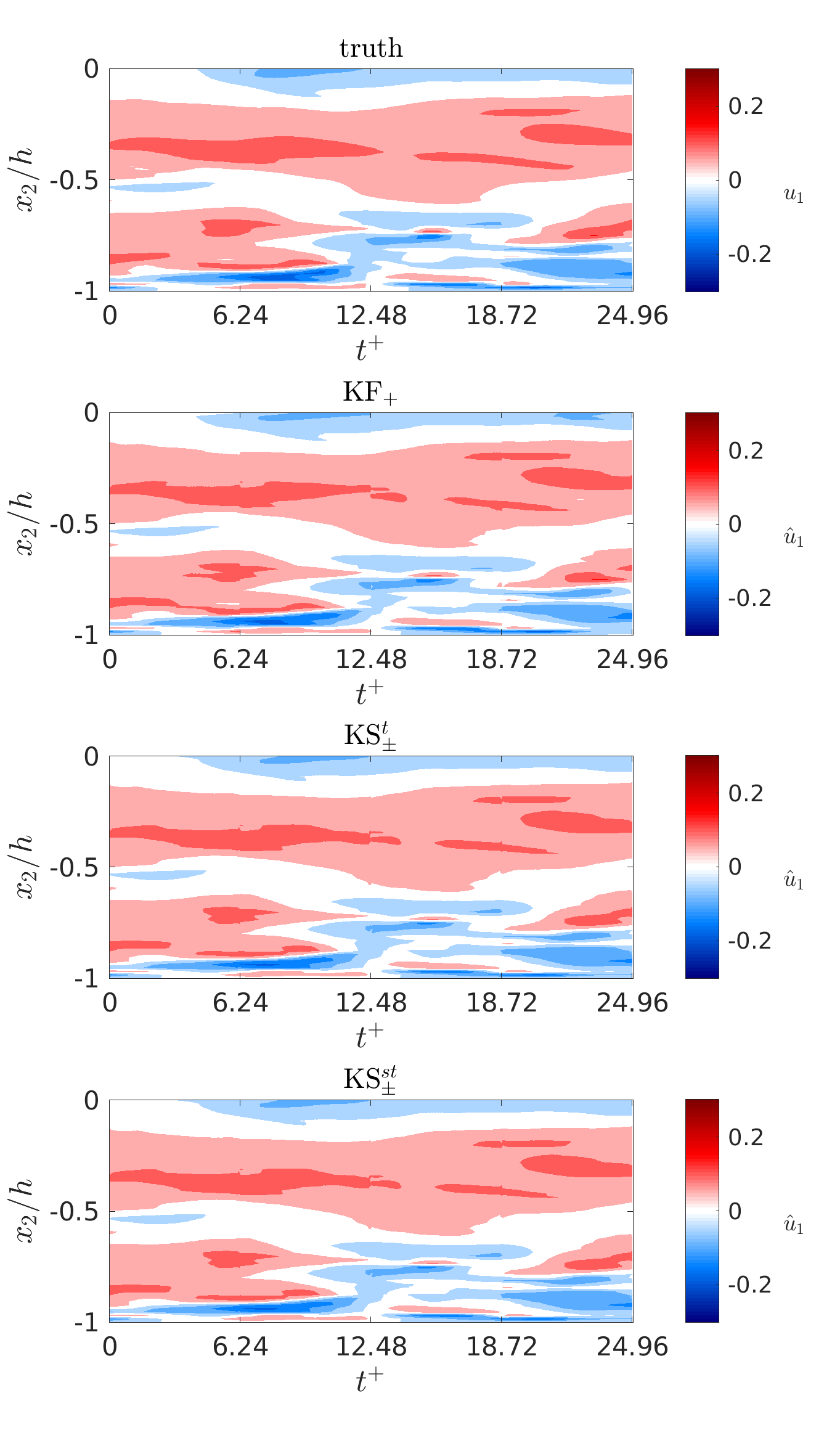}
        \end{minipage}
        \begin{minipage}[Wall-normal]{0.44\textwidth}
        \centering
        (b) $u_2$, $\hat{u}_2$
        \includegraphics[width=\linewidth]{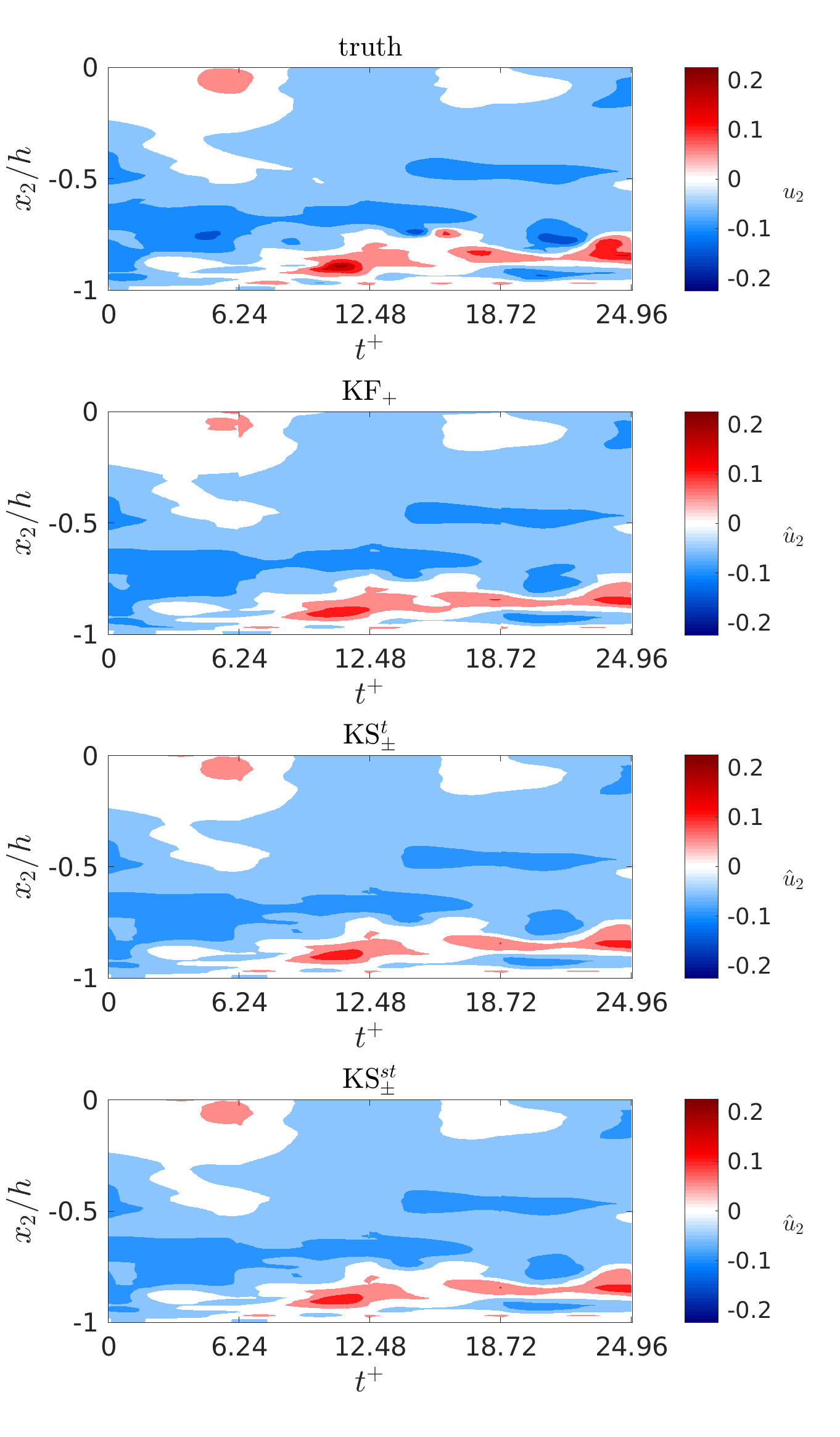}
        \end{minipage}\\
         \begin{minipage}[Streamwise]{0.44\textwidth}
        \centering
        (c) $\hat{u}_1-u_1$
        \vfill
        \includegraphics[width=\linewidth]{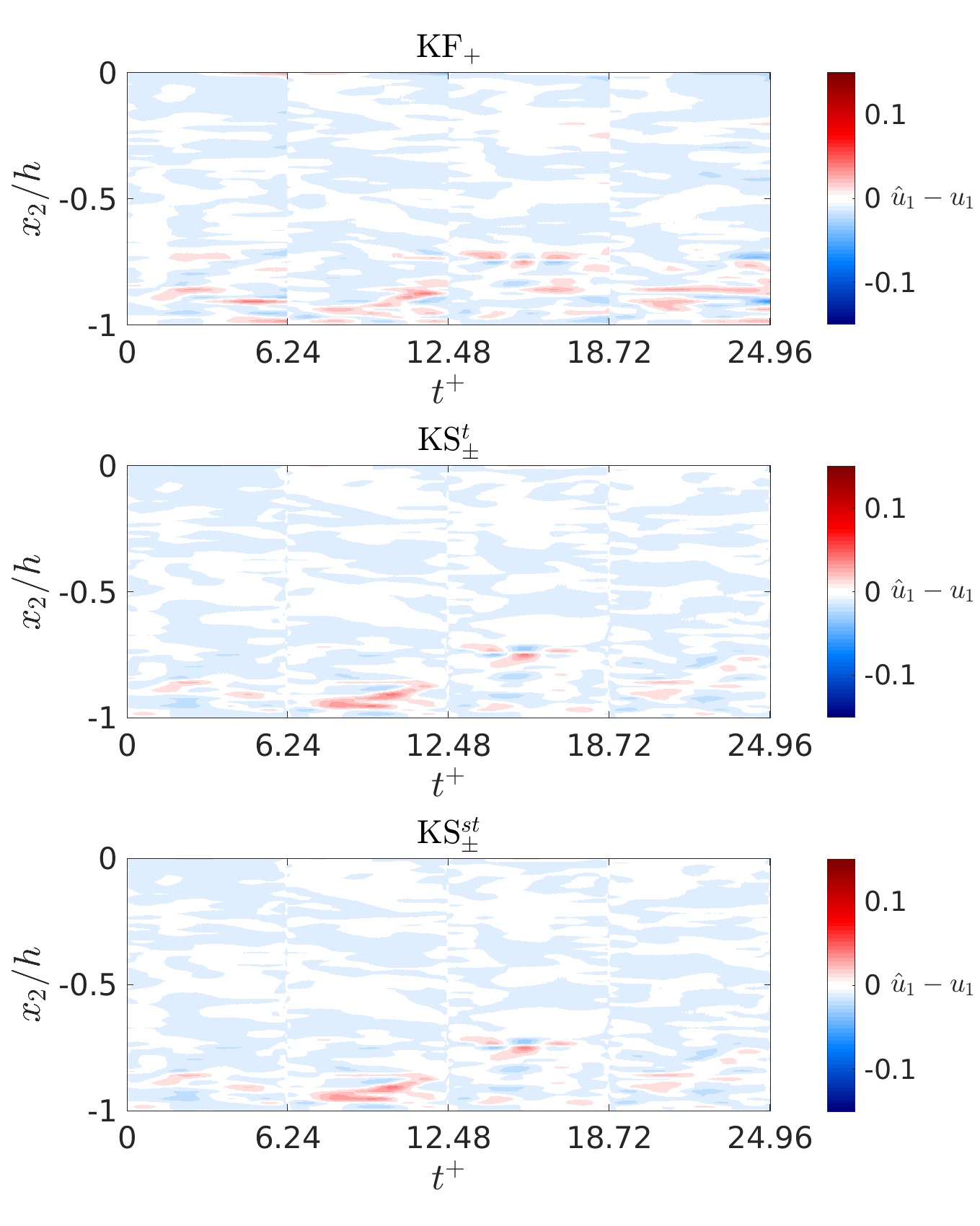}
        \end{minipage}
        \begin{minipage}[Wall-normal]{0.44\textwidth}
        \centering
        (d) $\hat{u}_2-u_2$
        \includegraphics[width=\linewidth]{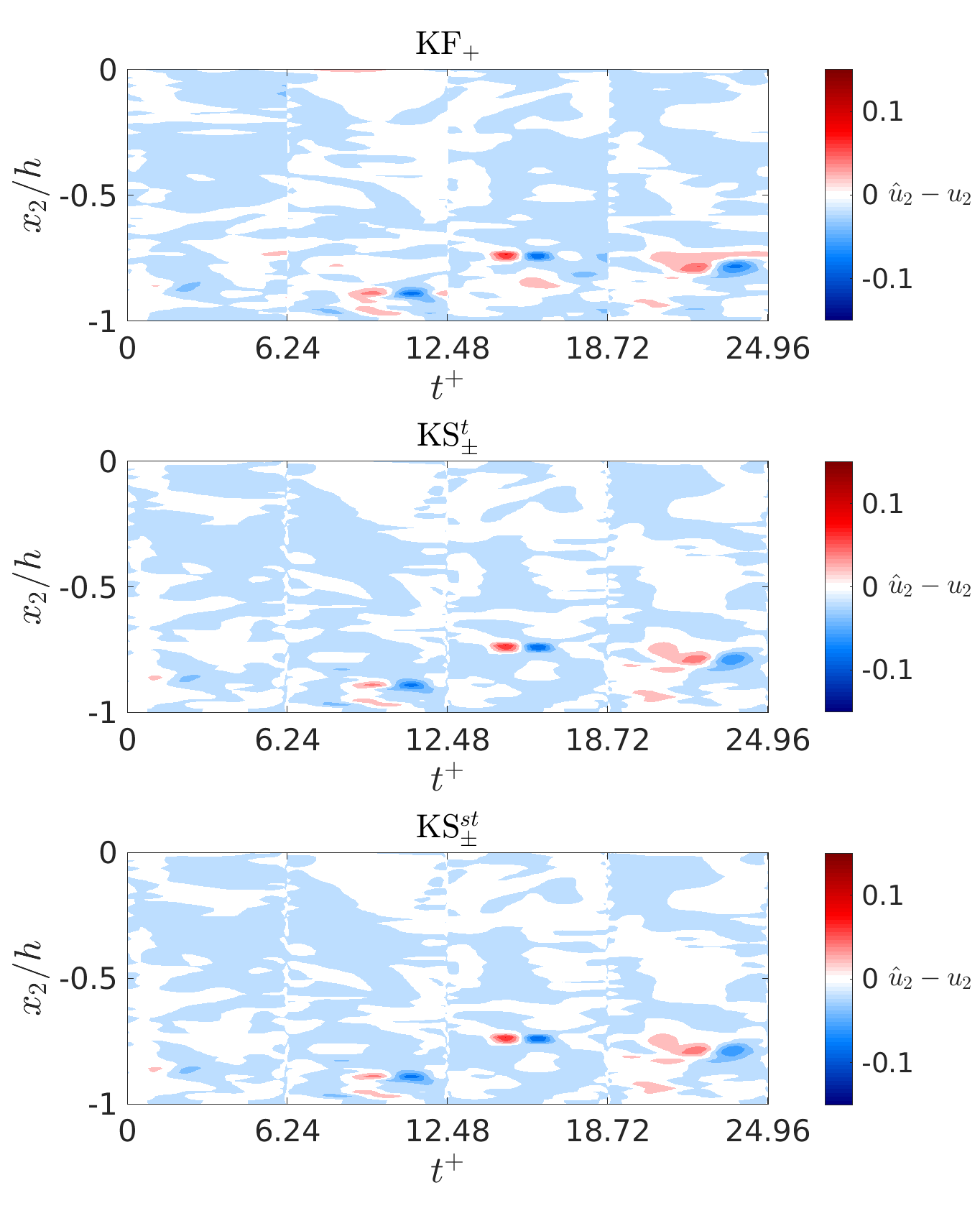}
        \end{minipage}
    \caption{The reconstructed flow field from multi-sensor fusion filters and smoothers compared with JHTDB ground truth along a vertical slice of the channel at $x_1=0.5h$. Reconstruction of streamwise velocity fluctuations are shown in~(a) and of wall-normal velocity fluctuations in~(b). Tiles~(c) and (d) show the corresponding reconstruction errors relative to the JHTDB ground truth.}
    \label{fig:QRrec}
\end{figure}

In the previous section, we reported on the error statistics of the flow reconstruction based on a Monte Carlo analysis over multiple realizations.
Now, we consider the flow field reconstruction for a single representative realization more closely.
We begin by comparing performance between the forward filter (KF$_+$)  and the two smoothers, one with the temporal weighting scheme (KS$_\pm^t$) and the other with the spatiotemporal weighting scheme (KS$_\pm^{st}$).
Figs.~\ref{fig:QRrec}(a)--(b) show the evolution of the ground truth and reconstructed velocity fluctuations  along a wall-normal slice of the channel at a streamwise station of $x_1=0.5h$.
These results correspond to four consecutive slow-sampling intervals and the case of 16 fast sensors, arranged according to the QR placement method described in Appendix~\ref{Sec:Sensor}. 

From visual inspection, it appears that both smoothers perform comparably, and capture many of the large and small scale features appearing in the ground truth.
The forward filter also captures dominant features in the flow, but finer details are not captured as well.
To better quantify and assess this performance, we also report the reconstruction errors in Figs.~\ref{fig:QRrec}(c)--(d).
These are obtained by subtracting the ground truth from each reconstruction.
%
As discussed when assessing performance in terms of RMSE, multi-sensor smoothing with either temporal or spatiotemporal weighting schemes perform comparably in a Monte Carlo study of RMSE.
Here too, we find that these two weighting schemes yield reconstructions with comparable error performance in time and space.
Both smoothers outperform the forward filter in terms of overall error, but the dominant errors appear to be concentrated at the same points in space-time, at least for this slice of the flow.

Note that the reconstruction errors for the fast filter tend to be smaller immediately after a slow snapshot is made available, then grow in time up until the next snapshot.
From Fig.~\ref{fig:QRrec}, it can be observed that flow field reconstruction from the forward filter is less smooth towards the end of each slow sampling interval, 
which leads to slightly larger reconstruction errors at the end of the interval.
This observation is consistent with the relatively large RMSE at the end of each interval, as reported in Fig.~\ref{fig:RMSEmc18}.
The flow field reconstruction from the smoothers do not exhibit the same degree of non-smoothness toward the end of the interval:
reconstruction error grows at first, but then decreases with time until the end of the interval.
These differences with the forward filter and the two smoothers are to be expected, since the smoother was designed to improve upon the forward filter results by taking a backward pass through the data using a backward filter.

\begin{figure}[h!]
        \begin{minipage}[Str]{0.5\textwidth}
        \centering
        (a)~$u_1$, $\hat{u}_1$
        \vfill
        \includegraphics[width=\textwidth]{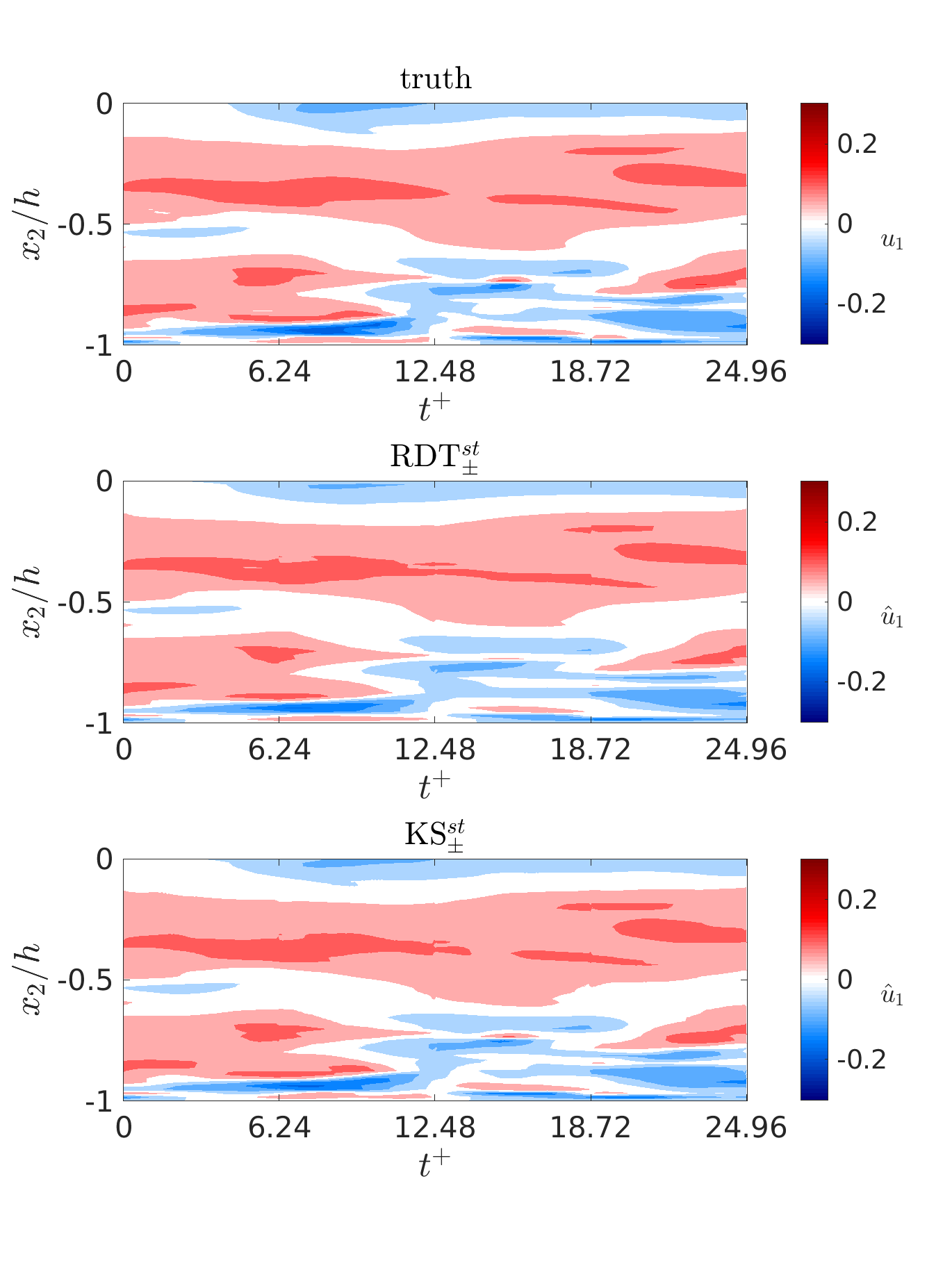}
        \end{minipage}\hfill
        \begin{minipage}[wln]{0.5\textwidth}
        \centering
        (b)~$u_2$, $\hat{u}_2$
        \vfill
        \includegraphics[width=\textwidth]{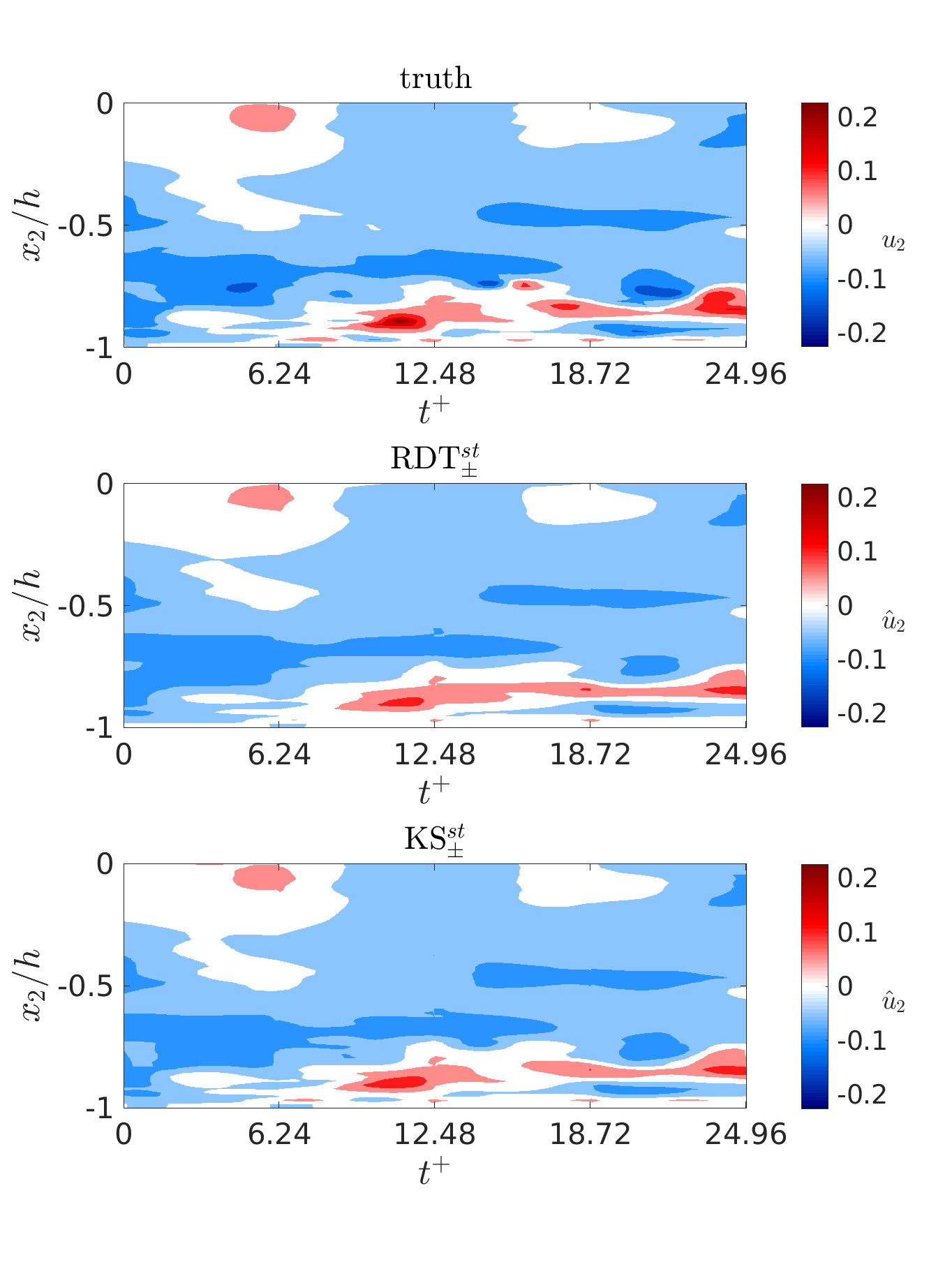}
        \end{minipage}\vfill
        \begin{minipage}[StrErr]{0.5\textwidth}
        \centering
        (c)~$\hat{u}_1-u_1$
        \vfill
        \includegraphics[width=\textwidth]{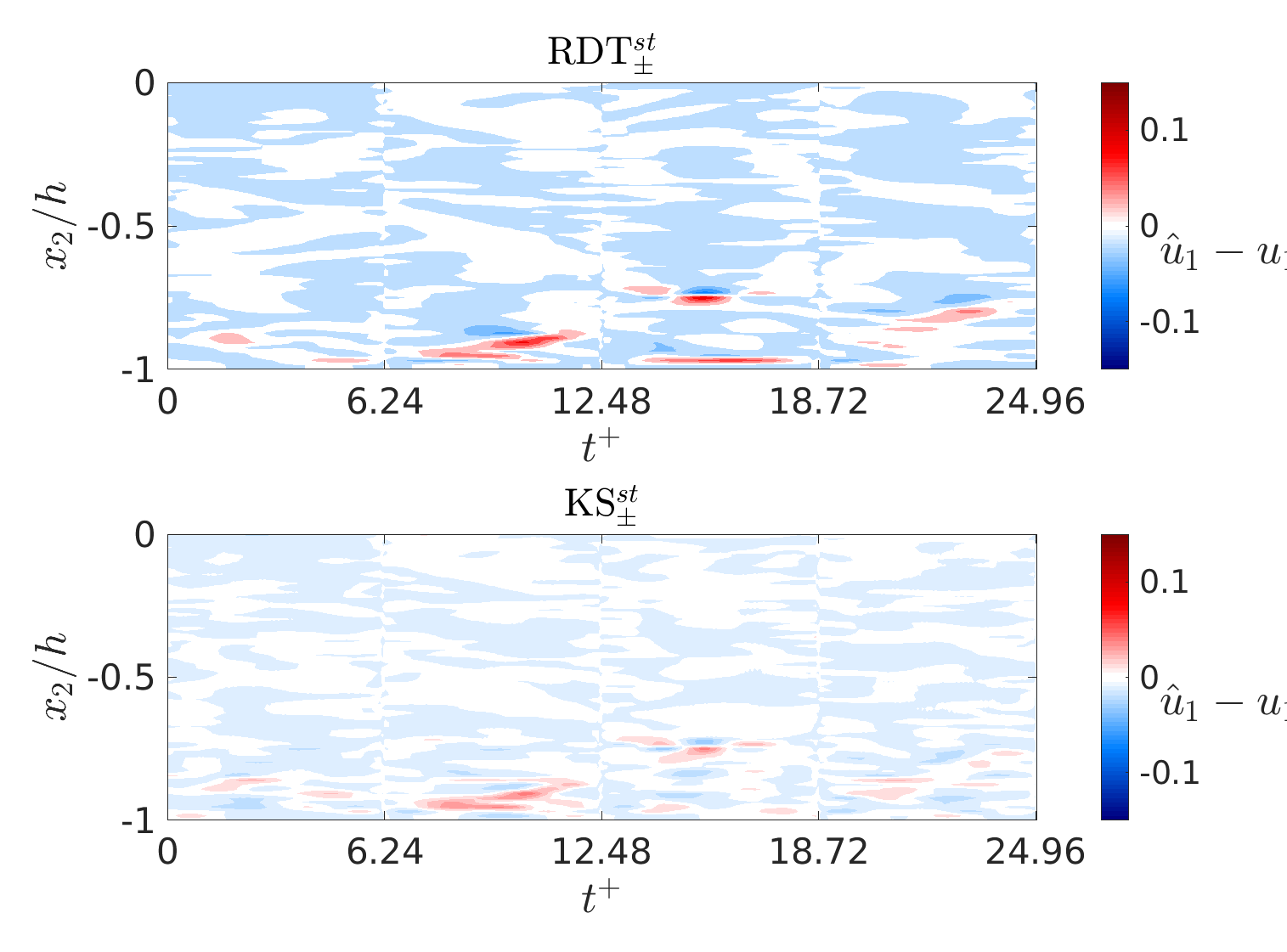}
        \end{minipage}\hfill
        \begin{minipage}[WlnErr]{0.5\textwidth}
        \centering
        (d)~$\hat{u}_2-u_2$
        \vfill
        \includegraphics[width=\textwidth]{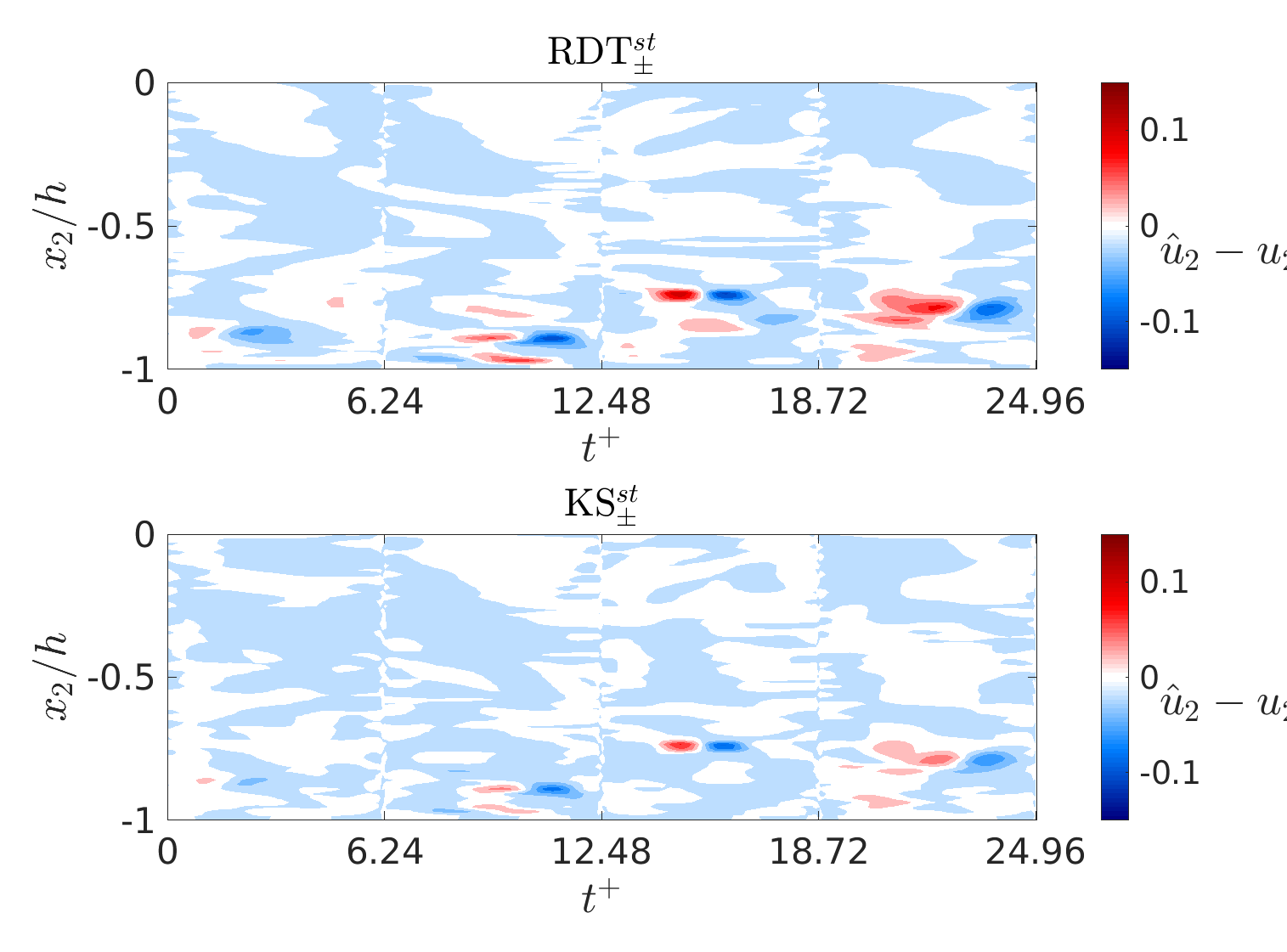}
        \end{minipage}
    \caption{The reconstruction flow field from multi-sensor fusion smoothing and model-predictions compared with JHTDB ground truth along a vertical slice of the channel at $x_1=0.5h$. Reconstruction of streamwise velocity fluctuations are shown in~(a) and of wall-normal velocity fluctuations in~(b). Tiles~(c) and (d) show the corresponding reconstruction errors relative to the JHTDB ground truth. }
    \label{fig:ModelComp}
\end{figure}

The results above compare flow reconstructions obtained by fusing measurements from slow and fast sensors.
The associated reconstruction performance can be bench-marked against the case of reconstructing the flow from model predictions based on only the slow measurements.
Figs.~\ref{fig:ModelComp}(a)--(b) report the flow reconstruction over the same wall-normal slice of the channel as before, but now showing results from spatiotemporally weighted forward and backward model predictions (RDT$_\pm^{st}$)---i.e.,~the model-prediction approach with the best reconstruction performance, based on the RMSE results reported in Fig.~\ref{fig:ModelCompRMSE}.
Results for the spatiotemporally weighted multi-sensor smoother (KS$_\pm^{st}$) and the ground truth are also reported here.
%
The reconstruction based on the multi-sensor fusion is notably closer to the ground truth than the model-prediction-based reconstruction.
These differences are more clearly seen in a visualization of the associated reconstruction errors shown in Figs.~\ref{fig:ModelComp}(c)--(d).
From this analysis, it is evident that multi-sensor fusion yields a lower reconstruction error than the model prediction approach, both in the streamwise and the wall-normal components of the velocity fluctuations.
Interestingly, even though the model-prediction-based reconstructions are based on both a forward and backward pass on the data, this approach tends to exhibit larger errors closer to the starts and ends of the slow sampling intervals than the multi-sensor approach.
Indeed, the fast sensors provide valuable information that benefits the reconstruction across the entire interval, including at times that the slow snapshots are available.

\begin{figure}[h!]
\begin{minipage}[StrSnap]{0.5\textwidth}
        \centering
        (a)~$u_1$, $\hat{u}_1$
        \vfill
        \includegraphics[width=\textwidth]{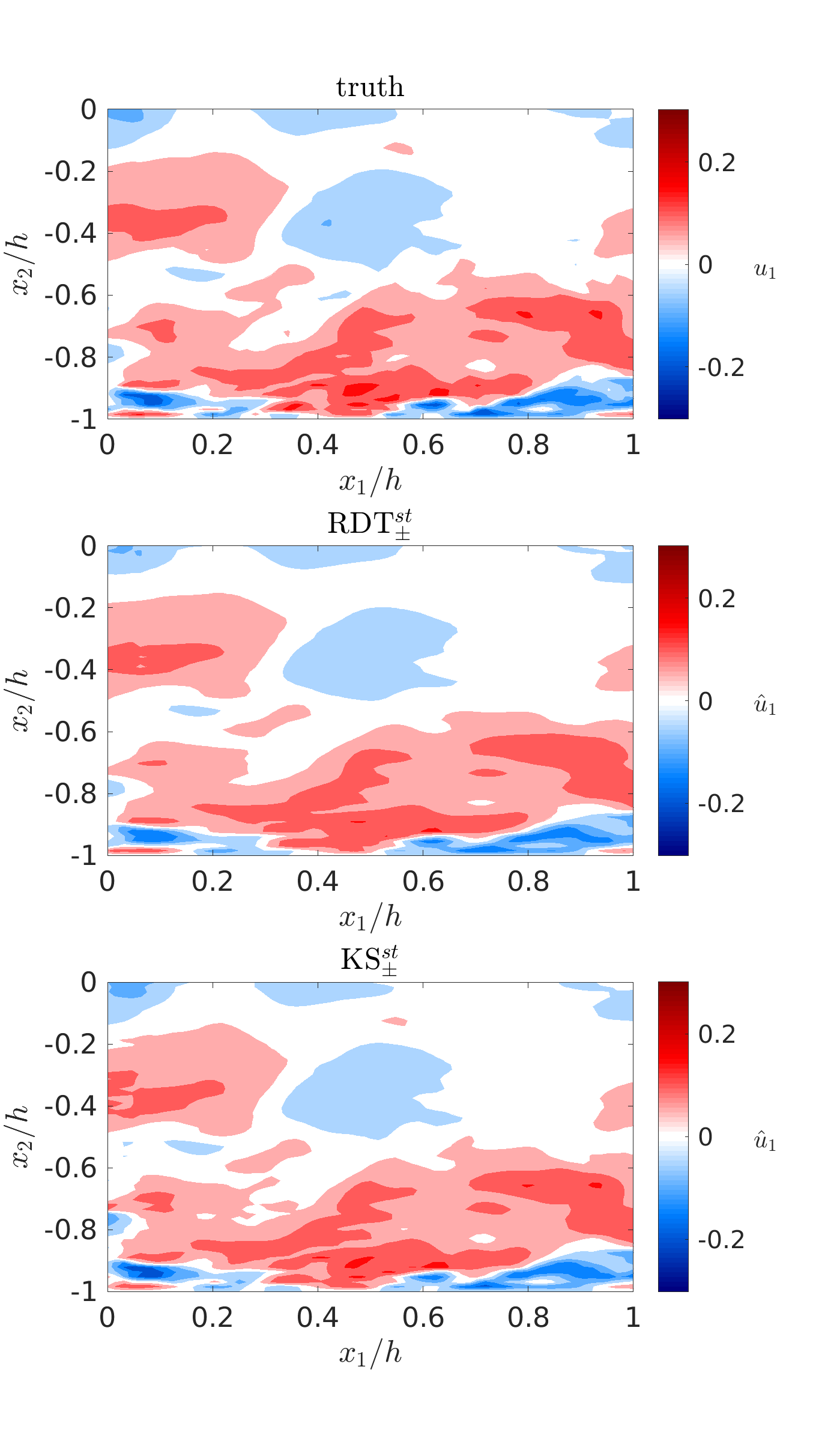}
        \end{minipage}\hfill
        \begin{minipage}[wlnSnap]{0.5\textwidth}
        \centering
        (b)~$u_2$, $\hat{u}_2$
        \vfill
        \includegraphics[width=\textwidth]{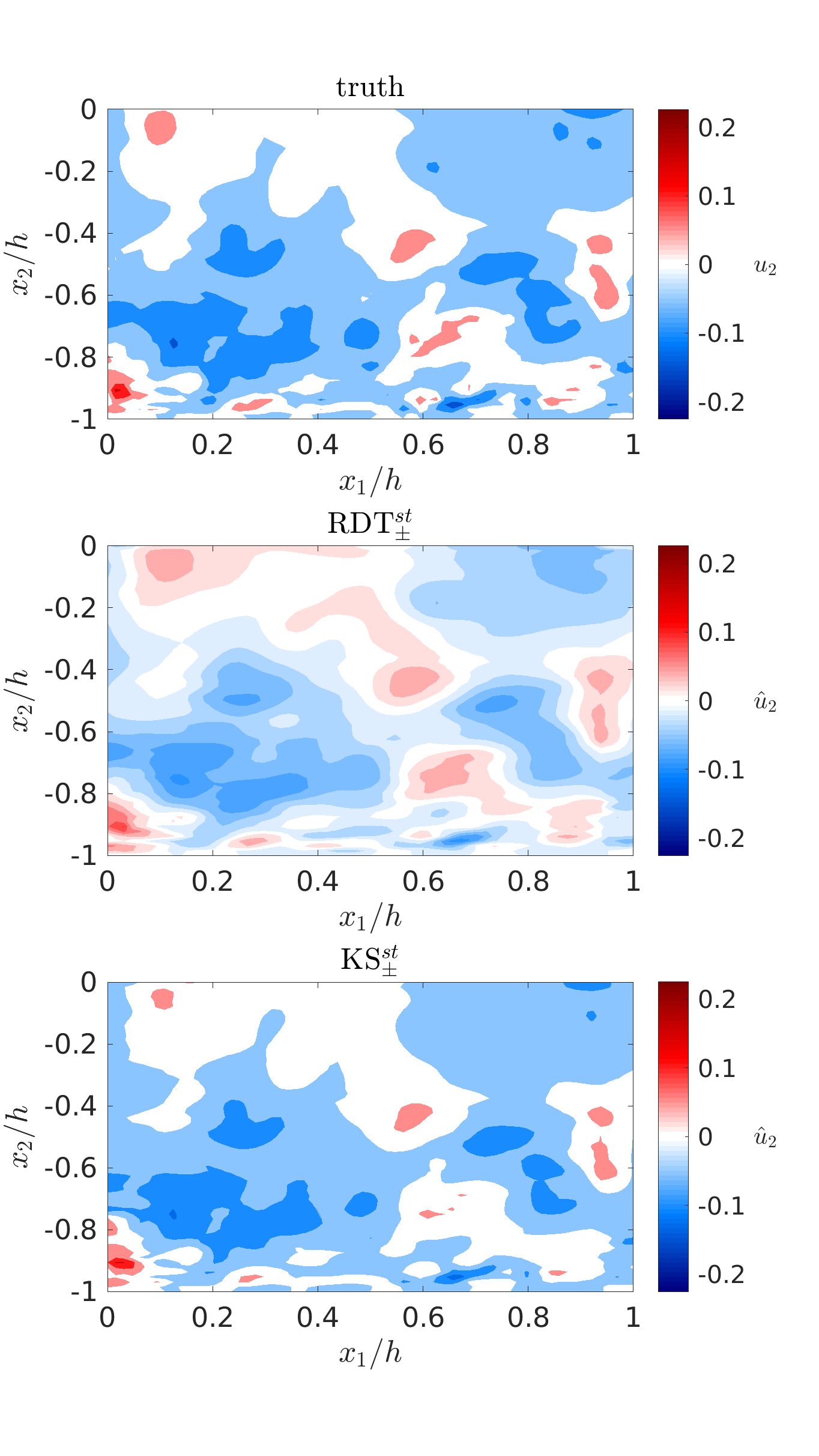}
        \end{minipage}
    \caption{Snapshots of the reconstructed velocity fluctuations at time $t^+ = 3.12$, where the largest RMSE tends to be observed. }
    \label{fig:VelosnapRec}
\end{figure}

Finally, we investigate the reconstructions from RDT$_\pm^{st}$ and KS$_\pm^{st}$ as a function of both spatial directions, focusing on a single time instant $t^+ = 3.12$ that is halfway between two slow snapshots (see Fig.~\ref{fig:VelosnapRec}).
Note that this is the time instant within an interval corresponding to relatively large RMSE in all of the reconstruction approaches (see Fig.~\ref{fig:RMSEmc18}). 
The most notable difference arises in the reconstruction of wall normal velocity fluctuations.
The multi-sensor fusion approach undoubtedly captures the dominant features and many of the finer features that can be seen in the ground truth, whereas the model-prediction-based reconstruction smears out many of these features.
The model-prediction approach performs better for reconstructing the streamwise velocity fluctuations than for wall-normal velocity fluctuations.
However, the multi-sensor fusion approach still captures finer features that the model prediction approach misses.
These observations are consistent with the fact that wall-normal reconstruction error $\epsilon_{x_2}$
was found to be greater than the streamwise reconstruction error $\epsilon_{x_1}$.
Further, these results suggest that  the dominant contributor to the RMSE improvements realized by the multi-sensor fusion approach can be attributed to improvements in reconstructing the wall-normal velocity fluctuations.
The fast sensors provide sufficient information that can be exploited by the fusion framework to improve model predictions based on RDT$_\pm^{st}$ from a maximum mean RMSE of $\epsilon\approx0.23$ down to $\epsilon\approx0.17$ for KS$_\pm^{st}$ (see Fig.~\ref{fig:ModelCompRMSE}).

\subsection{Effect of number and arrangement of fast sensors}\label{Sec:pEffect}
All of the multi-sensor fusion results reported to this point have corresponded to the case of $m=16$ fast sensors distributed in the channel based on the pivoted QR sensor arrangement described in  Appendix~\ref{Sec:Sensor}.
Here, we briefly investigate the influence of number of fast sensors $m$ and their arrangement on the quality of the flow reconstruction.
We consider sensor arrangements determined using three different approaches, all described in Appendix~\ref{Sec:Sensor}: (1)~uniformly distributed~~(UD) sensor placement, (2)~model-uncertainty-based~(MU) sensor placement, and (3)~pivoted QR sensor placement.
We only consider perfect squares $m=9$ and $m=16$ here in order to satisfy symmetry conditions for the UD placement.
Both values of $m$ are roughly three orders of magnitude less than the dimension of the flow state being reconstructed.
In the course of our study, we encountered difficulties in filter tuning when $m<9$ fast sensors were utilized, and so do not report on those results here.
Although reconstruction performance can be further improved through the use of a larger number of sensors (i.e.,~$m>16$), our ultimate interest is to leverage multi-sensor fusion for flow reconstruction in actual experiments where a larger number of fast sensors may be impractical and so do not report on such cases here either.

\begin{figure}[t]
\begin{tikzpicture}
\node[rotate=-90] at (0,0) {
\begin{minipage}{\textheight}
\begin{center}
\begin{minipage}[UD]{0.3\linewidth}
        (a)~UD  ($m = 9$)
        \vfill
        \includegraphics[width=\linewidth]{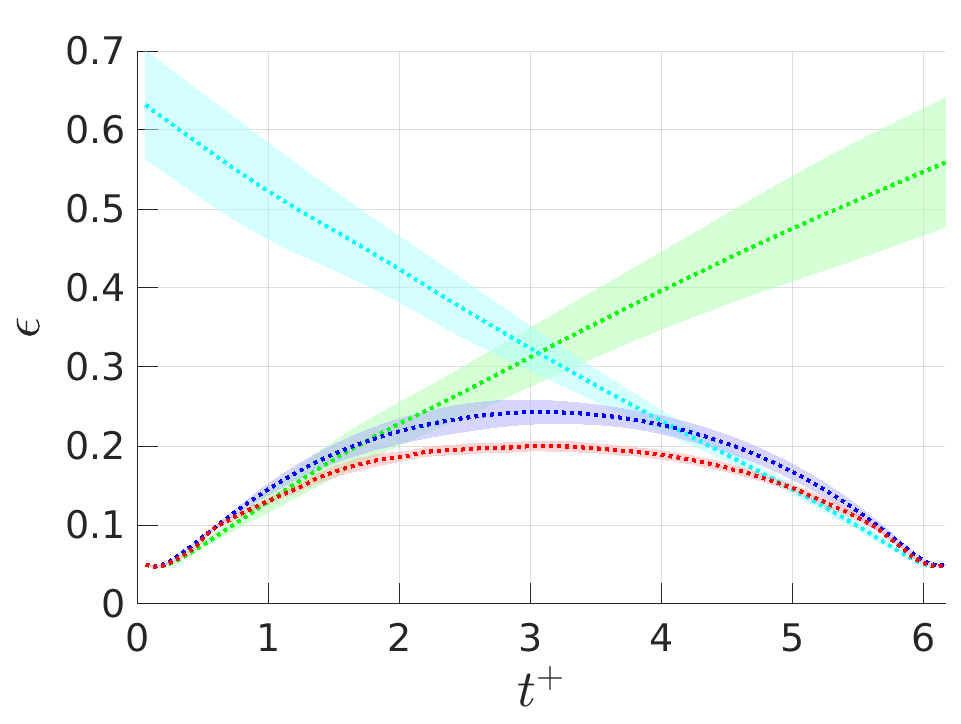}
        \end{minipage}\hfill
        \begin{minipage}[MU]{0.3\linewidth}
        \centering
        (b)~MU  ($m = 9$)
        \vfill
        \includegraphics[width=\linewidth]{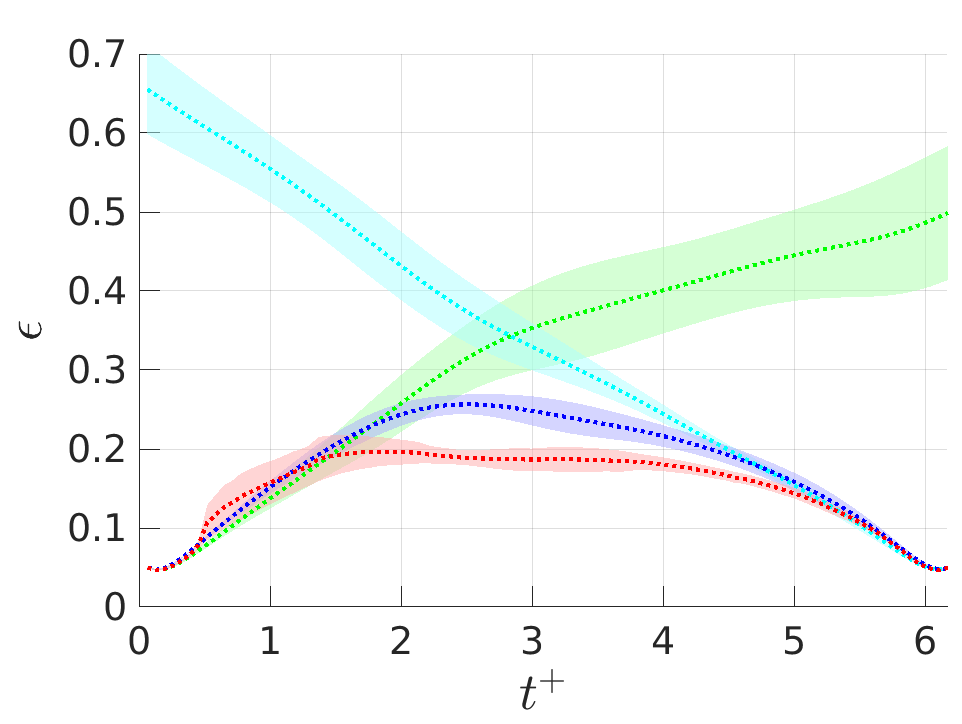}
        \end{minipage}\hfill
        \begin{minipage}[QR]{0.3\linewidth}
        \centering
        (c)~QR  ($m = 9$)
        \vfill
        \includegraphics[width=1.25\linewidth]{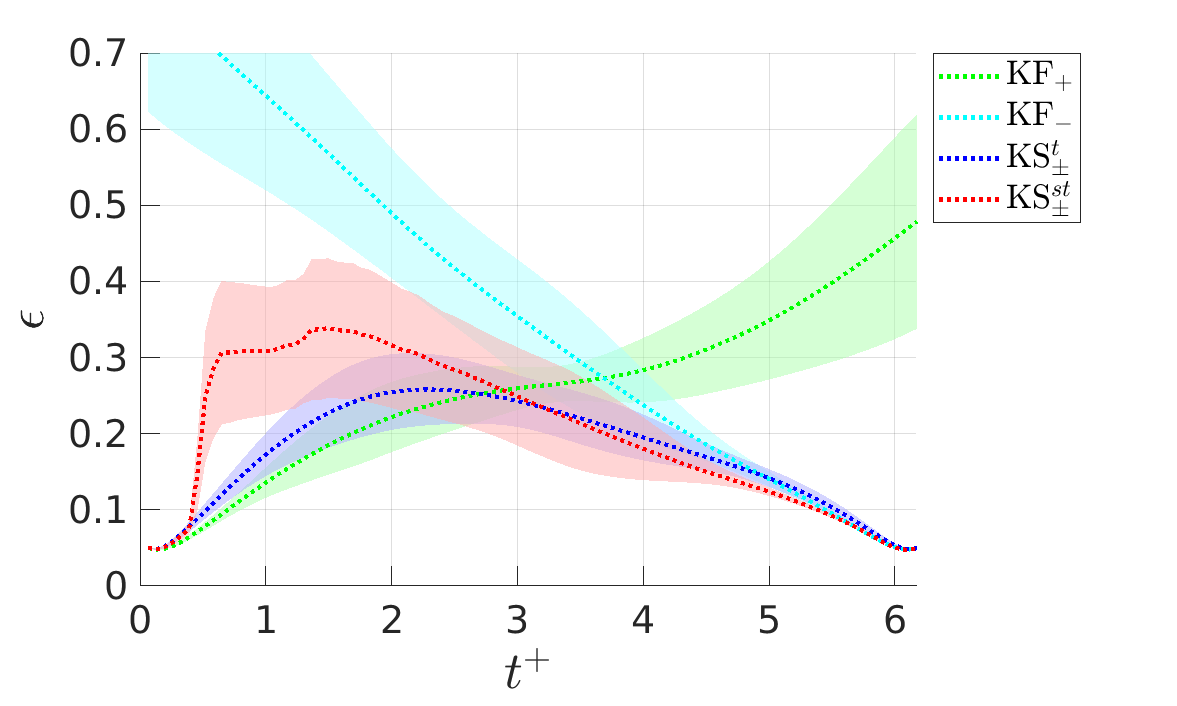}
        \end{minipage}\\
        \begin{minipage}[UD]{0.3\linewidth}
        (d)~UD  ($m = 16$)
        \vfill
        \includegraphics[width=\linewidth]{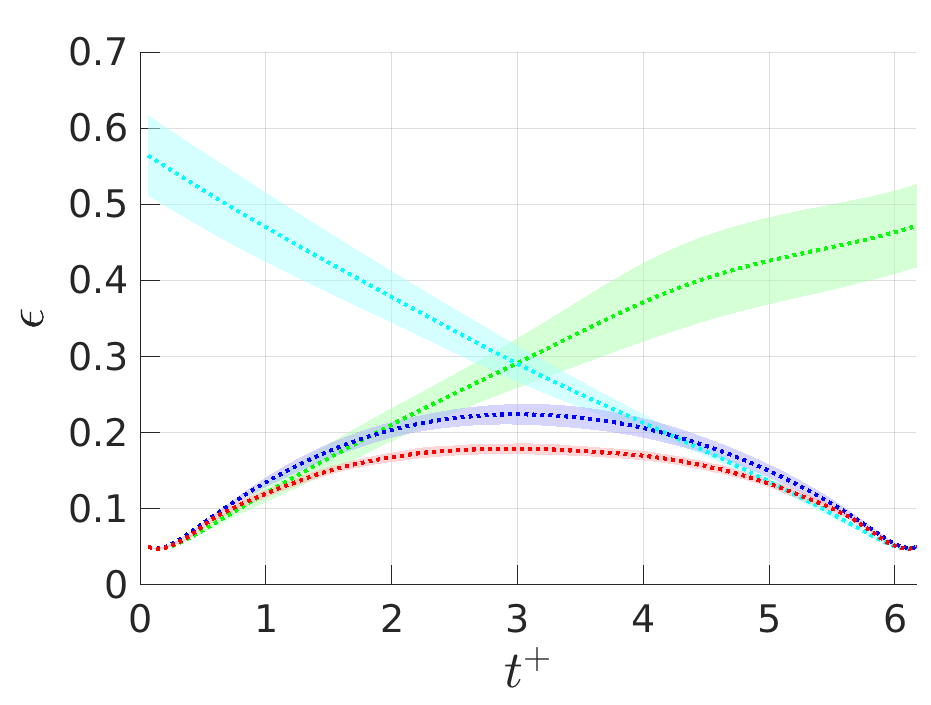}
        \end{minipage}\hfill
        \begin{minipage}[MU]{0.3\linewidth}
        \centering
        (e)~MU  ($m = 16$)
        \vfill
        \includegraphics[width=\linewidth]{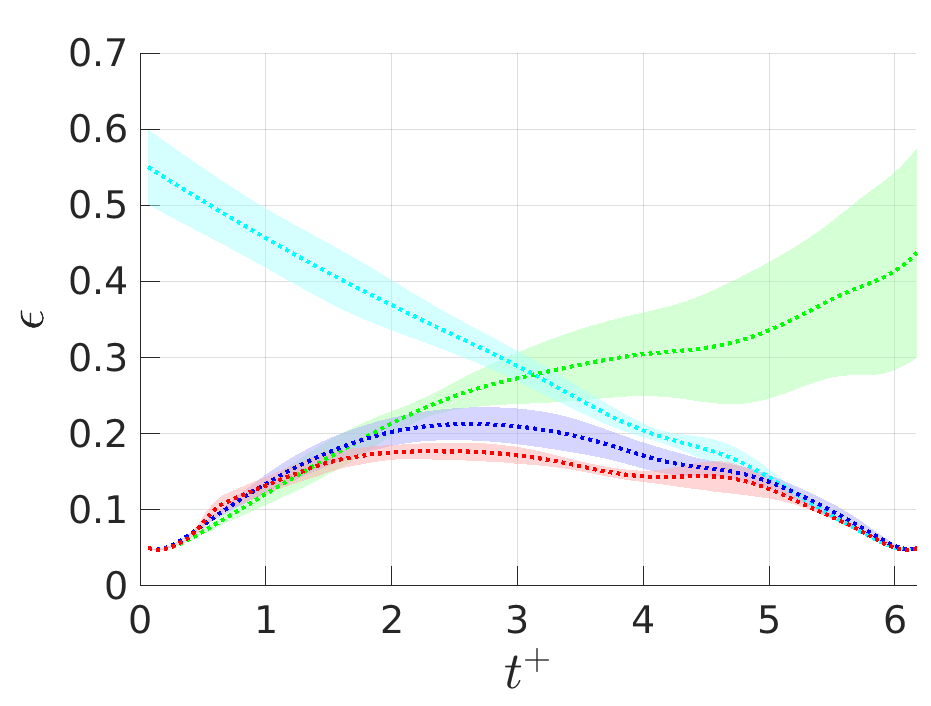}
        \end{minipage}\hfill
        \begin{minipage}[QR]{0.3\linewidth}
        \centering
        (f)~QR  ($m = 16$)
        \vfill
        \includegraphics[width=1.25\linewidth]{QR1Sigmap16.png}
        \end{minipage}
\end{center}
\captionof{figure}{ Mean and 1-$\sigma$ bounds for RMSE of flow reconstruction from 72 realizations using different sensor arrangements: (a)--(c) shows RMSE under UD, MU, and QR placement with 9 sensors, and (d)--(f) shows RMSE under UD, MU, and QR placement with 16 sensors. }
\label{fig:sensorCompRMSE}
\end{minipage}};
\end{tikzpicture}
\end{figure}

In Fig.~\ref{fig:sensorCompRMSE}, we report the RMSE for reconstruction using UD, MU, and QR sensor arrangements corresponding to $m=9$ and $m=16$ fast sensors.
Results are reported based on a Monte Carlo study with 72 realizations.
The mean RMSE from this Monte Carlo study is plotted for each case using a dotted line, and the shaded regions correspond to the 1-$\sigma$ (i.e.,~one standard deviation) bounds in the reconstruction resulting from the 72 realizations.
From these results, it is evident that a larger number of sensors results in a lower RMSE, regardless of the sensor arrangement---i.e.,~$m=16$ sensors yields a lower RSME on average than $m=9$ sensors for a given sensor placement strategy.
Similarly, a larger number of sensors tends to yield a lower variance as well.
This is especially evident in the QR placement results.
Our general finding that more sensors improves performance, both in terms of reducing RMSE and reducing variance, is to be expected.
However, it is interesting to note that the reconstruction variance is found to be least sensitive to the number of sensors when the UD sensor placement strategy is employed.
This indicates the UD sensor arrangement is less sensitive to the flow conditions and noise than the other sensor arrangements.
We also note that for both the UD and MU sensor arrangements, a spatiotemporal weighting scheme within the smoother improves the RMSE and variance of the reconstruction relative to a temporal weighting scheme. %
In contrast, in the case of the QR sensor arrangements, both the RMSE and variance of the reconstruction tend to be lower when a temporal weighting scheme is used, as opposed to a spatiotemporal weighting scheme, in the smoothing algorithm; however, this performance difference is less pronounced in the case of $m=16$ than it is for $m=9$.
The results here suggest that future work on determining an optimal sensor arrangement should consider the balance between RMSE and variance in the reconstruction error, and the sensitivities of each to specific sources of uncertainty.

\afterpage{\clearpage}

\begin{figure}[h!]
\begin{minipage}[StrCompErr]{0.45\textwidth}
        \centering
        (a)~$u_1$, $\hat{u}_1$ 
        \vfill
        \includegraphics[width=\textwidth]{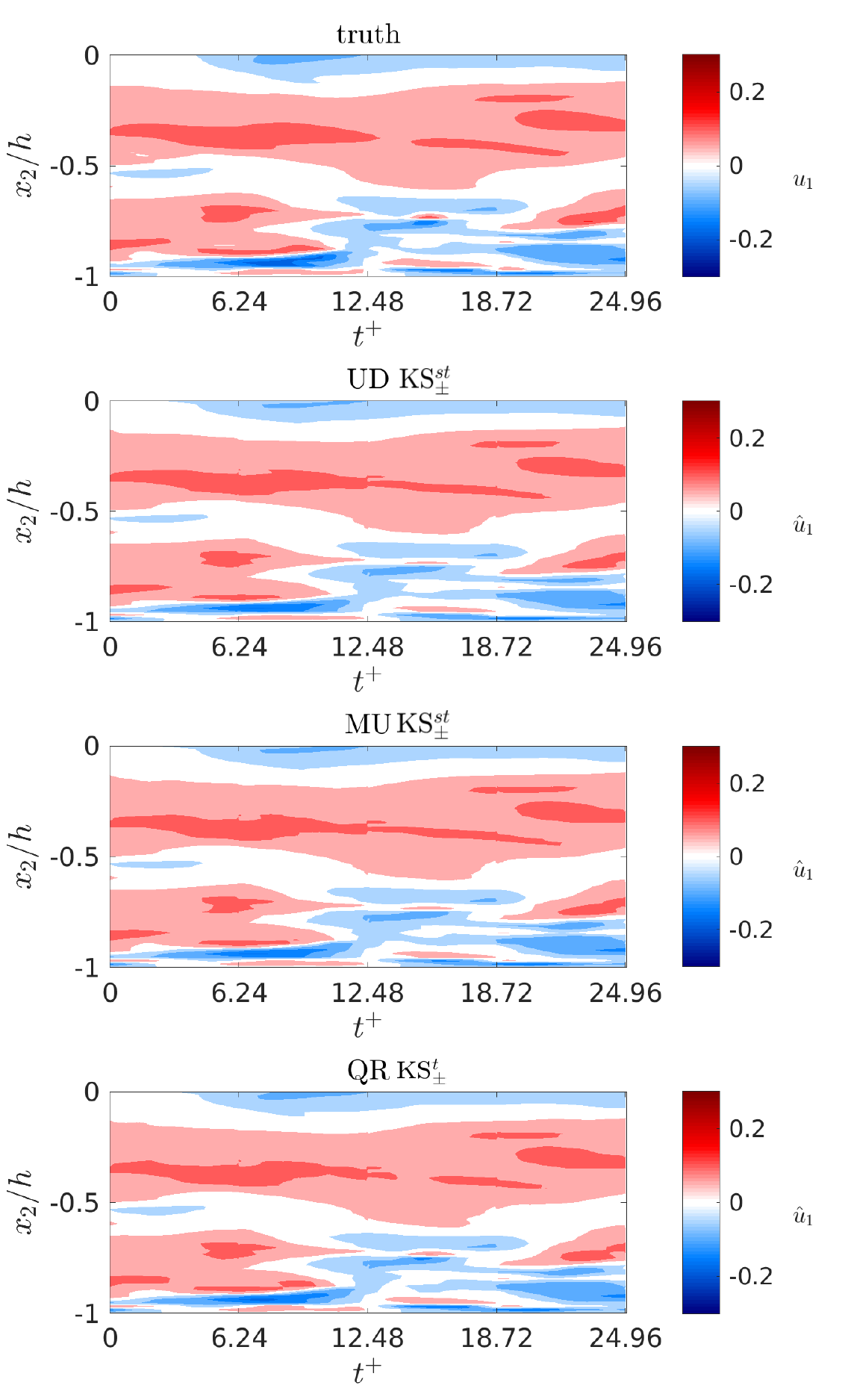}
        \end{minipage}\hfill
        \begin{minipage}[wlnCompErr]{0.45\textwidth}
        \centering
        (b)~$u_2$, $\hat{u}_2$
        \vfill
        \includegraphics[width=\textwidth]{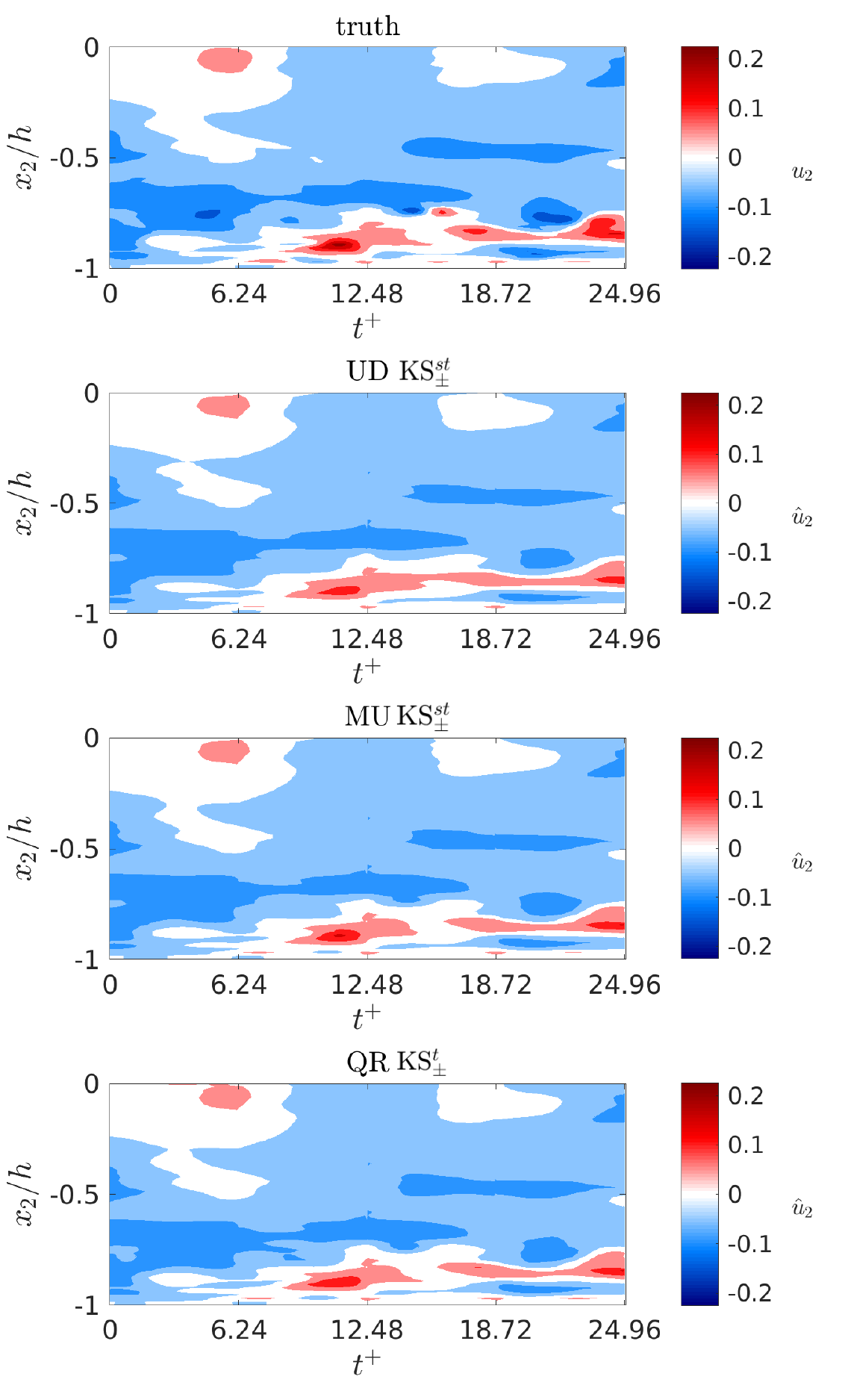}
        \end{minipage}\\
\begin{minipage}[StrCompErr]{0.45\textwidth}
        \centering
        (c)~$\hat{u}_1-u_1$
        \vfill
        \includegraphics[width=\textwidth]{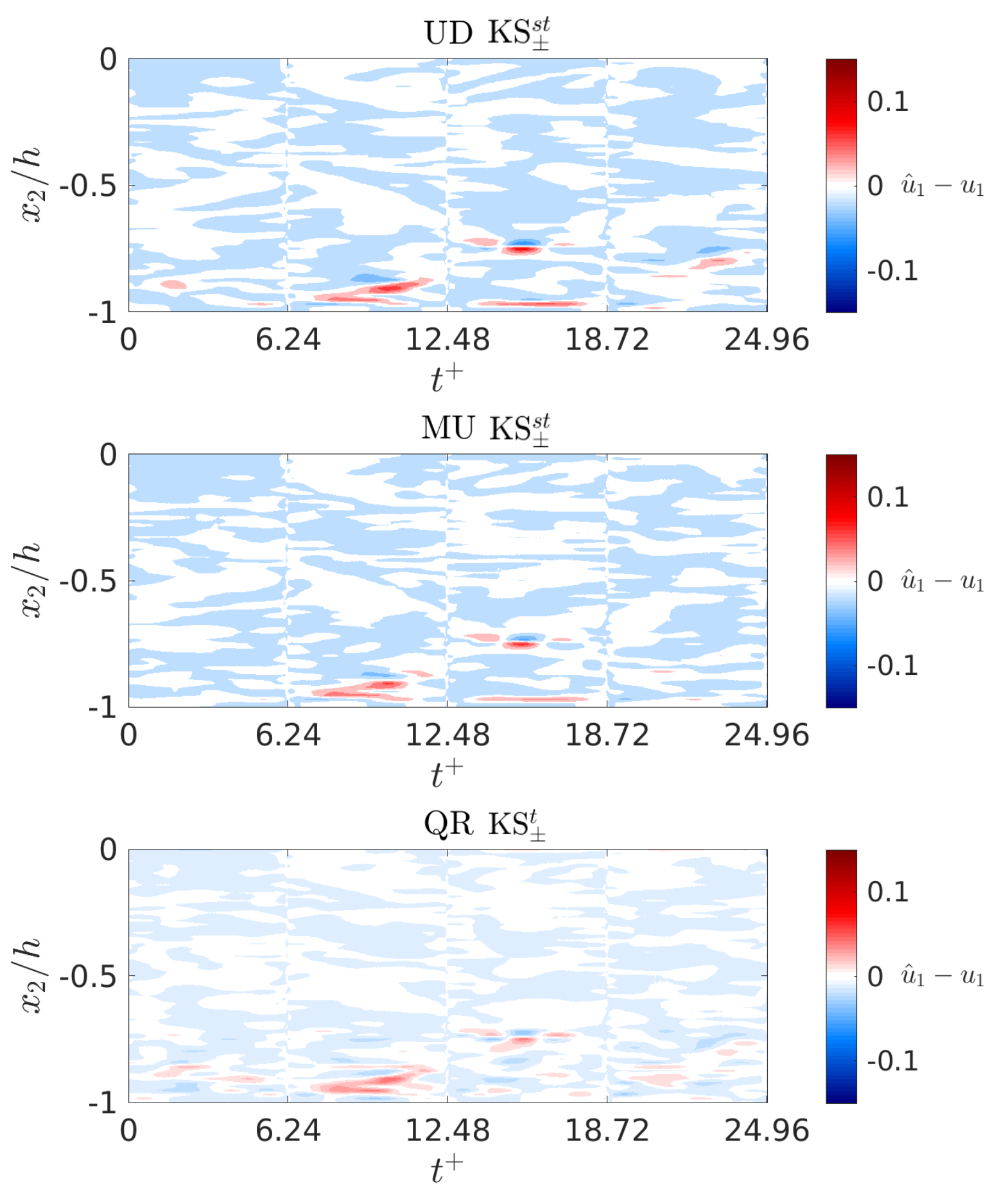}
        \end{minipage}\hfill
        \begin{minipage}[wlnCompErr]{0.45\textwidth}
        \centering
        (d)~$\hat{u}_2-u_2$
        \vfill
        \includegraphics[width=\textwidth]{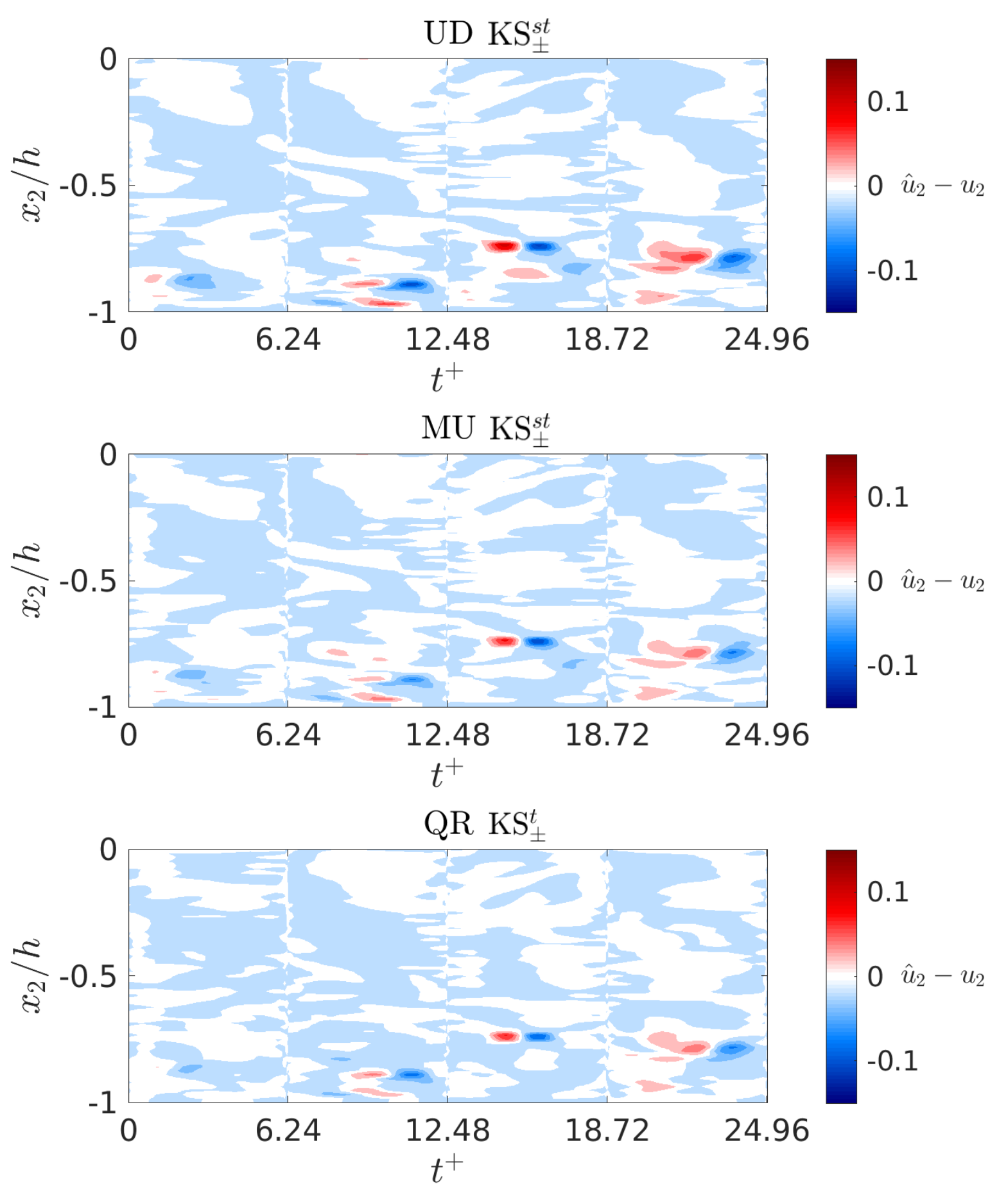}
        \end{minipage}
    \caption{The reconstructed flow field from multi-sensor fusion with UD, MU, and QR sensor arrangements compared with JHTDB ground truth along a vertical slice of the channel at $x_1 = 0.5h$. Reconstruction of streamwise velocity fluctuations are shown in~(a) and of wall-normal velocity fluctuations in~(b). Tiles~(c) and~(d) show the corresponding reconstruction errors relative to the JHTDB ground truth.}
    \label{fig:SensorCompErr}
\end{figure}

The results of the Monte Carlo study described here indicate that sensor fusion based on the QR sensor arrangement will yield the best reconstruction performance with a temporal weighting scheme in the smoother; whereas, for UD and MU sensor arrangements,  a spatiotemporal weighting scheme will yield the better performance. 
In Fig.~\ref{fig:SensorCompErr}(a)--(b), we report representative results for the reconstructed velocity-fluctuation-field based on each of these three sensor arrangements with $m=16$ and the corresponding weighting scheme yielding the best reconstruction performance.
It is evident that the reconstructed flow field from all three approaches captures the dominant flow features observed in the JHTDB ground truth.
The QR sensor arrangement appears to yield a better match to the ground truth than either the UD or MU arrangements, especially for the streamwise component of velocity.
Reconstructions from both the UD and MU arrangements appear strikingly similar to each other. 
To better quantify the performance, we report 
the associated reconstruction error relative to the JHTDB ground truth in Fig.~\ref{fig:SensorCompErr}(c)--(d).
These results confirm that the QR sensor arrangement outperforms the other two sensor arrangements by yielding a lower magnitude of reconstruction error.
To explain this difference, recall that the QR placement procedure (see Appendix~\ref{Sec:Sensor}) made use of a tailored basis of POD modes---extracted from 200 snapshots of the flow field---to determine the optimal arrangement of fast point sensors.
Although the snapshots used to determine the POD basis are distinct from the snapshots used to evaluate the sensor fusion algorithms, it still stands that the tailored basis of POD modes allows the QR pivoting procedure to determine the sensor locations that will best capture fluctuations in the turbulent kinetic energy. 
Indeed, in the next section we will investigate this idea further, and show that the QR sensor arrangement also outperform the other placement strategies in terms of capturing the second order statistics of the turbulent fluctuations.
In contrast, the UD and MU placement strategies do not consider the statistical characteristics of the turbulent fluctuations.
The UD strategy takes no physics into account at all.
The MU method only considers placement of sensors into high-error regions of the flow based on the model-based reconstruction.
The variance of these errors was not considered for the MU placement, but may be a useful heuristic to consider for fast point sensor placement in future investigations.

\subsection{Second order statistics comparison}\label{Sec:StatPSD}

\begin{figure}[h!]
    \centering
    \includegraphics[width=\textwidth]{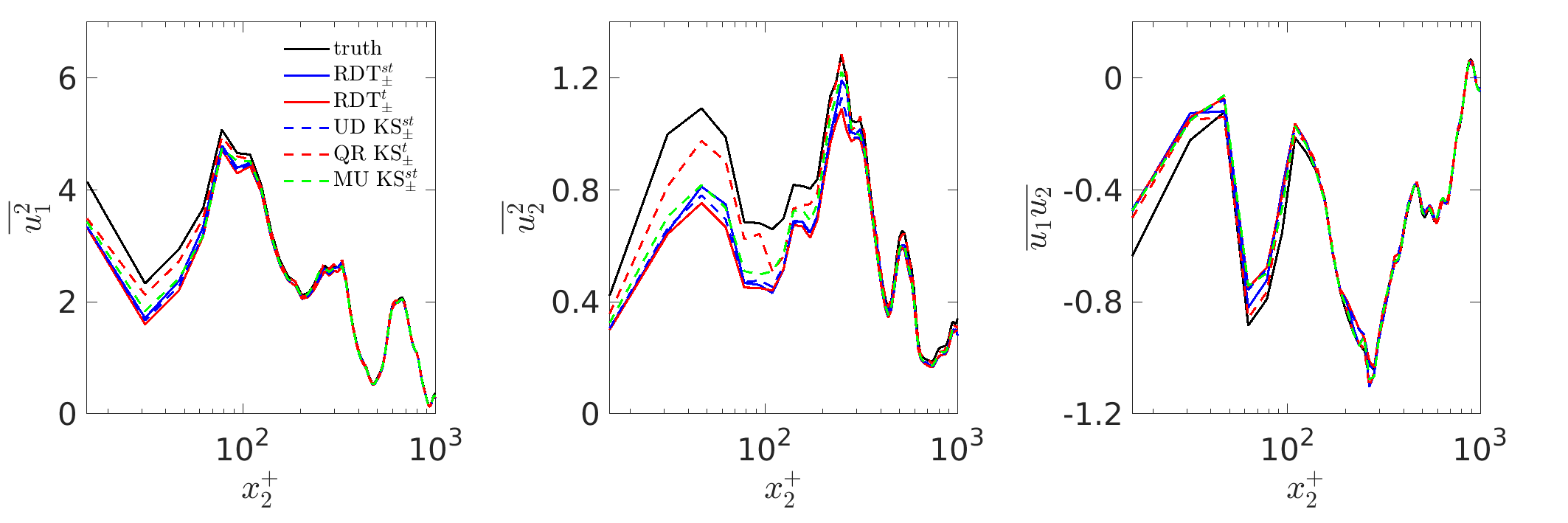}
    \caption{Comparison of reconstructed second-order statistics $\overline{u_1^2}$~(left), $\overline{u_2^2}$~(middle) and $\overline{u_1u_2}$~(right) spatially averaged along the wall-normal direction and over 72 realizations. 
    Note that the statistics do not resemble those for canonical channel flow due to the short time period being considered, i.e., they are not fully converged.
    }
    \label{fig:Stat}
\end{figure}

We now assess reconstruction performance of multi-sensor fusion in terms of capabilities for reproducing the second order statistics of the JHTDB ground truth flow.
In the previous section, we found that the QR arrangement of fast sensors modestly outperformed the UD and MU arrangements in terms of RMSE from a Monte Carlo study.
We expect that multi-sensor fusion based on the QR sensor arrangement will outperform reconstructions of the second order flow statistics based on UD and MU arrangement.
This is because the approach for determining the QR arrangement makes use of energy-optimal modes from the proper orthogonal decomposition~(POD).
Indeed, this turns out to be the case, as can be seen in Fig.~\ref{fig:Stat}.

Fig.~\ref{fig:Stat} shows the turbulence intensities $\overline{u_1^2}$ and ~$\overline{u_2^2}$ along with the Reynolds stress $\overline{u_1u_2}$ as a function of wall-normal station $x_2^+$ for the JHTDB ground truth and reconstructions based on model predictions (RDT$_\pm^{t}$, RDT$_\pm^{st}$) and multi-sensor fusion using on UD, MU, and QR sensor arrangements (UD KS$_\pm^{st}$, MU KS$_\pm^{st}$, QR KS$_\pm^{t}$). 
Far from the wall (approximately $x_2^+>100$), all methods capture the second order statistics comparably well.
Closer to the walls, the deviation between the reconstructed statistics and the ground truth is found to be larger.
This indicates that strong turbulence near the walls tends to degrade the reconstruction performance.
However, there is a notable difference in overall performance between multi-sensor fusion using the QR sensor arrangement and the other methods.
The temporally weighted smoother with QR sensor placement (dashed red line) demonstrates a markedly superior ability to capture the turbulence intensities $\overline{u_1^2}$ and $\overline{u_2^2}$ of the ground truth relative to the other reconstruction approaches for all $x_2^+$---exhibiting about half the deviation from the ground truth as the other approaches.
This is also true for reconstruction of the Reynolds stress $\overline{u_1 u_2}$; however, the ability to recover the Reynolds stress near the wall using the QR arrangement tends to be less pronounced.
In contrast, multi-sensor fusion with UD fast sensor arrangement yields comparable reconstruction of the second-order statistics as the model-prediction-based methods.

Considering the second-order statistical analysis of these three sensor arrangements with the corresponding weighting scheme yielding the best reconstruction performance, QR sensor placement with the temporal weighting scheme outperforms the other two placements with spatiotemporal weighting scheme, especially in reproducing the turbulence intensities. These results highlight the utility and related energy-optimal sensor placement strategies for turbulent flow reconstruction, even within the context of multi-sensor fusion.

\section{Conclusion}\label{Sec:conclusion}
In this paper, we have proposed and evaluated methods for model-based multi-sensor fusion for turbulent flow reconstruction.
We showed that non-time-resolved planar field measurements, time-resolved point measurements, and predictions from a simple physics-based model can be combined to reconstruct the velocity fluctuations in a turbulent channel flow from the Johns Hopkins Turbulence Database~(JHTDB).
This was achieved by using a ``fast filter'' to fuse fast point sensor measurements with RDT-based model predictions, then to use a ``slow filter'' to fuse these fast filter estimates with the non-time-resolved field measurements on a slower time-scale.
We further showed that the multi-rate filtering scheme can be used both forward and backward in time, and that physically motivated weighting schemes can be used to yield a smoothing algorithm with reduced error and variance in the flow reconstruction.
In addition, we found that these smoothing algorithms were also capable of reproducing the second order statistics of the underlying turbulent flow.

Monte Carlo simulations were used to investigate the statistical performance of the multi-sensor fusion methods.
The influence of the number and arrangement of fast point sensors was also investigated, with three sensor placement approaches studied: (1)~uniformly distributed~(UD) sensor placement, (2)~model-uncertainty-based~(MU) sensor placement, and (3)~pivoted QR sensor placement.
For all methods, reproduction quality improved with a larger number of sensors.
However, among the three methods,
the QR sensor placement strategy was found to have the best performance overall, yielding a lower RMSE and a better reproduction of the turbulence intensities.
The UD scheme is easier to implement \emph{a priori} and could also be applied in reconstruction with multi-rate PIV systems, e.g., high-rate low-resolution PIV and low-rate high-resolution PIV.

The results of our study draw attention to several avenues worthy of future research and investigation, especially with regards to enabling multi-sensor fusion for turbulent flow reconstruction in practice.
Firstly, the need for computationally efficient multi-sensor fusion algorithms cannot be understated.
The propagation of the covariance matrix within the slow and fast filters consumes the bulk of the computational time and effort.
The covariance matrix is of the order $N^2$, where $N$ is the state dimension.
Although the covariance matrix is symmetric---a structural property that can be exploited for computational speed-up---it stands that this will be the limiting computational bottleneck in larger-scale problems. 
In addition, sensor noise models used in this study were relatively simplified.
Future investigations must consider the role of sensor noise that are representative of the specific instrumentation being employed for flow reconstruction.
For example, the measurement uncertainty associated with the velocity field measured using a PIV system will necessarily be a function of the velocity magnitude.
Thus, simple additive Gaussian noise models for sensor noise do not account for these more realistic uncertainty profiles---even though they are standard choices for algorithm development in the literature on flow reconstruction. 
Finally, the role of the number and arrangement of fast sensors is central to achieving successful flow
reconstruction.
As we have seen in this study, some arrangements are better suited for reducing RMSE and improving the reconstruction of second-order statistics (e.g.,~pivoted QR placement); however, we have also seen that some arrangements yield lower variance in the flow reconstruction and thus exhibit less sensitivity/greater robustness to operating conditions (e.g.,~uniformly distributed sensors).
Further the weighting schemes used for smoothing between the forward and backward filter estimates did not directly account for the placement of sensors.
Future work must consider the placement of sensors and the weighting functions used for smoothing simultaneously.
We expect that by doing so, multi-sensor fusion will yield great improvements in flow reconstruction performance, in terms of minimizing reconstruction errors and variances, but also in terms of capturing a wider range of turbulent physics.

\begin{appendices}
\renewcommand{\thesection}{\Alph{section}}
\section{Sensor placement}\label{Sec:Sensor}

In this study, three different approaches are investigated for determining the arrangement of fast point sensors---encoded in the term $\Sp$---for multi-sensor fusion: (1)~uniformly distributed~~(UD) sensor placement, (2)~model-uncertainty-based~(MU) sensor placement, and (3)~pivoted QR sensor placement.
Each of these is described here.

\begin{figure}[ht!]
    \begin{center}
        \subfloat[UD $m = 9$]{
        \includegraphics[width=.48\textwidth]{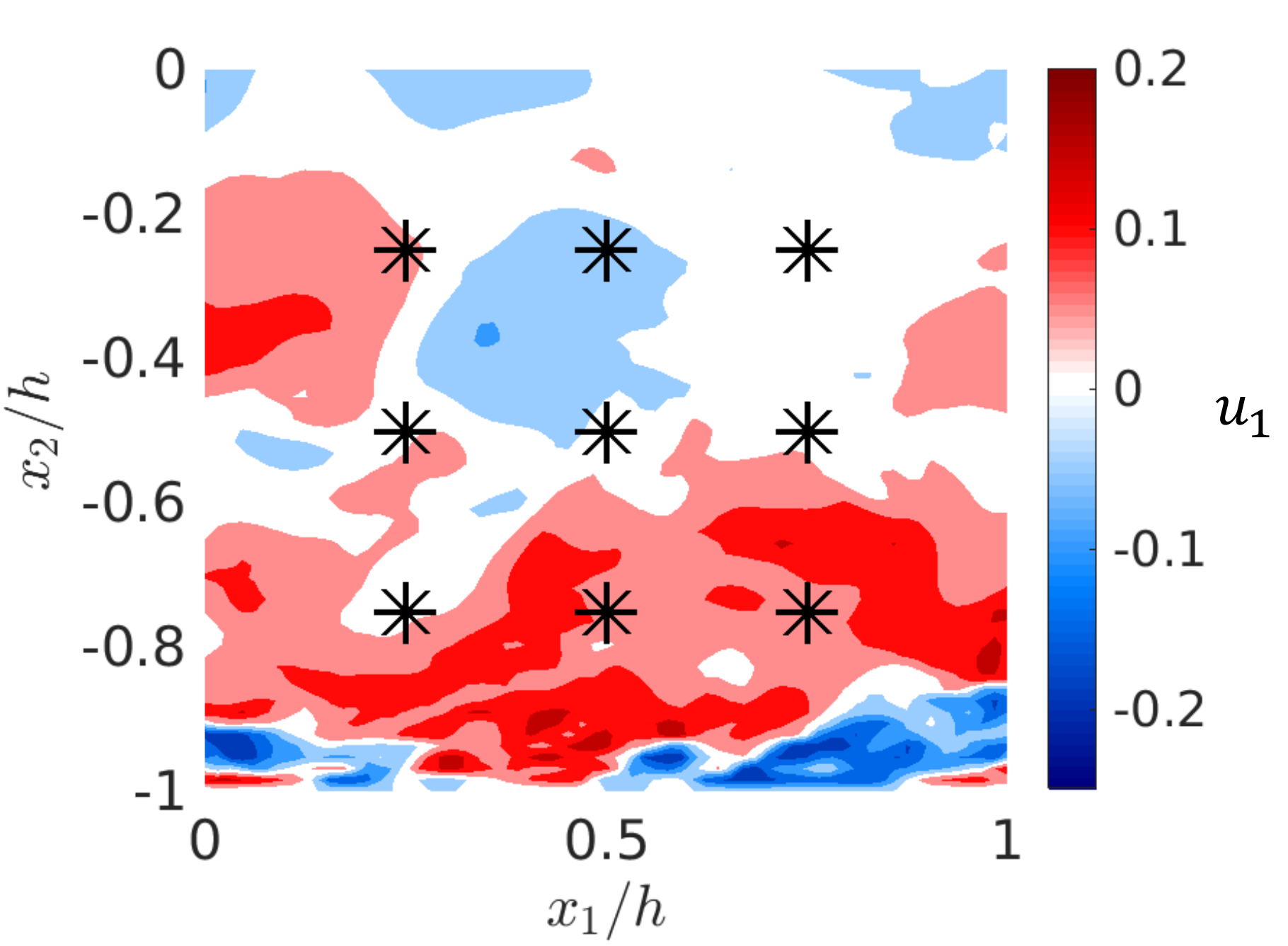}}
        \subfloat[UD $m = 16$]{
        \includegraphics[width=.48\textwidth]{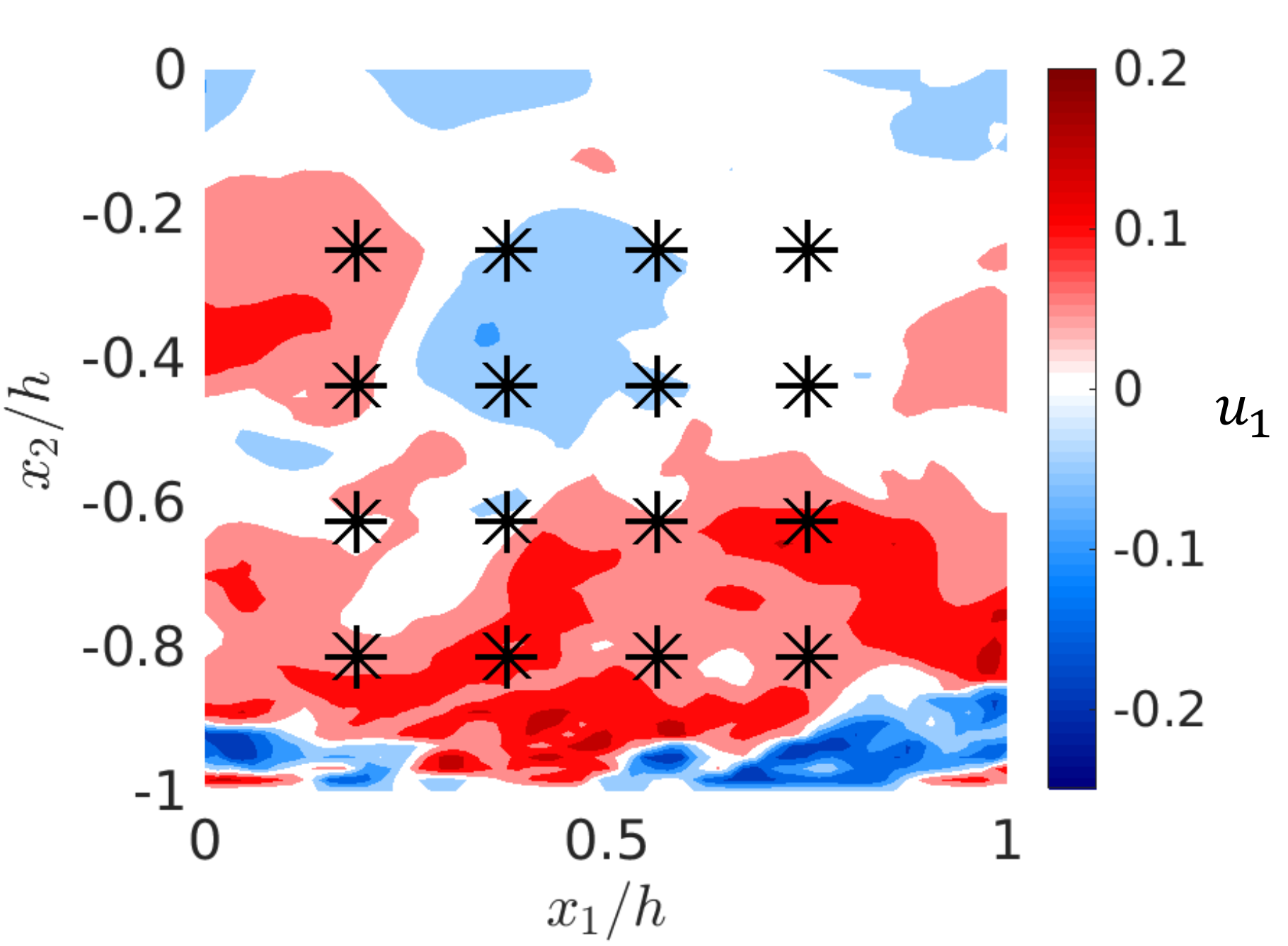}}  
    \end{center}
    \caption{Uniformly distributed (UD) sensor arrangements with $m=9$ and $m=16$ fast point sensors---drawn as black stars---overlaid on an arbitrary snapshot of streamwise velocity fluctuations.
    }
    \label{fig:UDsensor}
\end{figure}

The UD placement approach is the simplest of the three approaches considered. UD placement yields a uniform distribution of $m$ fast point sensors throughout the domain.
Sensors are evenly spaced from each other and from the domain boundaries along both the $x_1$ and $x_2$ directions.
Although this approach does not take into account any flow physics, it provides a reasonable benchmark for comparison.
The same UD approach was also considered in~\cite{van2015bayesian}.
In order to maintain symmetry in the resulting placement, the number of sensors $m$ is selected to be a perfect square.
Here, we consider only $m=9$ and $m=16$, which are values determined for by the QR placement approach---to be described momentarily---and used here for direct comparison.
The UD sensor placements for these two cases are plotted as black stars in Fig.~\ref{fig:UDsensor}, overlaid on a snapshot of the streamwise velocity fluctuations $u_1$ at an arbitrary instant in time.
%
%

\begin{figure}[ht!]
    \centering
    \includegraphics[width=0.6\textwidth]{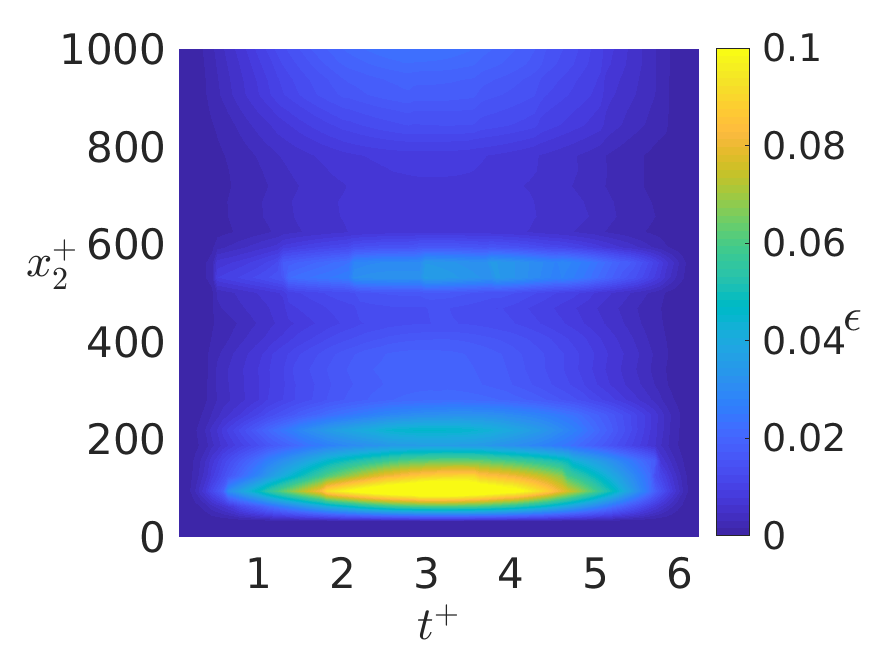}
    \caption{RDT-model-based reconstruction error (see Eq.~\ref{eqn:RMSEforRDT}) averaged over 50 realizations. The lower channel wall is located at $x_2^+ = 0$ and the center line is located at $x_2^+ = 1000$. The largest reconstruction error (the yellow area) arises at $x_2^+ \approx 100$, near the lower wall.}
    \label{fig:RDTfunctionError}
\end{figure}

\begin{figure}[ht!]
    \begin{center}
        \subfloat[MU $m = 9$]{
        \includegraphics[width=.48\textwidth]{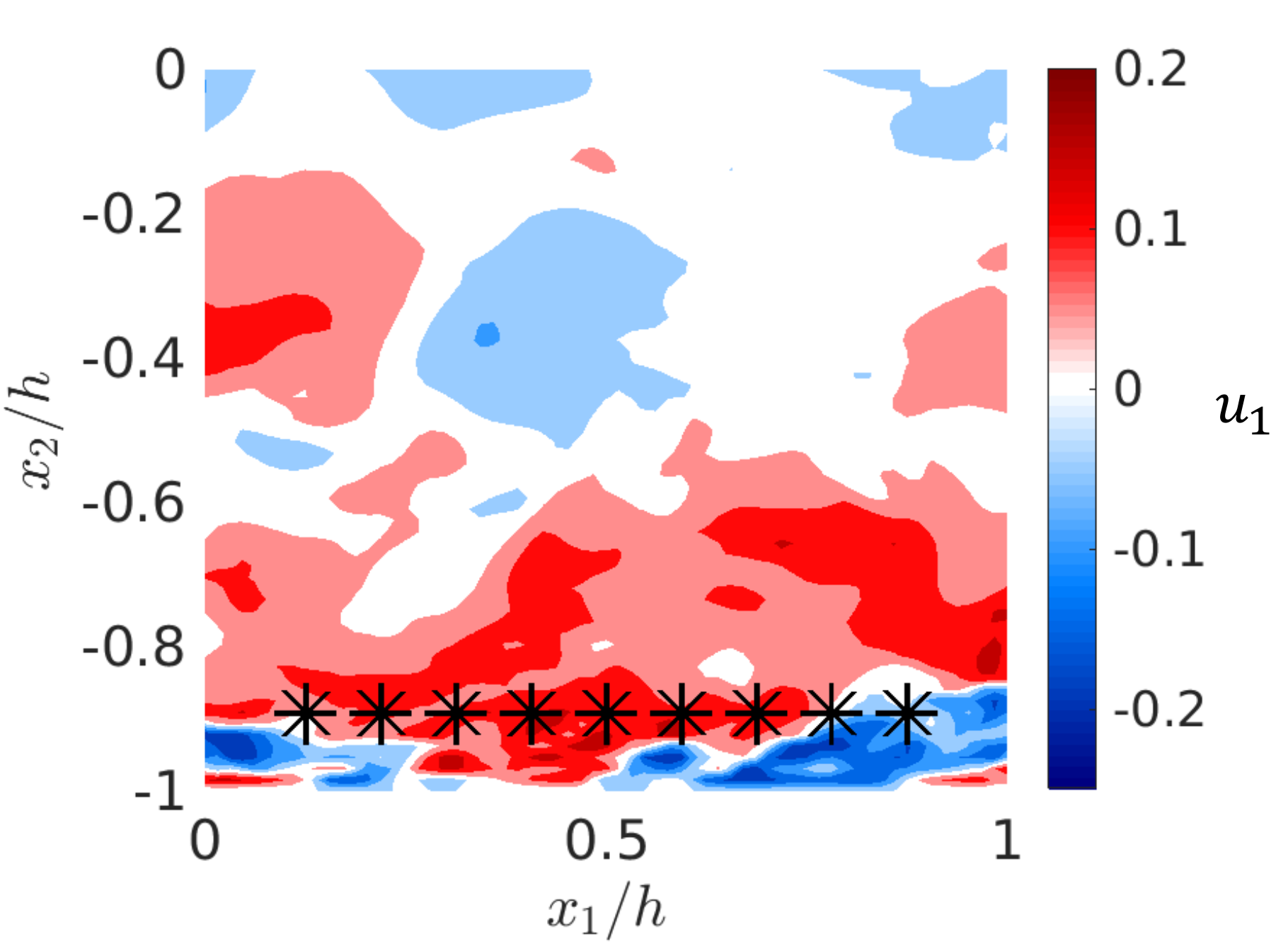}}
        \subfloat[MU $m = 16$]{
        \includegraphics[width=.48\textwidth]{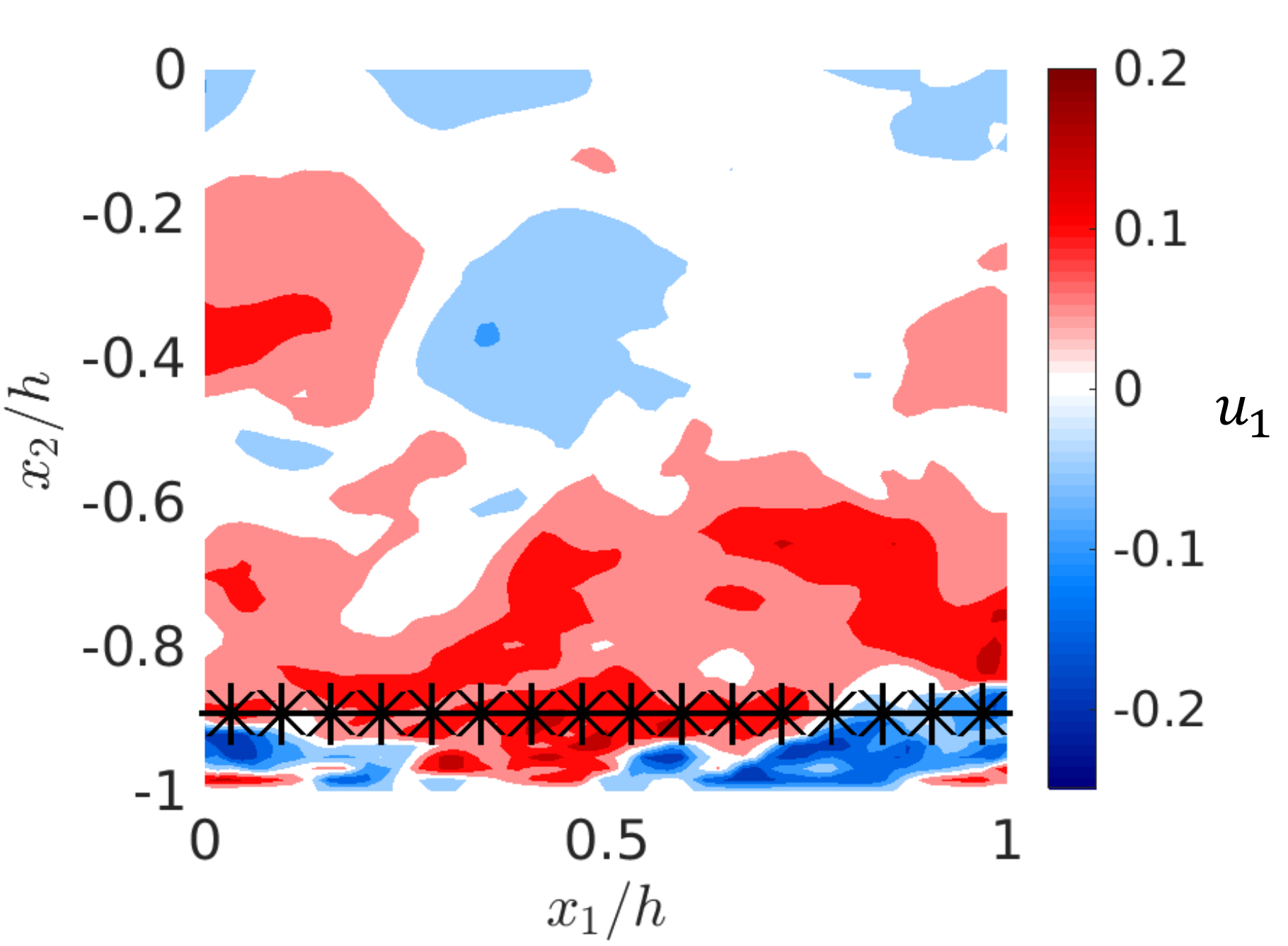}}  
    \end{center}
    \caption{Model-uncertainty-based (MU) sensor arrangements with $m=9$ and $m=16$ fast point sensors---drawn as black stars---overlaid on an arbitrary snapshot of streamwise velocity fluctuations.
    Fast point sensors are uniformly distributed in the streamwise direction at a wall-normal station of $x_2^+ = 100$, based on an analysis of the maximum model-based reconstruction error reported in Fig.~\ref{fig:RDTfunctionError}. }
    \label{fig:MUsensor}
\end{figure}

The UD method implicitly assumes that velocity fluctuations at all spatial points are equally important for flow reconstruction,
which is not true in general.
A simple means of addressing this shortcoming is to consider placing fast point sensors at locations with high reconstruction error, when a model-based reconstruction is used directly---i.e.,~a model-uncertainty-based (MU) placement strategy.
In Fig.~\ref{fig:RDTfunctionError}, we report wall-normal variation in error over time $\epsilon(x_2,t)$ between the RDT-based reconstruction and JHTDB ground truth averaged over 50 realizations of the flow.
The wall-normal variation in error over time is defined as
\begin{equation}
    \epsilon(x_2,t) = \frac{(\int_{x_1=0}^{h} \left(\left(u_1- \hat{u}_1\right)^2 + \left(u_2 - \hat{u}_2 \right)^2\right) dx_1)^{1/2}}{(\int_{x_1=0}^{h} \left((u_1)^2 + (u_2)^2 \right) dx_1)^{1/2}}.
    \label{eqn:RMSEforRDT}
\end{equation}
where $\hat{u}_1$ and $\hat{u}_2$ are the reconstructed velocity fluctuation in the streamwise and wall-normal components, and $u_1$ and $u_2$ are the velocity fluctuation components from the JHTDB ground truth.
From this analysis, we find that the peak reconstruction error arises in the near-wall region, roughly $x_2^+ \approx 100$ units from the lower channel wall.
Since this indicates a large uncertainty in the RDT-model-based reconstruction, we distribute $m$ fast point sensors uniformly along the streamwise direction at this wall-normal station, as shown in Fig~\ref{fig:MUsensor}.
%



\begin{figure}[ht!]
    \centering
    \includegraphics[width=0.5\textwidth]{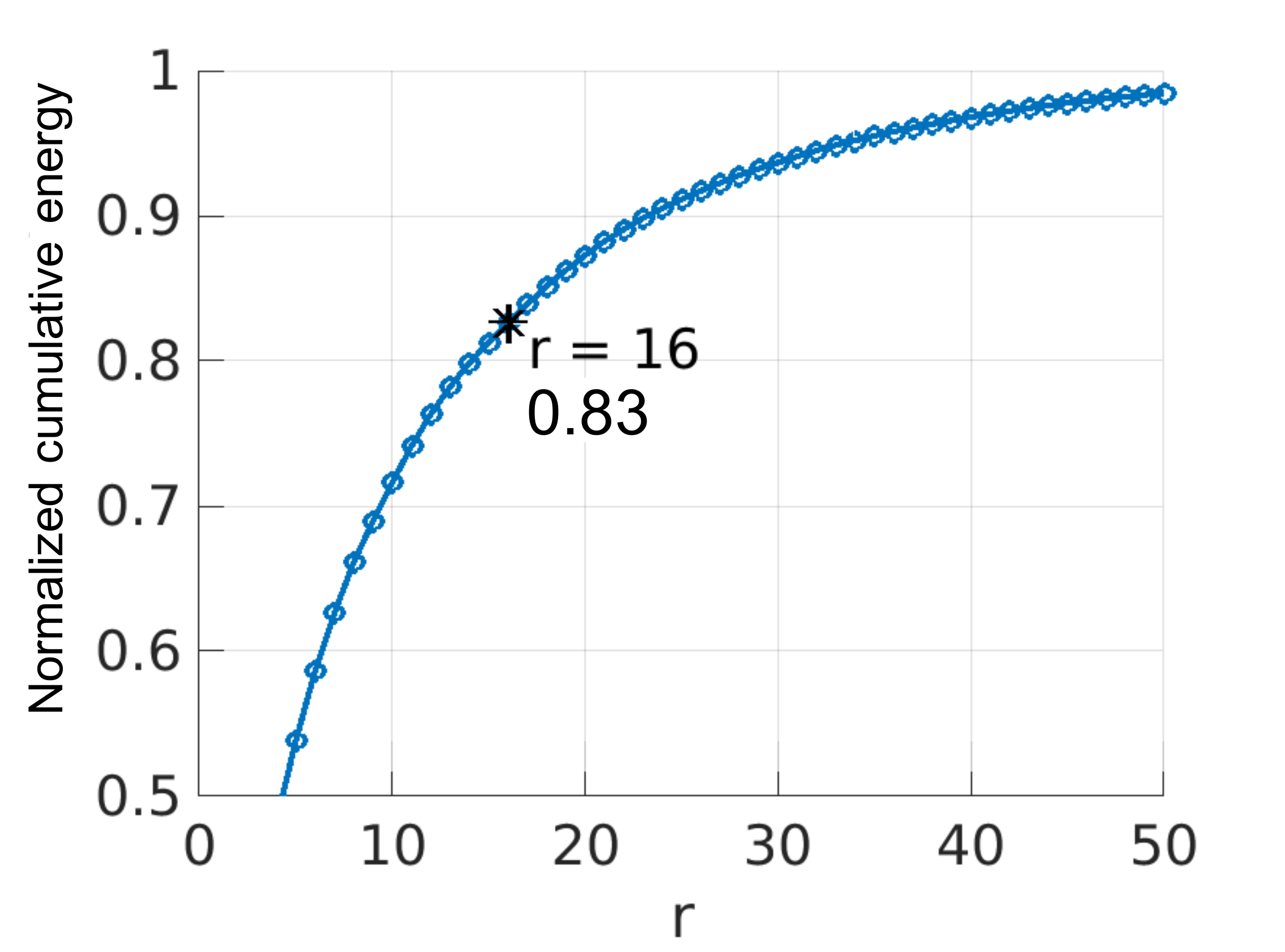}
    \caption{Cumulative contribution of the leading $r$ POD modes to the total energy of the training data. The leading 16 POD modes capture $83\%$ of the energy in the training data, and the leading 9 POD modes capture 68\%.}
    \label{fig:EnergyPerc}
\end{figure}

The final approach considered here is the sparse sensor placement algorithm based on a pivoted QR decomposition, as presented in~\cite{manohar2018data}. 
The QR approach uses snapshots of the flow field to obtain a tailored (reduced) basis of POD modes for capturing dominant signals in the flow.
Then, a column-pivoted QR factorization is used to construct $\Sp$, with the leading $m$ column-pivots indicating the $m$ sensor locations that will best approximate the training data in a least-squares sense.
The specific formulation and additional details of the approach can be found in~\cite{manohar2018data}. 
We emphasize that this QR approach requires training data for computing a tailored basis of POD modes, but that temporally resolved velocity field data will not be available in practice: this was the motivation for multi-sensor fusion in the first place.
As such, we collect 200 slow-in-time snapshots of the flow field to extract a POD basis.
The marginal contribution of each new POD mode to the energy in the training data is reported in Fig.~\ref{fig:EnergyPerc}.
This cumulative energy analysis indicates that a tailored basis of $r=16$ POD modes captures roughly 83\% of the energy.
Thus, a QR-based sensor placement on this tailored basis will require $m=16$ sensors.
We also consider a tailored basis of $r=9$ POD modes (captured using $m=9$ sensors) for comparison with the UD approach, which requires the number of sensors to be a perfect square to maintain symmetry.
We note that the specific sensor arrangements resulting from the QR approach were found to be sensitive to the training data.
This sensitivity was observed to be more significant with a larger number of sensors.
However, even with sensitivities and variations in the sensor arrangements, a larger number of sensors was found to improve flow reconstruction so long as appropriate filter tuning was performed within the multi-sensor fusion framework.
Due to these additional complexities, we only consider the cases of $m=9$ and $m=16$ sensors in this study. 

\begin{figure}[ht!]
    \begin{center}
        \subfloat[QR $m = 9$]{
        \includegraphics[width=.48\textwidth]{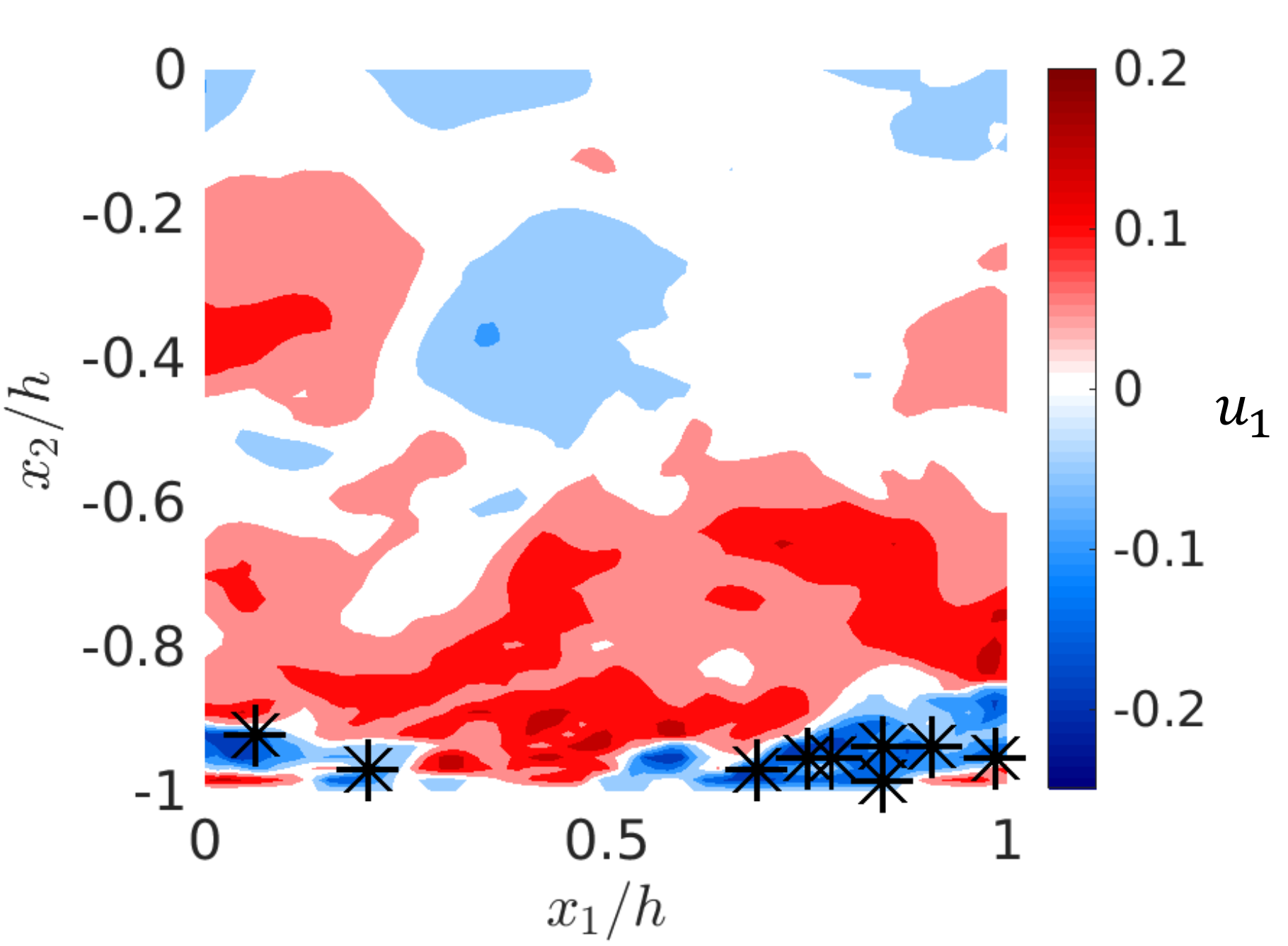}}
        \subfloat[QR $m = 16$]{
        \includegraphics[width=.48\textwidth]{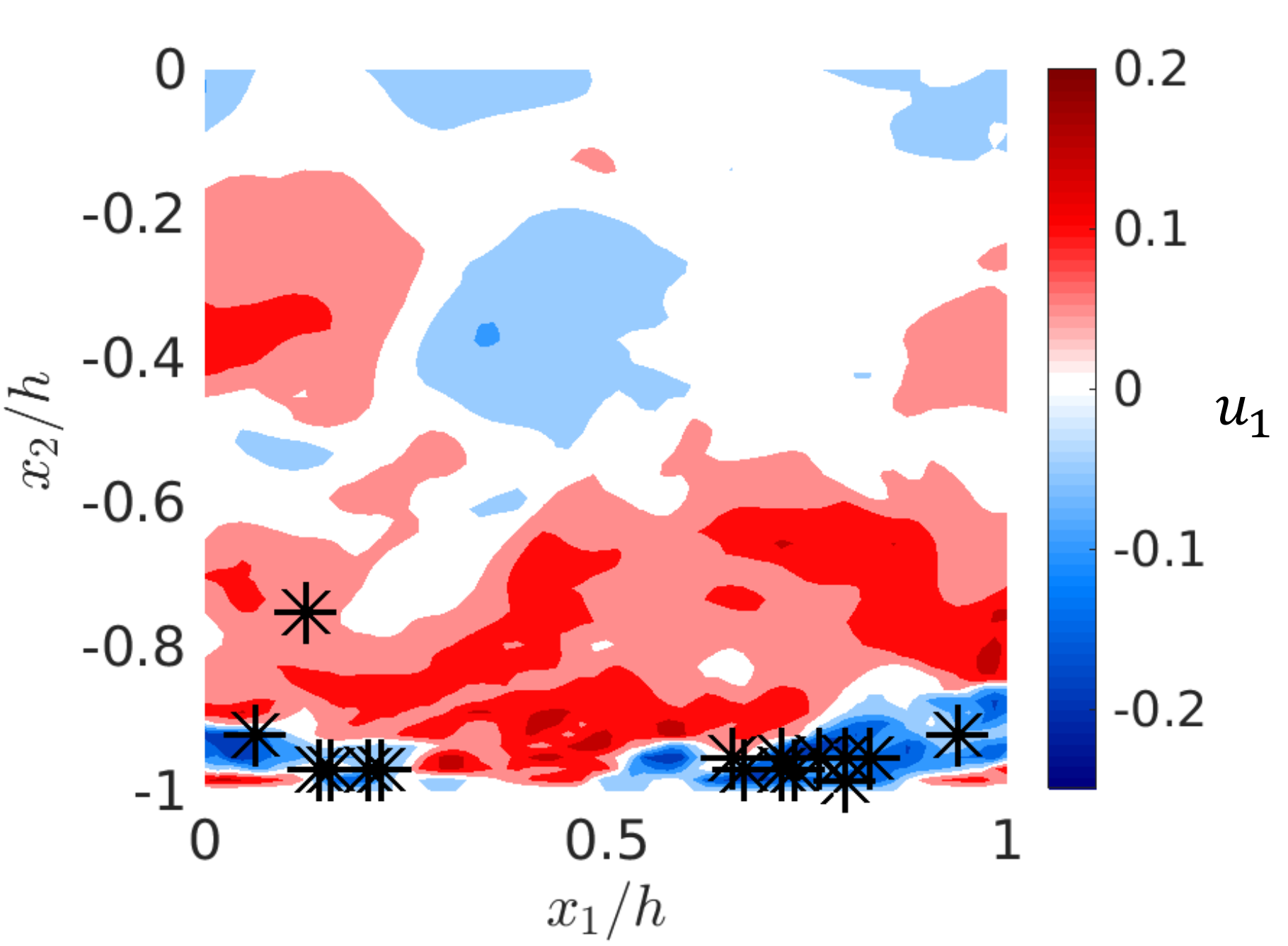}}  
    \end{center}
    \caption{
    QR sensor arrangements with $m=9$ and $m=16$ fast point sensors---drawn as black stars---overlaid on an arbitrary snapshot of streamwise velocity fluctuations.
    Sensor locations are chosen to capture maximal energy in the training data with a tailored basis of POD modes (see Fig.~\ref{fig:EnergyPerc}).}
    \label{fig:QRsensor}
\end{figure}

The sparse sensor arrangements resulting from the subsequent column-pivoted QR procedure for $m=9$ and $m=16$ sensors are shown in Fig.~\ref{fig:QRsensor}.
Most of the sensors in these arrangements are placed in the vicinity of the lower wall.
This confirms that the streamwise velocity fluctuations are a dominant feature of this flow that needs to be captured for flow reconstruction. 
Interestingly, near wall information was also found to be important using the MU placement approach.
However, the MU approach resulted in a uniform streamwise distribution of sensors at a wall-normal station of $x_2^+\approx 100$; in contrast, the QR placement tends to yield sensor locations with $x_2^+<100$ and an irregular streamwise distribution of sensors, which would not have been obtained using simple heuristics. 
 
\end{appendices}


%
%

\typeout{} 
\bibliographystyle{spmpsci}      
\bibliography{template}   

%



%
\setlength{\nomlabelwidth}{3cm}
\makenomenclature

\mbox{}

\nomenclature[28]{$\boldsymbol{u}=[u_1,u_2,u_3]$}{Turbulent flow velocity fluctuation}
\nomenclature[18]{$\boldsymbol{U}=[U(x_2),0,0]$}{ Wall-bounded turbulent flow mean profile}
\nomenclature[23]{$p$}{pressure fluctuation}
\nomenclature[33]{$(x_1,x_2,x_3)$}{Three dimensional coordinate system}
\nomenclature[24]{$q$}{System states}
\nomenclature[34]{$y_f$}{``Fast'' point measurements}
\nomenclature[35]{$y_s$}{``Slow'' field measurements}
\nomenclature[14]{$\mathbb{S}$}{Subsampling matrix from states to point measurements}
\nomenclature[01]{$\Delta t_f^{+}$}{Non-dimensional sampling time of ``fast'' point measurements}
\nomenclature[02]{$\Delta t_s^{+}$}{Non-dimensional sampling time of ``slow'' field measurements}
\nomenclature[19]{$\ell$}{Sampling time ratio of field measurements to point measurements}
\nomenclature[13]{$N$}{State dimension}
\nomenclature[20]{$m$}{Number of point measurements}
\nomenclature[09]{$A_-$}{System dynamics operator backward in time}
\nomenclature[10]{$A_+$}{System dynamics operator forward in time}
\nomenclature[32]{$w\sim \mathcal{N}(0, Q)$}{Process noise with Gaussian distribution of zero mean and covariance matrix $Q$}
\nomenclature[31]{$v_s\sim \mathcal{N}(0, R_s)$}{Field measurement noise with Gaussian distribution of zero mean and covariance matrix $R_s$}
\nomenclature[30]{$v_f\sim \mathcal{N}(0, R_f)$}{Point measurement noise with Gaussian distribution of zero mean and covariance matrix $R_f$}
\nomenclature[11]{$G_+$}{Weighting factor of the forward estimate}
\nomenclature[10]{$G_-$}{Weighting factor of the backward estimate}
\nomenclature[12]{$L_{x_1}$}{Streamwise window size of the PIV snapshot.}
\nomenclature[18]{$h$}{Half channel height}
\nomenclature[17]{$T^+$}{Total time duration of reconstruction}
\nomenclature[27]{$t^+$}{DNS time step}
\nomenclature[29]{$u_\tau$}{Friction velocity}
\nomenclature[08]{$\nu$}{Kinematic viscosity}
\nomenclature[05]{$\epsilon$}{2-D Root mean square error~(RMSE)}
\nomenclature[06]{$\epsilon_{x_1}$}{Streamwise RMSE}
\nomenclature[07]{$\epsilon_{x_2}$}{Wall-normal RMSE}
\nomenclature[26]{$\hat{q}_+$}{Forward-in-time estimate of states}
\nomenclature[25]{$\hat{q}_-$}{Backward-in-time estimate of states}
\nomenclature[21]{$n_{x_1}$}{Streamwise grid point of PIV snapshots}
\nomenclature[22]{$n_{x_2}$}{Wall-normal grid point of PIV snapshots}
\nomenclature[03]{$\Delta x_1^+$}{Non-dimensional streamwise spatial resolution}
\nomenclature[04]{$\Delta x_2^+$}{Non-dimensional wall-normal spatial resolution}
\nomenclature[15]{$SNR_f$}{Point measurement signal to noise ratio}
\nomenclature[16]{$SNR_s$}{Field measurement signal to noise ratio}

\printnomenclature

\end{document}